\documentclass[a4paper,11pt]{article}
\pdfoutput=1 

\usepackage{jinstpub}

\usepackage{lineno}

\usepackage{graphicx}
\usepackage{subcaption}
\usepackage{booktabs} 
\usepackage[export]{adjustbox}
\usepackage{array}
\newcolumntype{P}[1]{>{\centering\arraybackslash}p{#1}}
\usepackage{floatrow}
\graphicspath{ {./images/} }

\usepackage[utf8]{inputenc}
\usepackage{siunitx}
\usepackage{gensymb}
\usepackage{wasysym}
\usepackage{upgreek}

\title{Performance of a First Full-Size WOM-Based Liquid Scintillator Detector Cell as Prototype for the SHiP Surrounding Background Tagger}

\author[f]{J.~Alt}
\author[k]{, O.~Bezshyyko}
\author[m]{, M.~B\"ohles}
\author[b,*]{, A.~Brignoli}
\author[b]{, A.~Conaboy}
\author[m]{, P.~Deucher}
\author[b]{, C.~Eckardt}
\author[b]{, A.~Ernst}
\author[f]{, H.~Fischer}
\author[m,*]{, A.~Hollnagel}
\author[f]{, M.~Jadidi}
\author[b,*]{, H.~Lacker}
\author[f,*]{, F.~Lyons}
\author[f]{, T.~Molzberger}
\author[f]{, S.~Ochoa}
\author[k]{, V.~Orlov}
\author[b]{, A.~Reghunath}
\author[a]{, F.~Rehbein}
\author[j]{, M.~Schaaf}
\author[b]{, C.~Scharf}
\author[b]{, J.~Schmidt}
\author[f]{, M.~Schumann}
\author[b]{, A.~Vagts}
\author[m]{, M.~Wurm}

\emailAdd{brignoli@physik.hu-berlin.de}
\emailAdd{annika.hollnagel@uni-mainz.de}
\emailAdd{lacker@physik.hu-berlin.de}
\emailAdd{fairhurst.lyons@physik.uni-freiburg.de}

\affiliation[a]{Rheinisch-Westf\"alische Technische Hochschule Aachen, Physikalisches Institut 3A, 52074 Aachen, Germany}
\affiliation[b]{Humboldt-Universit\"at zu Berlin, Institut f\"ur Physik, 12489 Berlin, Germany}
\affiliation[f]{Albert-Ludwigs-Universit\"at Freiburg, Physikalisches Institut, 79104 Freiburg, Germany}
\affiliation[j]{Forschungszentrum Jülich GmbH, Zentralinstitut für Engineering, Elektronik und Analytik -- Engineering und Technologie, 52428 Jülich, Germany}
\affiliation[m]{Johannes Gutenberg-Universit\"at Mainz, Institut f\"ur Physik \& Exzellenzcluster PRISMA$^+$, 55128 Mainz, Germany}
\affiliation[k]{Taras Shevchenko National University, 01601 Kyiv, Ukraine}
\affiliation[*]{Corresponding authors}

\date{October 2023}

\abstract{As a prototype detector for the SHiP Surrounding Background Tagger (SBT), we constructed a cell ($120\, \mathrm{cm} \times 80\, \mathrm{cm} \times 25\, \mathrm{cm}$) made from corten steel that is filled with liquid scintillator (LS) composed of linear alkylbenzene (LAB) and 2,5-diphenyloxazole (PPO). The detector is equipped with two Wavelength-shifting Optical Modules (WOMs) for light collection of the primary scintillation photons. Each WOM consists of an acrylic tube that is dip-coated with a wavelength-shifting layer on its surface. Via internal total reflection, the secondary photons emitted by the molecules of the wavelength shifter are guided to a ring-shaped array of 40 silicon photomultipliers (SiPMs) coupled to the WOM for light detection. The granularity of these SiPM arrays provides an innovative method to gain spatial information on the particle crossing point. Several improvements in the detector design significantly increased the light yield with respect to earlier proof-of-principle detectors.

We report on the performance of this prototype detector during an exposure to high-energy positrons at the DESY II test beam facility by measuring the collected integrated yield and the signal time-of-arrival in each of the SiPM arrays. The resulting detection efficiency and reconstructed energy deposition of the incident positrons are presented, as well as the spatial and time resolution of the detector. These results are then compared to Monte Carlo simulations.}

\begin{document}
\maketitle

\section{Introduction}
\label{Sec:Introduction}
In the SHiP experiment proposed for a future CERN SPS Beam Dump Facility (BDF)~\cite{SHiP:2015vad,SHIP:2021tpn}, the large evacuated decay volume of the Hidden Sector (HS) detector is surrounded by a veto detector enabling the tagging of muons in the energy range between about 1 GeV and up to 400 GeV entering the vessel from the sides, as well as reactions of muons and neutrinos in the decay vessel walls and their vicinity. For active detector material, it is foreseen to surround the decay vessel with liquid scintillator (LS), since only this will be able to provide optimal hermeticity. The name of this veto detector is thus the SHiP Surrounding Background Tagger (SBT), or LS-SBT -- indicating the usage of liquid scintillator (LS). 
As a benchmark, the SBT should be able to detect with high efficiency the energy depositions of minimum ionising particles (MIPs) that have minimum path length through the liquid scintillator layer of the LS-SBT. Background rejection studies of muon or neutrino deep-inelastic reactions in SHiP have used this energy threshold for the LS-SBT and assumed a $>\SI{99}{\percent}$ detection efficiency for such an energy deposition~\cite{SHiP-LoI}. Additional requirements for the SBT are a time resolution in the nanosecond range and good spatial resolution in order to locate the origin of possible background reactions in time and space~\cite{SHiP-LoI}.
Future studies will explore how to further exploit the time and spatial information to identify the nature of background events (muon or neutrino-induced) which can help to validate the prediction of the background, and how to avoid the possible false veto of a true signal due to SBT activity originating from a proton interaction in the beam dump target that is close in time with the proton interaction that produced the signal.

The steel support structure of the SHiP decay vessel naturally provides a cell-like segmentation for the SBT. Each of these cells will be completely filled with LS and equipped with dedicated photosensors. The latter are inspired by the proposal of Wavelength-shifting Optical Modules (WOMs) for the upgrade of the water(ice)-Cherenkov neutrino telescope IceCube: Low-cost large-area photodetectors that also exhibit low noise~\cite{Bastian-Querner:2021uqv,Hebecker:2016mrq}. This will be the first application of such WOMs for a liquid scintillator detector, while the first IceCube WOM modules are going to be installed during the next South Pole deployment season~\cite{IceCube:2021mdb}.

The LS-WOM detector technology has been tested and further advanced in three test beam campaigns at CERN and DESY, studying prototype cells of varying size and with different readout technologies~\cite{Ehlert:2018pke,SHIP:2021tpn}. 
With a previous detector cell of volume $120\, \mathrm{cm} \times 80 \, \mathrm{cm} \times 25 \, \mathrm{cm}$ constructed from stainless steel and filled with a LS of linear alkylbenzene (LAB) and 2~g/l \hbox{2,5-diphenyloxazole (PPO)}, for particles crossing the detector at the largest distances from the WOM light sensor, the required detection efficiency of \SI{99}{\percent} could not be reached~\cite{SHIP:2021tpn}.

The following measures for increasing the light yield were thus identified in order to fulfil the requirements for the Surrounding Background Tagger defined by the SHiP LoI~\cite{SHiP-LoI}:
\begin{enumerate}
    \item Purification of the LS to increase the light attenuation length, particularly within the fluor emission range below 400~nm (PPO emission maximum at $\sim$360~nm). 
    \item Improving the inner wall reflectivity of the LS detector cells for the scintillation photons.
    \item Increasing the absorption probability of the scintillation photons in the WOM coating.
    \item Optimising the optical coupling between SiPM arrays and WOMs.
\end{enumerate}

These improvements have since been implemented and will be described in more detail below. The performance of a new full-size prototype detector has been studied in a dedicated positron exposure campaign in October 2022 at the DESY II test beam facility \cite{Diener:2018qap}, and the results of these measurements are compared to a Monte Carlo simulation of the detector cell response. Even though the response of the LS-SBT to the used positron beam of various energies is different from MIPs, the study is well-suited for an initial detector characterisation. 
We further complement these studies with measurements conducted at a dedicated small-scale laboratory setup, allowing us to compare the performance of different WOM coating procedures. Here, also the capability of reconstructing the location of the primary light source within the detector based on the integrated yield distribution measured by the SiPM array is demonstrated.

\section{General design of the SHiP LS-SBT and basic concept of WOM-LS detector cells}
\label{sec:DetectorCellConcept}

\begin{figure}[ht]
    \centering
\includegraphics[width=1.\textwidth]{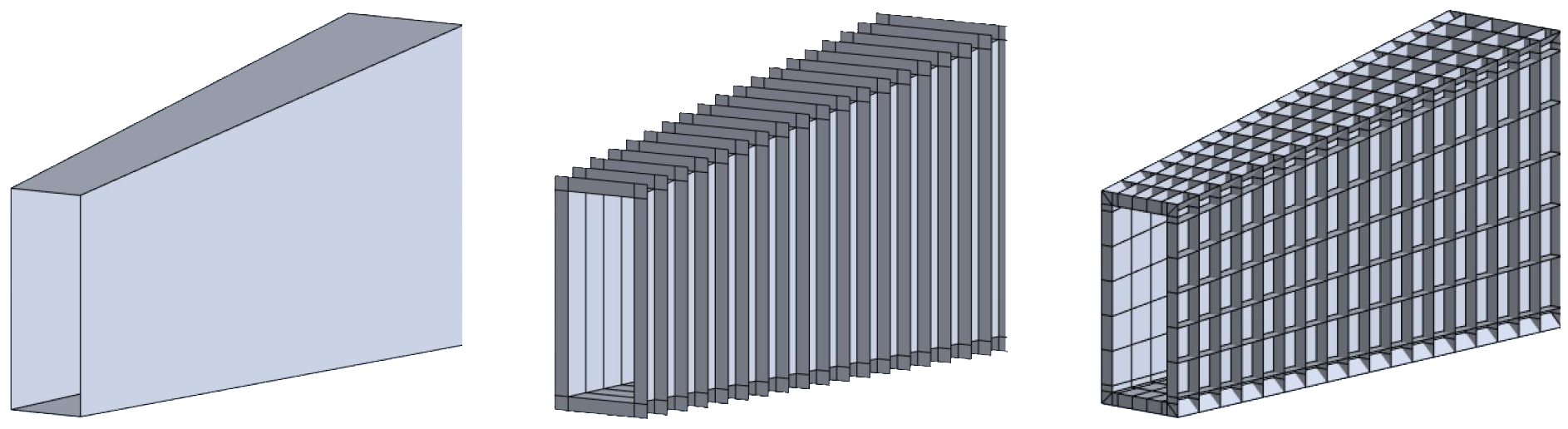}
    \caption{Schematic illustration of the LS-SBT cell segmentation defined by the the SHiP decay vessel structure (not to scale). \textit{Left:} Inner steel walls of the SHiP decay vessel. \textit{Middle:} Addition of "vertical" stiffening members at distances of 80\,cm.
\textit{Right:} Addition of "horizontal" stiffening members, creating the SBT cell structure. The cells are then closed from the outside by corten steel sheets (not shown). \textit{Courtesy A. Miano}}
    \label{fig:SBTsegmentation}
\end{figure}

The SBT veto detector surrounding the SHiP HS decay volume has to fulfil two primary criteria: It needs to provide an active medium with versatile shape that perfectly fills the gaps in the support structure of the vacuum vessel, combined with a high veto efficiency above \SI{99}{\percent} in detecting energy depositions typical for minimum ionising particles. To this end, the LS-SBT concept will comprise the following elements:

\paragraph{SHiP decay vessel and segmented geometry of the LS-SBT:} 
Along the original SPS proton beam direction, the frustrum-shaped SHiP decay vessel will have a length of $\sim$50.0\,m, with a rectangular aperture of $1.0 \, \mathrm{m} \times 2.7 \,\mathrm{m}$ at its front and $4.0 \, \mathrm{m} \times 6.0 \, \mathrm{m}$ at its rear end, respectively. It is currently planned to be constructed from 20\,mm-thick S355JO(J2/K2)W corten steel sheets. To avoid bending of the evacuated structure, the decay vessel walls (on the sides, as well as on the top and bottom) will be reinforced on the outside by a regular pattern of stiffening members. These stiffening members are made from 30\,mm-thick corten steel, 20\,cm in height, and oriented either perpendicular to the beam direction (called “vertical”) or approximately along the beam direction (called quasi-”horizontal”). In beam direction, the vertical stiffening members will be parallel to each other with a distance of 80\,cm, while the distance of the horizontal stiffening members will slightly increase along the length of the decay vessel. The cuboid structures defined by the (inner) decay vessel wall and the stiffening members will be welded closed with (outer) corten steel sheets of 20\,mm thickness. The resulting cells are then filled with liquid scintillator -- thus composing the LS-SBT enveloping the SHiP decay vessel and covering the complete holding structure with active detector material. The thickness of the LS layer surrounding the decay vessel is defined by the stiffening member height of 20\,cm, and the typical area of the cells is $80 \, \mathrm{cm} \times \mathcal{O}(1 \,\mathrm{m})$ in “vertical” and “horizontal” directions, respectively. An overall number of $\sim$1500 cells will contain a volume of about $150\,\mathrm{m}^3$ liquid scintillator. Fig.~\ref{fig:SBTsegmentation} illustrates the segmented cell-like structure of the decay vessel.                                                  

\begin{figure}[ht]
    \centering
\includegraphics[width=0.5\textwidth]{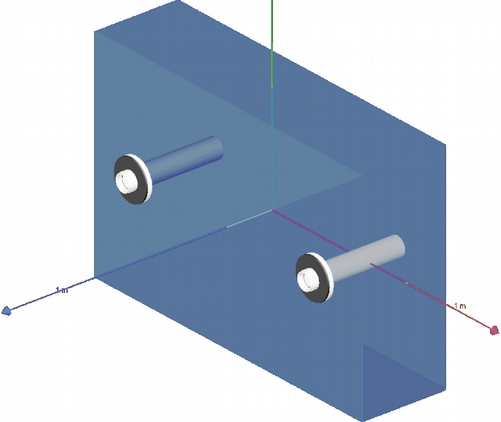}
    \caption{Schematic drawing of one LS-SBT cell showing the positions of the two WOM tubes
    inside the cell.}
    \label{fig:WOMpositions}
\end{figure}

Each LS-SBT cell is instrumented with two wavelength-shifting optical modules inserted into the cell from the outside at equal distance from the two vertical stiffening members of the cell (see Fig.~\ref{fig:WOMpositions} for a schematic drawing). The cylinder-symmetric geometry of the WOM tubes ensures efficient collection of scintillation light emitted from any point within the cell. Signals from two individual WOM tubes will provide better timing and spatial information on the interaction of a particle with the active detector material, and also better uniformity in detector response over the cell area than a single WOM could deliver.

\paragraph{Liquid scintillator:} Organic liquid scintillators (LS) are an inexpensive and versatile detector medium that can be employed to realise large active volumes. In recent decades, they have often been used in low-energy neutrino observatories (300~t $\to$ 20\,000~t scale). Chemical purification techniques enable the creation of scintillators of ultra-high radio and optical purity, thus allowing placement of the light sensors on the outer edges of the volume even for large detectors~\cite{KamLAND,Borexino,SnoPlus,JUNO}.

The setup presented for the SHiP LS-SBT consists of comparatively small LS cells equipped with optical sensors that are most sensitive in the near-UV range, with a sharp drop in acceptance at the visible blue spectrum. The liquid scintillator employed consists of only two components: The solvent linear alkylbenzene (LAB), which is widely used in present-day neutrino detectors due to its high transparency and chemical inertness~\cite{SnoPlus,JUNO}, and the fluor 2,5-diphenyloxazole (PPO) esteemed for its high quantum efficiency, fast fluorescence time of 1.6\,ns (direct excitation), and good chemical solubility. The combination of LAB with 2~g/l of PPO provides a favourable primary light yield of $\sim10^4$ photons per MeV. The PPO emission spectrum peaks at $\sim$360~nm, which is in a range where the light attenuation length of raw LAB is of the order of meters. Due to absorption by both solvent and fluor, the effective spectrum of scintillation light shifts to $\sim$390~nm after travelling several meters. This self-absorption can be mitigated by optical purification of the LAB (see Section~\ref{Sec:LiquidScintillatorPurification}).

\paragraph{WOM light sensors:}
A Wavelength-shifting Optical Module (WOM) is a transparent tube (typically made of quartz glass or PMMA) that is coated with a layer of wavelength-shifting (WLS) dye (e.g. by using dip-coating), thus providing a large active surface~\cite{Hebecker:2016mrq,Bastian-Querner:2021uqv}. The coating can be on only the outside
of the WOM tube, or on both outside and inside. For the SHiP LS-SBT, we have chosen to employ a coating procedure that applies the coating to both outer and inner WOM PMMA tube walls, in order to maximise the absorption probability of the scintillation photons. When a UV photon passes the WLS layer, it will get absorbed by a WLS molecule with a probability that depends on the WLS layer thickness, according to the law of Beer-Lambert. Subsequent to absorption of the UV photon, a secondary photon of longer wavelength -- typically in the visible range of the electromagnetic spectrum -- will be emitted. If the WOM is surrounded by a medium of much smaller refractive index (e.g. air), most of the isotropically emitted secondary photons will fulfil the condition of total reflection when reaching the WOM walls. For the SHiP LS-SBT, the WOMs are placed inside transparent PMMA vessels, separating the WOMs from the LS and creating a layer of air around them. By using a double-walled vessel structure enclosing the WOM tube on both its outside and inside by concentric PMMA vessel walls, the LS can still fill a part of the inner WOM volume (see Section~\ref{Sec:LiquidScintillatorDetectorCell}, Fig.~\ref{fig:WOM_vesselsketch} for details).

Neglecting absorption and scattering losses, up to \SI{74.6}{\percent} of the secondary photons emitted from a molecule inside the WLS layer will be reflected at the WOM tube surface and guided towards the ends of the WOM tube~\cite{Hebecker:2016mrq}. Photosensors, such as photomultiplier tubes (PMTs) or arrays of silicon photomultipliers (SiPMs), thus only have to cover this exit area (i.e. the WOM tube diameter or ring surface) to collect the majority of secondary photons.

In the case of the SHiP LS-SBT, the WOM tubes will have a length of 200\,mm, an outer diameter of 60\,mm, and and a wall thickness of 3\,mm. Each of these WOMs is instrumented at one end with a ring-shaped array of 40 $3\, \mathrm{mm} \times 3\, \mathrm{mm}$-SiPMs.

\begin{figure}[ht]
    \centering
    \includegraphics[width=0.49\textwidth]{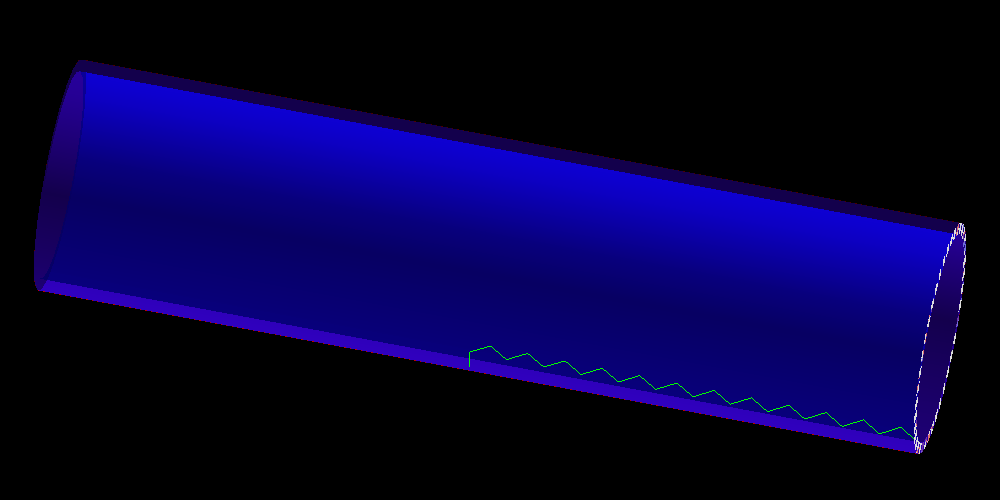} \hspace{1mm}
    \includegraphics[width=0.49\textwidth]{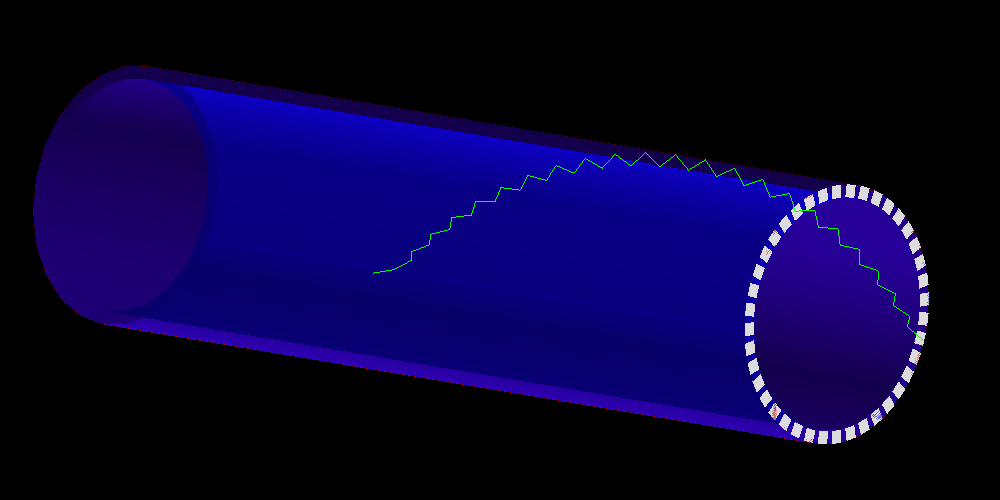} \\
    \vspace{3mm}
    \includegraphics[width=0.49\textwidth]{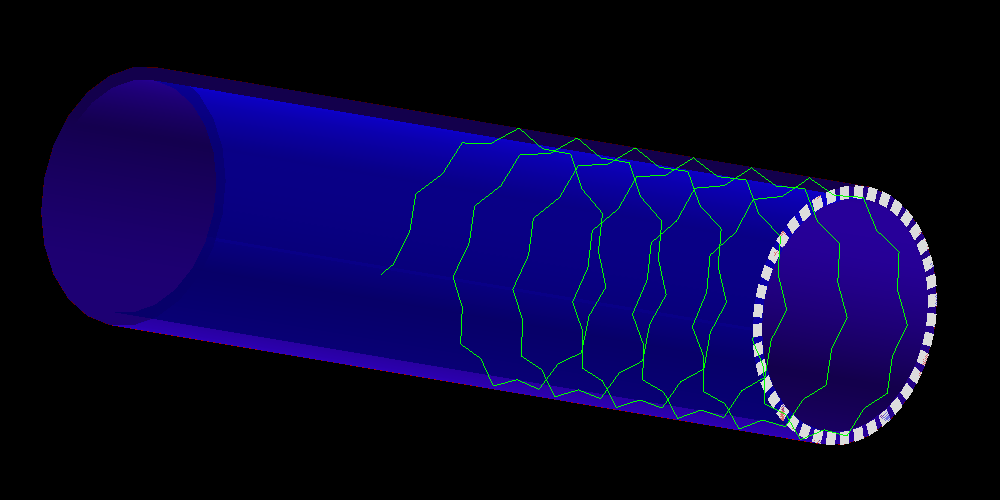} \hspace{1mm}
    \includegraphics[width=0.49\textwidth]{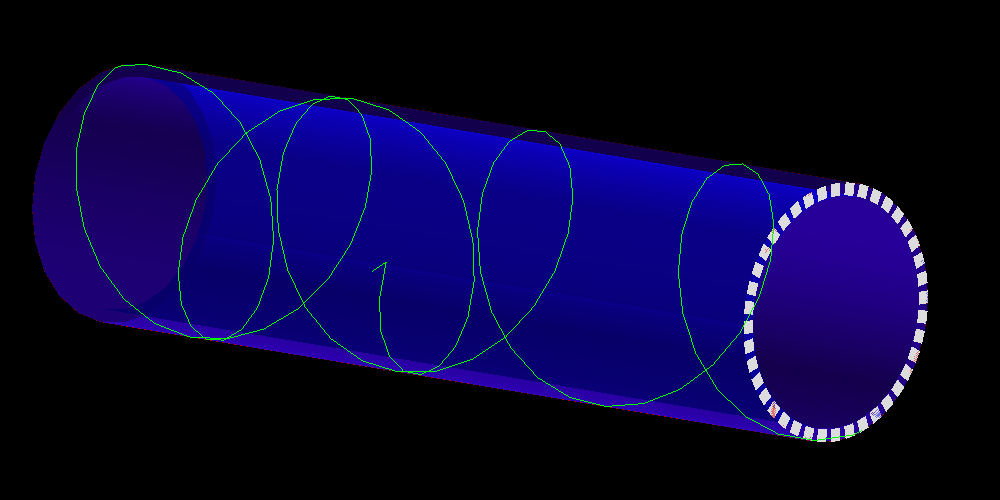}
    \caption{WOM working principle, illustrated by a Geant4 photon transport simulation. A UV photon is absorbed in the WLS layer, followed by the emitted secondary photons (green) taking exemplary paths until reaching the end of the WOM tube instrumented with a ring-array of 40 SiPMs (white).
\hbox{\textit{Top left:} Secondary} photon travelling via total internal reflection 'directly' towards the SiPM array.
\hbox{\textit{Top right:} Secondary} photon spiralling $\sim$\SI{180}{\degree} in azimuth angle w.r.t. the WOM symmetry axis before reaching the SiPM array.
\hbox{\textit{Bottom left:} Secondary} photon spiralling several times around the WOM symmetry axis before reaching the SiPM array. 
\hbox{\textit{Bottom right:} Secondary} photon emitted close to the non-instrumented end of the WOM, spiralling around the WOM symmetry axis several times before being totally reflected at the non-instrumented end of the WOM and then spiralling back, finally reaching the SiPM array.}
    \label{fig:WOMprinciple}
\end{figure}

To illustrate the WOM working principle, Fig.~\ref{fig:WOMprinciple} shows examples of secondary photon paths obtained from a Geant4-based Monte Carlo simulation~\cite{Agostinelli}: A primary UV photon hits the WOM surface and is absorbed inside the WLS layer, then a secondary photon of wavelength $>$400\,nm is re-emitted from the point of absorption and guided towards the end of the WOM tube by total internal reflection. Fig.~\ref{fig:WOMprinciple}, \textit{top left} shows a rare case where the secondary photon is detected at almost the same azimuth angle (defined w.r.t. the symmetry axis of the WOM) as the point of absorption of the primary photon: Detecting such secondary photons can provide information also on the azimuth angle of the primary light source -- the capability of extracting spatial information from the light yield distribution on the SiPM array will be discussed in Section~\ref{Sec:CosmicsTeststand}. Fig.~\ref{fig:WOMprinciple}, \textit{top right} and Fig.~\ref{fig:WOMprinciple}, \textit{bottom left} illustrate more frequent cases where the secondary photon spirals around the WOM symmetry axis: Spatial information about the primary light source position will be diluted in the detection of these photons. Fig.~\ref{fig:WOMprinciple}, \textit{bottom right} shows an example where the secondary photon is first emitted towards the non-instrumented WOM end (where it gets totally reflected) but is eventually guided to the instrumented end of the tube -- the fraction of such events is non-negligible.

\section{Improving the scintillation light collection}
\label{Sec:LightCollectionImprovements}

In this section, we describe the different steps taken to increase the light collection measured by the SiPM array.

\subsection{Purification of the liquid scintillator}
\label{Sec:LiquidScintillatorPurification}

The liquid scintillator serving as active detector material for the LS-SBT is composed of two components: The solvent linear alkylbenzene (LAB) and the fluor 2,5-diphenyloxazole (PPO), which is added at a concentration of 2.0~g/l. While primary light yield and fluorescence time mostly depend on the fluor concentration and have already been optimised based on past experience with other experiments~\cite{Borexino,SnoPlus,JUNO}, the light collection can still be significantly improved by removing optical impurities from the solvent, reducing the loss of scintillation photons before they can reach the WOMs.

\begin{figure}[ht]
    \begin{subfigure}[c]{0.17\textwidth}
        \includegraphics[width=1.0\textwidth]{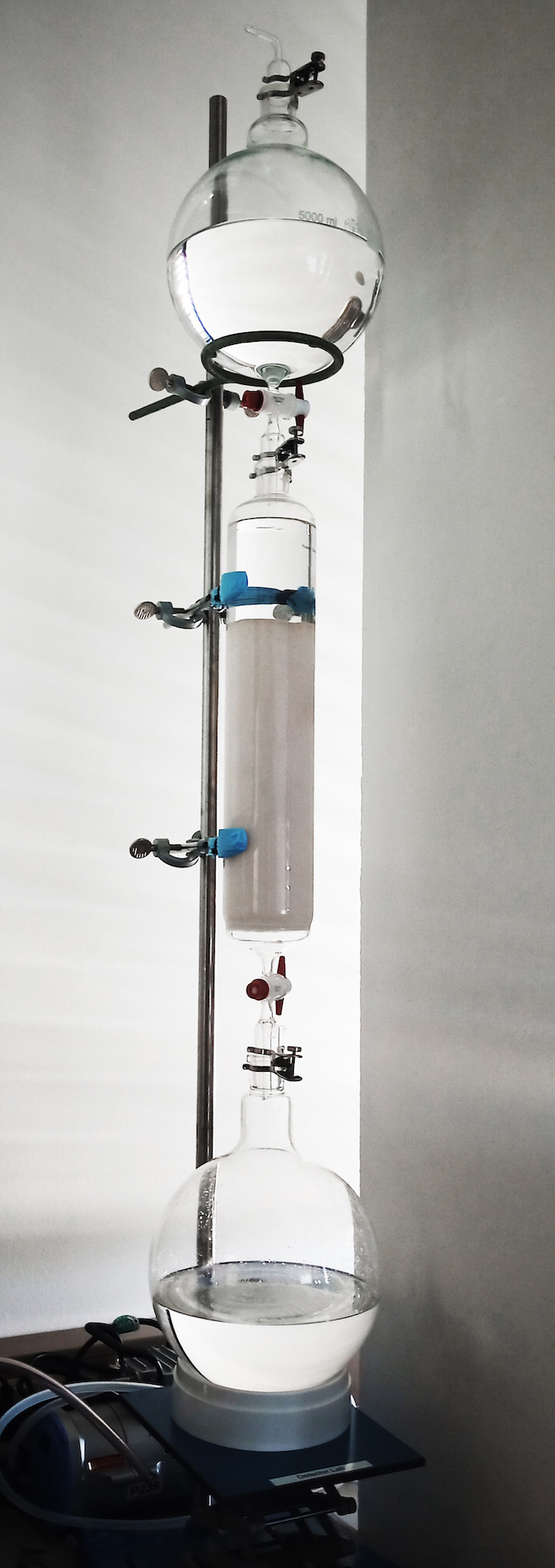}
        \label{fig:al2o3-column_2022}
    \end{subfigure}
    \hspace{1.0cm}
    \begin{subfigure}[c]{0.75\textwidth}
        \includegraphics[width=1.0\textwidth]{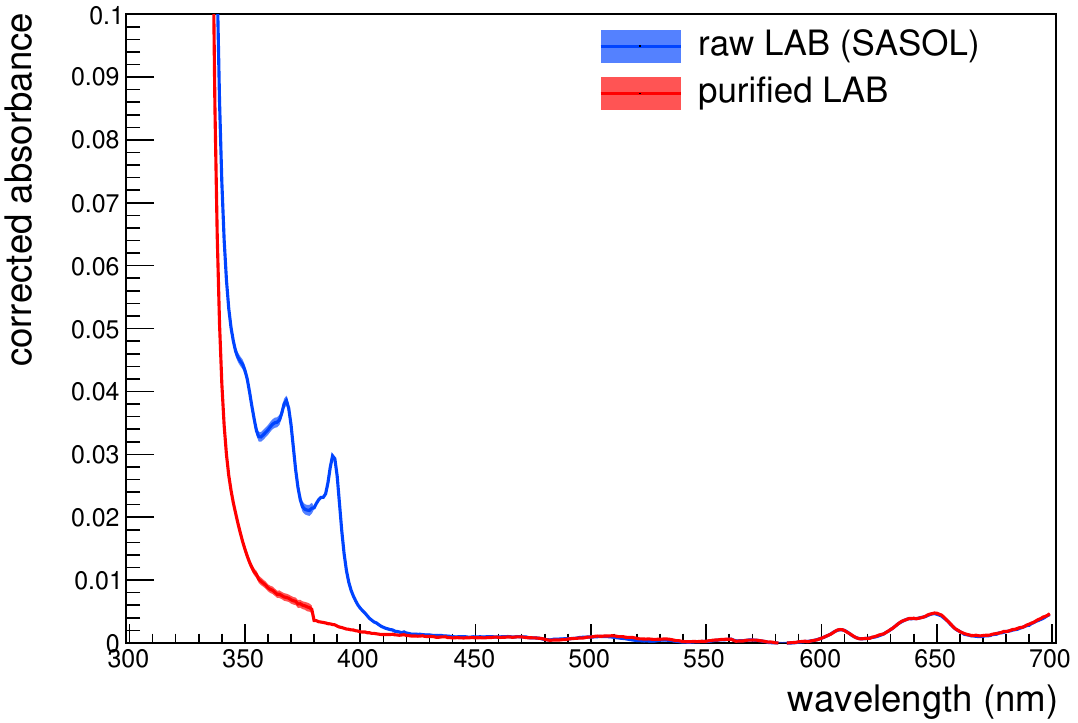}
        \label{fig:absorbance_2023}
    \end{subfigure}
    \caption{{\it Left:} Alumina column employed to purify the LAB scintillator solvent. {\it Right:} UV-Vis absorbance measurements of raw and purified LAB samples in a 10\,cm cuvette: Alumina column purification is very effective in removing organic impurities that absorb scintillation photons in the near-UV (360\,nm -- 400\,nm).}
    \label{fig:aluminafiltration}
\end{figure}

Studies performed in the context of the JUNO experiment have demonstrated the high optical purification efficiency of both fractional distillation and column filtration with alumina (Al$_2$O$_3$) as bed material~\cite{JunoColumn}. Alumina filtration shows equal or better performance to distillation and is at small scales technically more practical to implement. For the LS-SBT prototype detector, the filtration of \SI{250}{l} LAB was conducted using a 2-litre glass column (\SI{80}{mm} diameter, \SI{400}{mm} height) filled with a bed of \SI{1.6}{kg} of alumina powder\footnote{Merck 90 active acidic, activation level 1, grain size 0.063\,mm -- 0.20\,mm} atop a P3\footnote{$16 \, \mathrm{\mu m} - 40\,  \mathrm{\mu m}$ pore size} PTFE frit. Flow speed was increased to about \SI{0.8}{l/h} using a small vacuum pump connected below the frit. \SI{20}{l} -- \SI{25}{l} of LAB\footnote{SASOL Italy S.p.A., Hyblene 113 / Sasolab C12H} could be passed through the column before the bed material needed to be exchanged. The column laboratory setup can be seen in the \textit{left} panel of Fig.~\ref{fig:aluminafiltration}.

The right panel of Fig.~\ref{fig:aluminafiltration} shows the light absorbance for both raw LAB and Al$_2$O$_3$-purified LAB in the relevant wavelength range of \SI{300}{nm} -- \SI{700}{nm}. The wavelength-dependent absorbance $A(\lambda)$ was obtained by UV-Vis spectroscopy using a PerkinElmer $\Lambda850$ spectrophotometer, the LAB samples were inserted into a Suprasil glass cuvette of $l=\SI{100}{mm}$ path length. The absorbance is defined as the logarithm of the ratio of incident intensity $I_0$ and attenuated intensity $I(l)$ after traversing the cuvette:
\begin{equation}
A=\log_{10}\left(\frac{I_0}{I(l)}\right) = \log_{10}\left(\exp\left(\frac{l}{L_{\rm att}}\right)\right),
\end{equation}
which can be related to the attenuation length $L_{\rm att}(\lambda)$ based on the law of Beer-Lambert. Results have to be corrected for Fresnel reflection on the outer glass -- air transitions and focusing effects (only statistical uncertainties are shown). The column purification proves to be very effective in removing two characteristic absorption lines caused by organic impurities in raw LAB: At the peak emission of PPO at $\lambda\sim$\SI{360}{nm}, the observed reduction in absorbance from $A\sim$0.035 to $\sim$0.008 roughly translates to an increase in light attenuation length from $L_{\rm att}\sim$1.2\,m to $\sim$5.0\,m (this value is slightly reduced after addition of the fluor PPO). Note that the attenuation lengths $L_{\rm att}$ measured with the spectrophotometer are very long compared to the path length of the cuvette: Especially for the more transparent region of wavelengths $> 400 \, \mathrm{nm}$, the absolute values feature large systematic uncertainties which are hard to quantify, and a dedicated laboratory setup will be needed to provide more precise measurements.

\subsection{Increasing the cell wall reflectivity}
\label{Sec:CellWallReflectivity}

Due to the geometry of the detector, the relative solid angle under which any point of light emission within the scintillator volume can be observed will be comparatively small. The light collection efficiency can thus be substantially increased by improving the reflectivity of the inner detector cell walls, permitting several reflections of photons to occur before the light reaches a WOM.

\begin{figure}[ht]
    \begin{subfigure}[c]{0.45\textwidth}
        \includegraphics[width=1.0\textwidth]{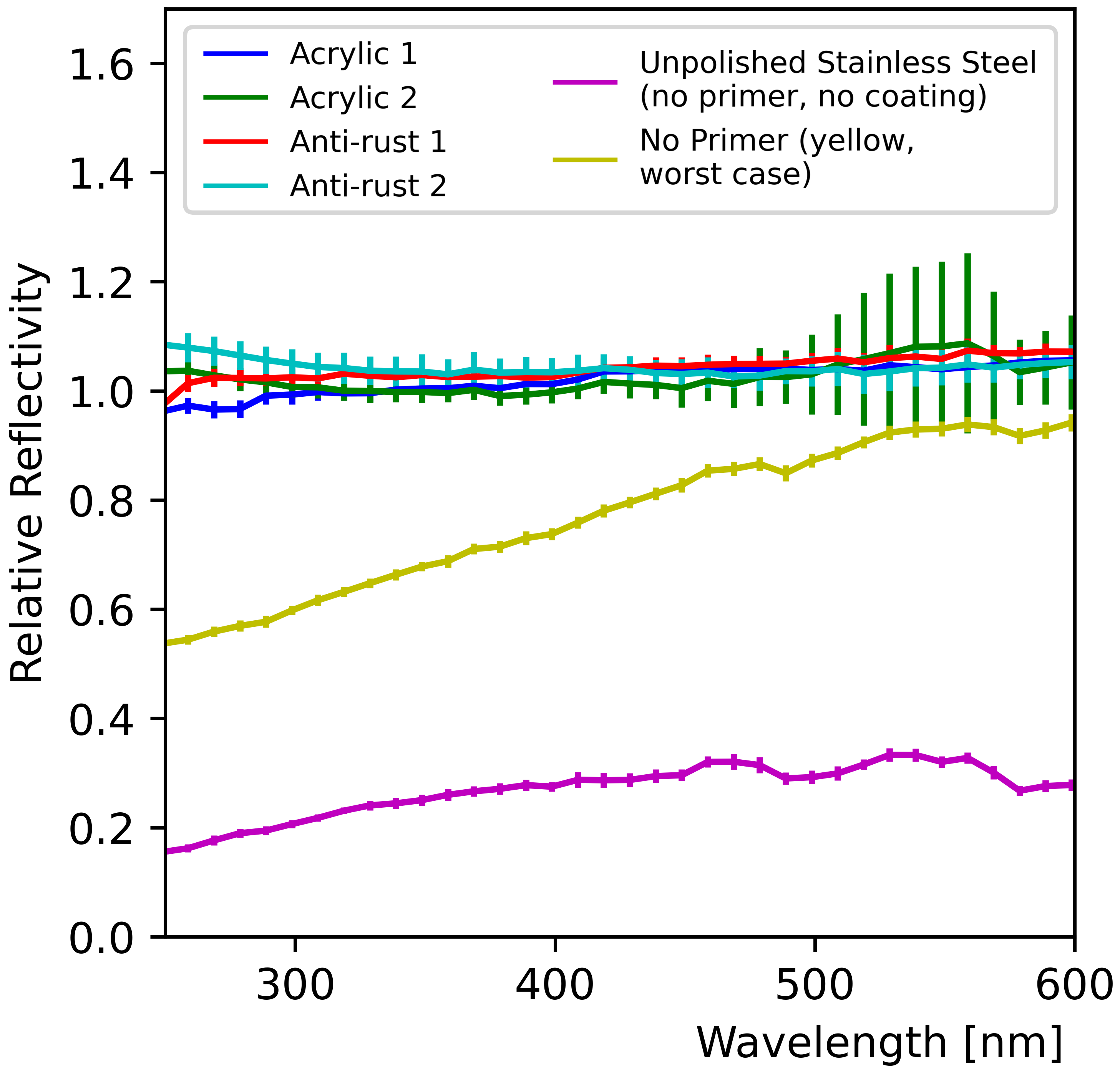}
        \label{fig:reflectivity-01}
    \end{subfigure}
    \hspace{1.0cm}
    \begin{subfigure}[c]{0.47\textwidth}
        \includegraphics[width=1.0\textwidth]{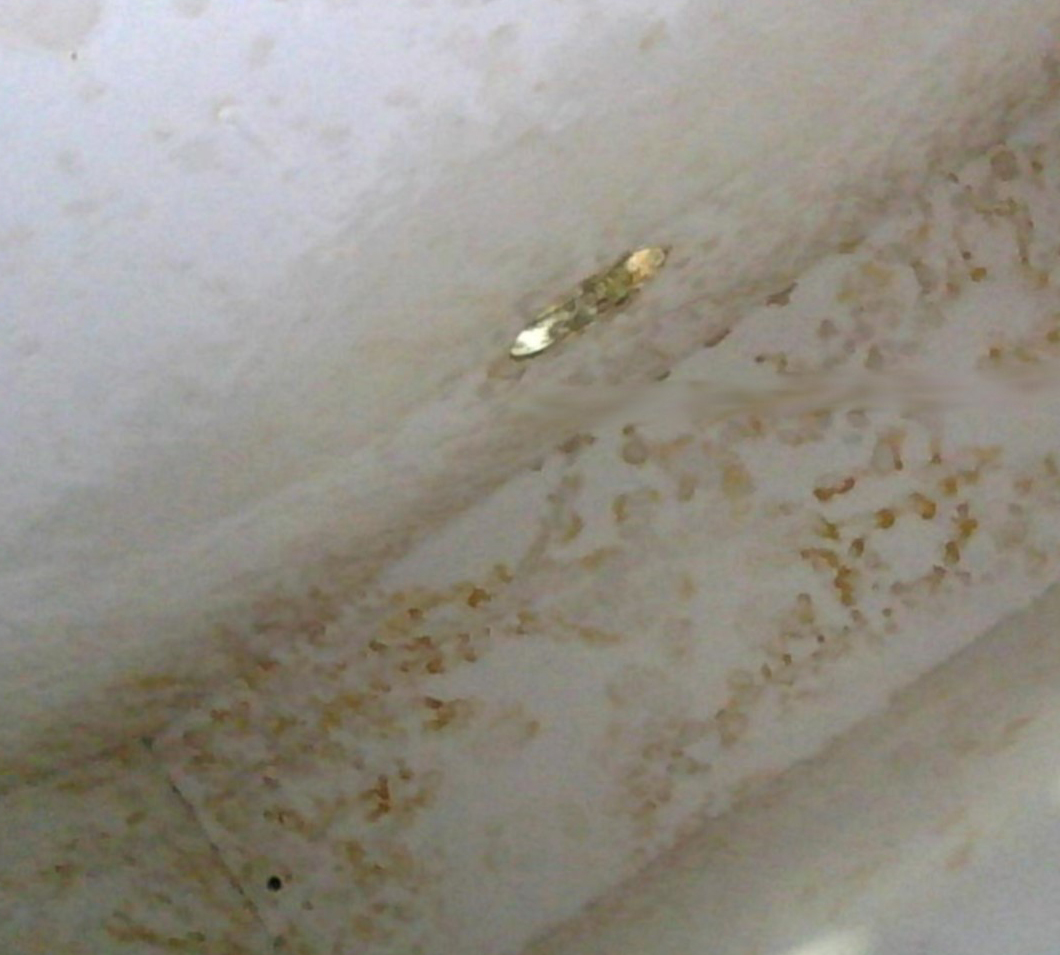}
        \label{fig:reflectivity-02}
    \end{subfigure}
    \caption{{\it Left:} Wavelength-dependent diffuse reflectivity, measured relative to a BaSO$_4$ standard with $R(\lambda \geq \SI{300}{nm})\geq \SI{95}{\percent}$. The standard was produced by applying the BaSO$_4$ coating to stainless steel. The Corten steel samples coated with combinations of acrylic primer and BaSO$_4$ paint exhibit high reflectivity, while a sample coated only with paint but no primer results in worse performance. A raw stainless steel surface provides only $\sim$\SI{25}{\percent} relative diffuse reflectivity at 360~nm in comparison. The error bars display the statistical uncertainty.  {\it Right:} View of the inside of the detector cell after coating with primer and BaSO$_4$ paint. Despite the primer, weak rust stains become visible.}
    \label{fig:reflectivity}
\end{figure}

Corten steel (and also stainless steel) is a poor reflector in the near-UV range of the PPO emission spectrum. Hence, we investigated reflective paints containing pigments of titanium-oxide (TiO$_2$) and barium-sulfate (BaSO$_4$). While the widely-used TiO$_2$ exhibits only weak reflectivity for $\lambda<\SI{400}{nm}$, BaSO$_4$ paint can reach a reflectivity of $R(\lambda)\geq \SI{95}{\percent}$ over a broad range of UV wavelengths. These expectations were cross-checked with a custom-built laboratory setup for measuring the diffuse reflectivity of a surface sample as a function of wavelength (Fig.~\ref{fig:reflectivity}, \textit{left}). The surfaces studied in this measurement were small rectangular samples of: Corten steel coated with different primers and BaSO$_4$ paint, corten steel coated with BaSO$_4$ paint only (no primer), and bare stainless steel (unpolished). The samples coated with both primer (any) and BaSO$_4$ paint provided the expected high reflectivities, while the sample with BaSO$_4$ paint only (no primer) developed a slightly yellow tint. Bare unpolished stainless steel exhibits significantly worse diffuse reflectivity.

For the final SHiP decay vessel, the inner surface of the individual LS-SBT cells can only be coated after the cell walls have been welded closed. Thus, we had to devise a technique allowing both primer and reflective paint to be applied via the circular openings created for the WOMs. This was achieved by means of a commercial compressed-air spray gun equipped with straight or bent (90$^\circ$) nozzles, respectively. Two layers of primer, followed by two layers of reflective BaSO$_4$ paint\footnote{Berghof Fluoroplastic Technology GmbH, Optopolymer OPRC, CAS-Number 7727-43-7} were applied sequentially over the course of several days, allowing each layer to dry for at least 24 hours. Several days into this process, it was observed that the corten steel was reacting with remaining humidity in the water-based primer, creating visible rust stains in the paint (Fig.~\ref{fig:reflectivity}, \textit{right}). These stains not only persisted through the coating process, but also intensified over time, thus reducing the overall reflectivity of the inner cell surface. We assume that the difference to the (stain-free) samples investigated during the material tests preceding the prototype cell coating was caused by poor air circulation (and hence increased drying times of both primer and paint) when applied within the welded cell. Coating tests and measurements currently performed in Mainz seem to confirm this hypothesis. 
Even stained, the coated cell walls represent a distinct improvement in reflectivity when compared to bare steel, with $R\sim\SI{65}{\percent}$, corresponding to simulations (see Section~\ref{Sec:LightCollectionandDetectorResponse}).

\subsection{Increasing the WOM WLS dye photon absorption probability}
\label{Sec:WOMAbsorptionProbability}

Building on the experience in WOM development for the IceCube upgrade~\cite{Hebecker:2016mrq}, the WOMs considered for SHiP originally employed the following WLS dye composition (hereafter referred to as the WLS 'standard' dye): \SI{77.3}{\percent} toluene and \SI{22.3}{\percent} Paraloid B723 (also called PEMA) as base material, and the two wavelength shifters \SI{0.13}{\percent} bis-MSB and \SI{0.27}{\percent} p-terphenyl (PTP). The WLS layer is produced  utilising a ND-DC dip-coater from Nadetech Innovations S.L. company. The standard dip-coating procedure with the standard WLS dye was to immerse the WOM tube inside the liquid for 80\,s and then withdraw it at a speed of 93\,mm/min.

\begin{figure}[ht]
    \centering
    \includegraphics[width=1.0\textwidth]
    {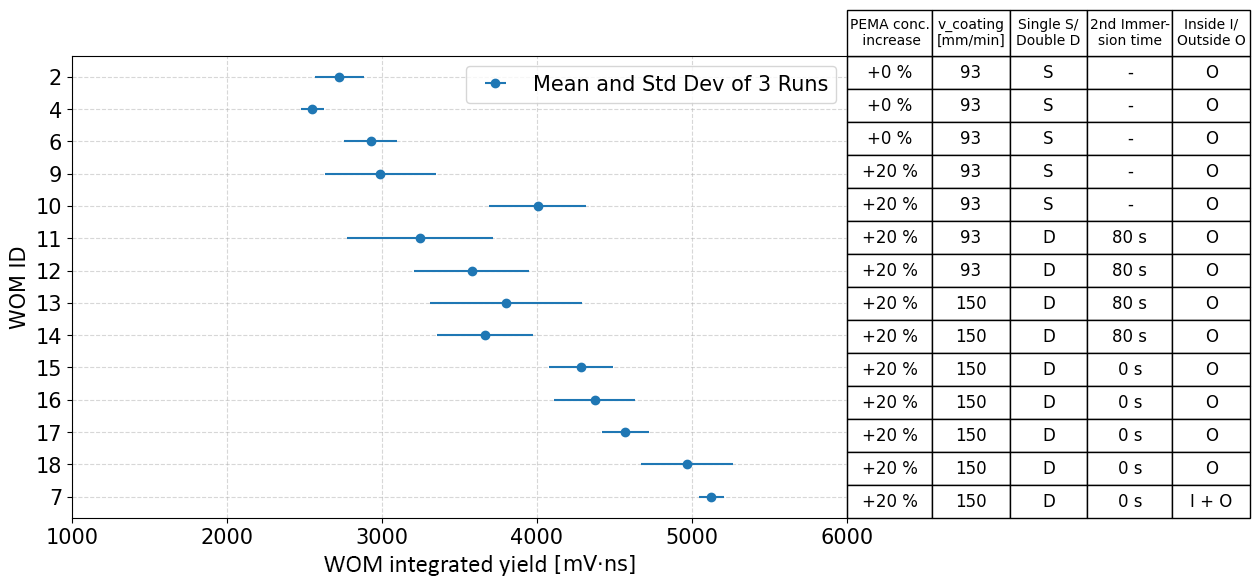}
    \caption{Integrated yield obtained with a single WOM, measured as a time integral over the SiPMs waveforms, with WLS layers produced using different dip-coating parameters and procedures, with the WOM placed inside a LS-filled cell as described in Section~\ref{Sec:CosmicsTeststand} using events fulfilling a cosmics trigger. For each WOM, the mean of three measurements is displayed. Each measurement was taken after dismounting and remounting of the SiPM array in order to quantify the systematic uncertainty in the optical coupling of the SiPM array to the WOM. The error bar is the standard deviation obtained from the three measurements. The different WOMs, labelled with numbers, are grouped according to the coating procedure outlined in the table on the right-hand side of the plot to quantify systematic uncertainties of the coating procedure.}
    \label{fig:womMPVs}
\end{figure}

For the study presented in this paper, an optimised WLS dye and a more refined coating procedure were developed and applied for the production of the WOMs used in the new LS-SBT detector prototype~\cite{Schmidt}. The performance of improved dye and dip-coating procedure were quantified by measuring the increase in absorption probability in coated slides (made from glass or PMMA) using a PerkinElmer Lambda 950 transmission spectrometer, and by the light yield of cosmic muons obtained with the new WOMs as photodetectors in a dedicated small-scale LS detector setup at HU Berlin (Section~\ref{Sec:CosmicsTeststand}). The results are summarised in Fig.~\ref{fig:womMPVs}.

\begin{figure}[ht]
    \centering
    \includegraphics[width=1.0\textwidth]{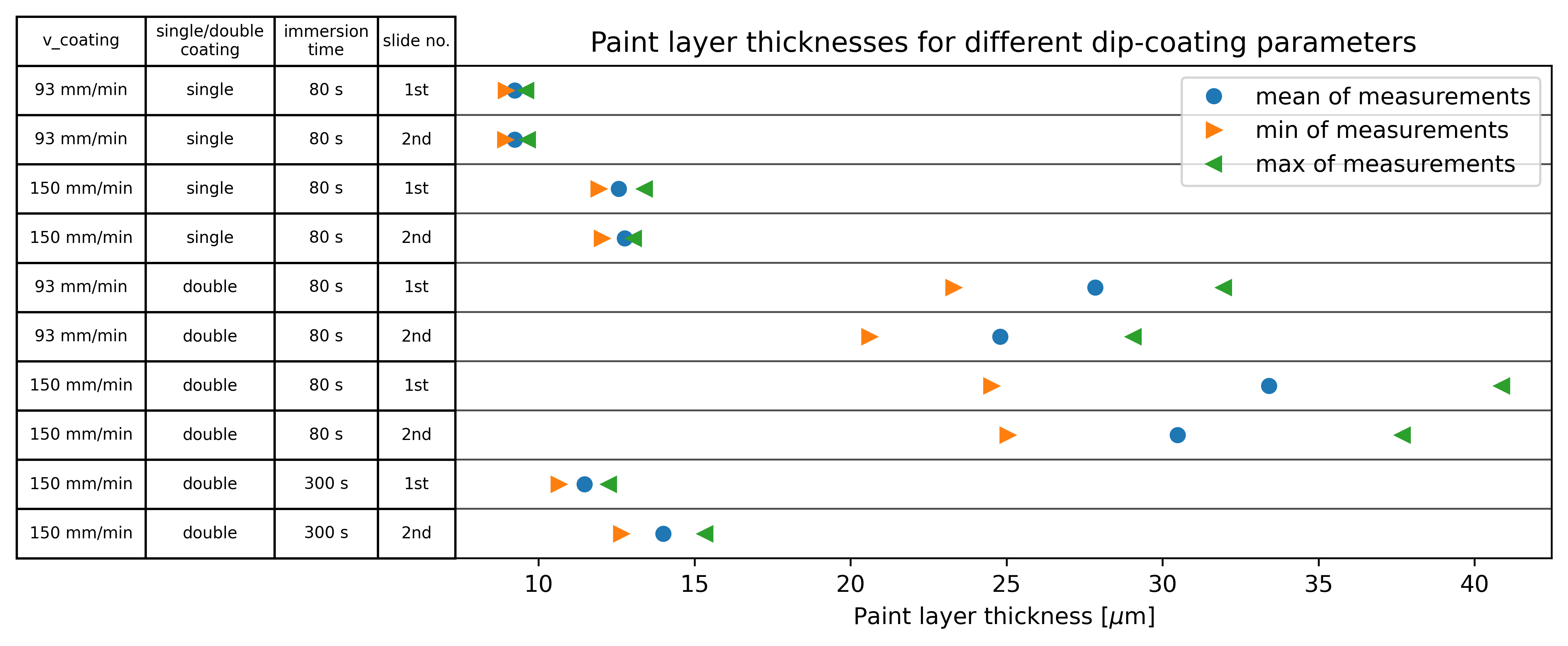}
    \caption{WLS layer thickness on dip-coated glass slides for different dip-coating methods and parameters.}
    \label{fig:WLSthickness}
\end{figure}

For coated glass slides, it was possible to measure the WLS layer thickness using a DektakXT stylus profilometer from Bruker company by scratching the coated surface. The results are summarised in Fig.~\ref{fig:WLSthickness}. This allowed us to correlate the improvement in performance directly with the increase in WLS layer thickness. For coated PMMA slides, neither this technique is applicable (since PMMA and PEMA are chemically very similar) nor can the coating thickness be determined by comparing the weight of a slide before and after coating: The toluene partly dissolves the PMMA surface, which could even result in a smaller total slide weight after coating. However, for the same type of dye and the same dip-coating procedure, we found similar transmittance values for coated glass slides and coated PMMA slides (after applying corrections accounting for the different light transmission in uncoated glass w.r.t. PMMA). Hence, we expect the improvements observed for glass slides to also hold for slides made from PMMA.

\begin{figure}[ht]
    \centering
    \includegraphics[width=1.0\textwidth]{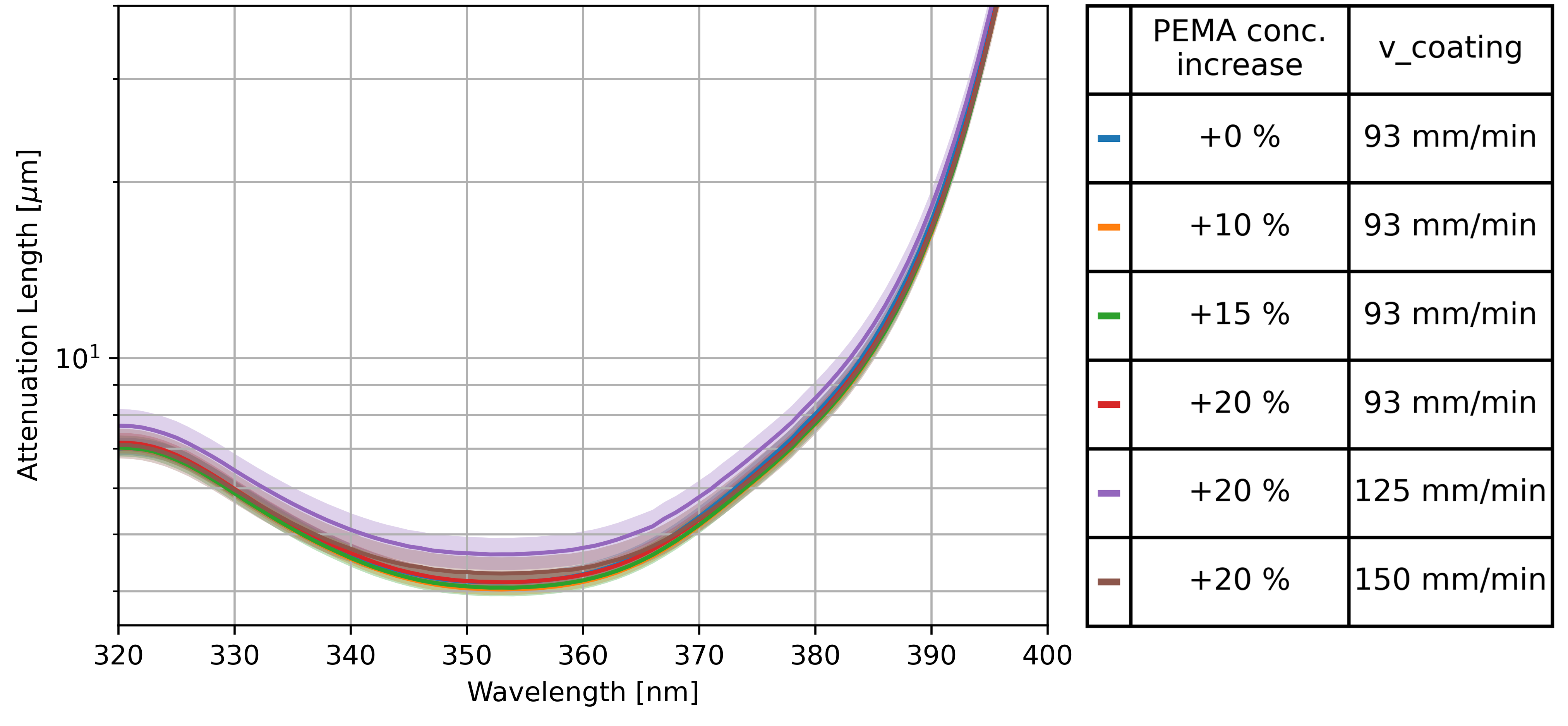}
    \caption{Attenuation length as a function of wavelength of the WLS layer measured with dip-coated glass slides for different coating methods and parameters, as explained in the text.}
    \label{fig:WLSattenuationlength}
\end{figure}

The transmission measurements with coated glass slides were used to determine the light attenuation length of the WLS layer by correcting for the light transmission measured with uncoated slides and using the WLS layer thickness measurements obtained with the profilometer. Fig.~\ref{fig:WLSattenuationlength} shows the attenuation length as a function of wavelength: Within the uncertainties, the attenuation lengths obtained using different coating parameters and methods are in agreement -- this is expected if the different dip-coating procedures only affect the WLS layer thickness, but not the WLS concentration in the dye.

Of particular relevance are attenuation lengths in the emission range of PPO (from $\sim$350~nm to about 430~nm): Between 350~nm and 385~nm, the measured attenuation length is at most 10~$\upmu$m, while above 385~nm it rapidly increases with the wavelength. Since a significant part of the PPO emission spectrum is in the wavelength range above 385~nm, it is thus important to obtain WLS layer thicknesses significantly above 10~$\upmu$m in order to guarantee a large absorption probability of the scintillation photons in the WLS layer -- as the attenuation length cannot be further reduced by increasing the WLS concentration.

In studies with coated glass slides and coated PMMA slides, several parameters were varied with the aim of enhancing the absorption probability of the WLS layer on the WOM surface~\cite{Schmidt}. Increasing the concentration of the wavelength shifters by significantly more than \SI{10}{\percent} proved to be impossible, since the mixture was already saturated with WLS molecules, resulting in an observed crystallisation of WLS molecules when further raising the WLS concentration in the dye.

The standard WLS dye was mixed at a temperature of \SI{110}{\celsius}. Such high temperature, which allows fast dissolving of the WLS in toluene, could potentially destroy the molecular structure of some of the WLS molecules (in particular the double bonds in bis-MSB) and thus decrease the maximal absorption probability. However, neither significant increase nor decrease in absorption probability were observed for the glass slides when lowering the mixing temperature to \SI{55}{\celsius}. As a precaution, \SI{55}{\celsius} was chosen as the new standard mixing temperature for the WLS dye, still guaranteeing fast dissolving of the WLS in toluene.

\begin{figure}[ht]
    \centering
    \includegraphics[width=1.0\textwidth]{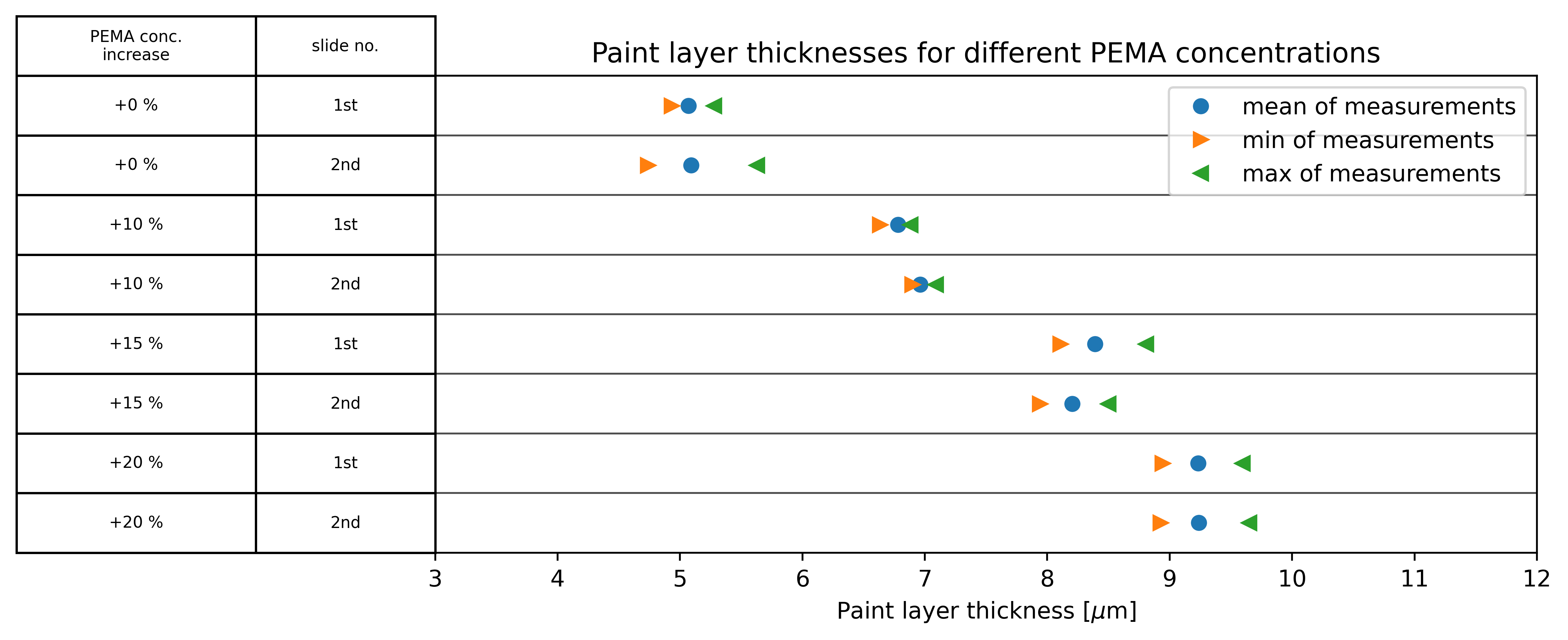}
    \caption{WLS layer thickness on dip-coated glass slides for PEMA concentration increase using toluene evaporation.}
\label{fig:WLSthicknessPEMAconcentration}
\end{figure}

The coating layer thickness $d$ is expected to scale with $\sqrt{\eta}$, where $\eta$ is the viscosity of the dye~\cite{IceCube:2021mdb,Brinker}. Hence, evaporation of toluene after mixing but before dip-coating can increase the concentration of PEMA inside the dye before drying and as a result its viscosity. The WLS layer thickness, measured at six different points on coated glass slides, could be almost doubled compared to dip-coating with the standard dye from about 5~$\upmu$m to about 9~$\upmu$m, when about \SI{20}{\percent} of the toluene is evaporated after the mixing~\cite{Schmidt}. This is shown in Fig.~\ref{fig:WLSthicknessPEMAconcentration} presenting the average thickness values and the minimum and maximum value of the six measurements as a function of relative PEMA concentration increase by toluene evaporation.

The velocity, $v_{coating}$,  with which the tube is extracted from the dye, also has a significant impact on the coating thickness: $d\propto\sqrt{v_{coating}}$~\cite{IceCube:2021mdb,Brinker}. This was qualitatively confirmed by the layer thickness measurements on glass slides with the profilometer: the WLS layer thickness could be increased from about 9~$\upmu$m at a coating velocity of 93\,mm/min to about 13~$\upmu$m at a coating velocity of 150\,mm/min (compare the first two and the second two rows in Fig.~\ref{fig:WLSthickness}).

One naively expects to increase the absorption probability by coating the WOM a second time. However, since the toluene dissolves PMMA, it also (partially) dissolves the already existing WLS layer. Hence, the immersion time of the WOM in the WLS dye before the second coating should be sufficiently small. Fig.~\ref{fig:WLSthickness} shows that one cannot increase the layer thickness by double-coating if the immersion time of the WOM inside the dye before the second coating is quite long (300~s). Double coating significantly increases the WLS layer thickness up to $\sim30~\upmu$m with a large variation between about 20~$\upmu$m and 40~$\upmu$m, depending on where the thickness is measured on the glass slide. Since the immersion time is already sizeable during the dip-coating process itself given a maximum dip-coating velocity of 150\,mm/min, it was decided to set the immersion time before starting the second dip-coating to zero. This in fact improved the WOM light collection significantly, as can be seen in Fig.~\ref{fig:womMPVs}.

In the standard coating procedure, the WOMs are only coated outside. The procedure was modified by fixing the tube only at the top so that the outside and the inside could be coated simultaneously. This also resulted in an increase of the WOM light collection in the cosmics test detector setup (Section~\ref{Sec:CosmicsTeststand}) compared to WOMs that were only coated outside as shown in Fig.~\ref{fig:womMPVs}.

One should mention, however, that this procedure results in dried dye remainders at the tube end opposite to the SiPM array, changing its planar and polished surface geometry. From this point of view, the procedure for dip-coating the tubes from outside and inside is not yet optimal, because a significant part of WLS photons collected by the SiPM array are photons that were travelling first to the non-instrumented WOM end where they are totally reflected and then guided back to instrumented WOM end. Geant4 photon transport simulations show that depending on the light attenuation length inside the WOM and on the position where the primary photon is absorbed in the WOM, the fraction of such photons detected by the SiPM array varies between about \SI{25}{\percent} and \SI{42}{\percent}. The simulation is confirmed by the following two observations: 1) adding a specular reflector to the non-instrumented WOM end does not result in a significant increase in light collection, 2) gluing a black paper at the end of the non-instrumented WOM end decreases the detected signal by about \SI{30}{\percent}. As a result, changing the planar and polished surface of the non-instrumented WOM end by dried WLS dye remainders will result in a loss of detected photons. It is planned to modify the dip-coating setup and procedure in the future to avoid this problem when WOMs are dip-coated from outside and from inside.

In summary, for the LS-SBT detector prototype for the DESY test beam measurements, we produced WOMs with a WLS dye that was mixed at \SI{55}{\celsius} and in which about \SI{20}{\percent} of the toluene was evaporated after dye mixing. The WOM was immersed 80~s before the first dip-coating and extracted with a velocity of 150\,mm/min. After two days of drying the WOM was dip-coated a second time with the same coating speed and without any significant immersion time before start of the dip-coating. While the dip-coating procedure has been optimised, it is not known exactly which WLS layer thickness was realised on the coated WOMs. Comparing transmission for coated glass slides and coated PMMA slides show up to about \SI{10}{\percent} difference in transmission for PMMA slides which translates into a about \SI{10}{\percent} difference in WLS thickness on PMMA slides. Also the results obtained on slides might not be directly transferable to a WOM. For the simulations used in this work, we assume a WLS thickness of 20~$\upmu$m, keeping in mind that the actual value might significantly differ.

\subsection{Optimising the optical coupling between SiPM array and WOM}
\label{Sec:Optimisi ngOpticalCoupling}

The WOM wall diameter of 3\,mm exactly fits the SiPM side length of 3\,mm. It is therefore important to ensure that the WOM is positioned in a well-controlled way inside the PMMA vessel and that the SiPM array is well-positioned with respect to the WOM position. Any misalignment would cause a reduction of detected photons. For the
detector cell prototype, a new PMMA vessel design plus mechanics was developed that guaranteed a well-controlled positioning of the WOM and the SiPM printed circuit board (PCB) with respect to each other, minimising potential light losses as described in more detail in Section~\ref{Sec:LiquidScintillatorDetectorCell}.

\subsection{SiPM detection efficiency}
\label{Sec:SiPMDetectionEfficiency}

In earlier test beam measurements with LS-SBT prototypes, Hamamatsu S13360-3075 SiPMs with a large pixel pitch of $75~\upmu$m were used to obtain a large detection efficiency. We operated these SiPMs at about 4~V overvoltage, for which the datasheet quotes a typical efficiency of around \SI{55}{\percent} at a wavelength of 450~nm~\cite{Hamamatsu-S13360}. For the detector prototype described in this work, we used Hamamatsu S14160-3050HS SiPMs~\cite{Hamamatsu-S14160}, since one can reach similar photon detection efficiencies with a pixel pitch of $50~\upmu$m and similar overvoltages but at a significantly lower price per SiPM, which becomes relevant for a large-area detector like the SHiP LS-SBT instrumented with several thousands of WOMs. In the test beam measurement, we operated the S14160-3050HS SiPMs at an overvoltage of about 3.7~V, for which the datasheet quotes a typical efficiency of \SI{55}{\percent}. 

\section{Performance studies of WOMs with a small-scale detector in a cosmics test setup}
\label{Sec:CosmicsTeststand}

\subsection{The cosmics test setup}
A test setup of a small stainless steel box with outer dimensions of $50.4 \times 50.4 \times 25$\,cm$^3$ filled with purified LAB (Section~\ref{Sec:LiquidScintillatorPurification}) plus 2~g/l PPO, which can be equipped with one WOM in the centre of the box, was built in Berlin. The WOM was coupled to a SiPM array of 40 Hamamatsu S14160-3050Hs SiPMs, subdivided into eight groups of five SiPMs. The signals from the eight SiPMs groups were amplified by an eMUSIC chip~\cite{MUSIC-ASIC}, mounted on a PCB produced by the Scientifica
company.

\begin{figure}[ht]
    \centering
    \includegraphics[width = 1.0\textwidth]{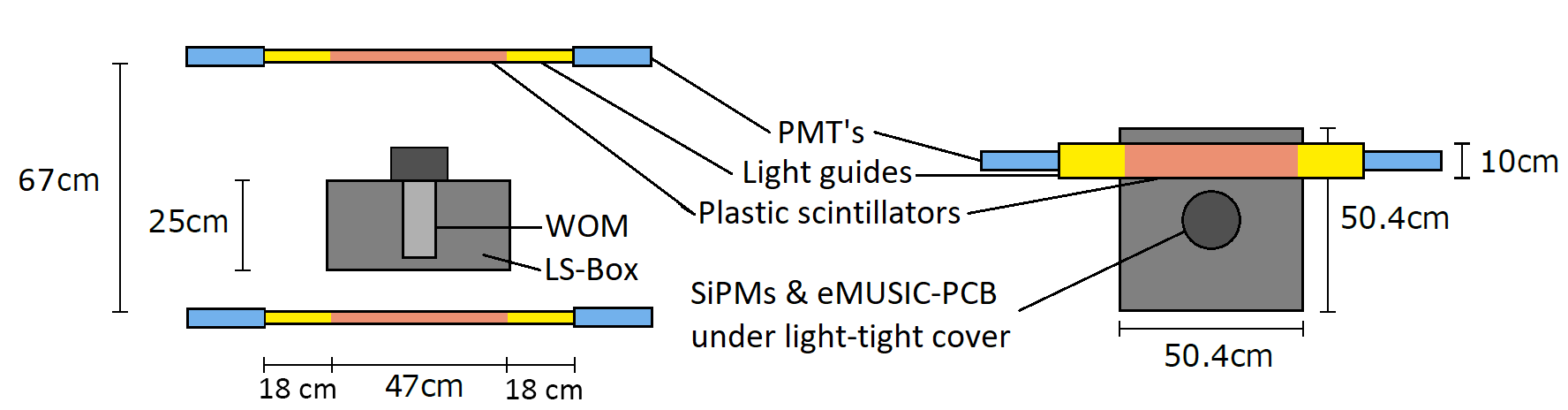}
    \caption{Sketch of the small-scale cosmics test setup. \textit{Left:} Side view. \textit{Right:} Top view.}
    \label{fig:CosmicsTestSetup}
\end{figure}

The box was sandwiched in between two long plastic scintillators (made from NE110) of dimensions 8$45 \times 10 \times 1$9\,cm$^3$, which allows one to trigger on the crossing of a muon through both scintillators. The two scintillators have a light-guide on each end, to which a Philips XP2008 PMT is optically coupled. The positions of the two scintillators can be varied sideways, perpendicular to the long side of the scintillator, so that different areas of the LS box can be covered. 
A schematic of the LS-filled box and plastic scintillator placement is shown in Fig.~\ref{fig:CosmicsTestSetup}. The PMT signals together with the amplified WOM SiPM signals were digitised by a 16-channel WaveCatcher digitiser~\cite{WaveCatcher}
using a sampling rate of 3.2 GS/s (Giga Samples per second).
The digitised waveforms were recorded as an event when
a trigger condition on a defined set of WaveCatcher input signals were fulfilled.
For the trigger, a coincidence of the PMT signals within a time window of 15 ns was required.

For the light collection tests of different WOMs described in Section~\ref{Sec:WOMAbsorptionProbability}, the two scintillators were always kept in the same position and only events triggered by a coincidence of the four PMT signals of the scintillators were registered.

\subsection{Information on the direction of primary photons hitting the WOM using the integrated yield in the SiPM groups}
\label{Sec:ReconstructionAzimuthalAngle}

As already discussed in Section \ref{sec:DetectorCellConcept}, many of the wavelength-shifted secondary photons are guided in spiralling paths to the SiPM array, completely diluting information of where the primary photon hit the WOM. At first glance, it seems unlikely that any information can be extracted about the light coming from a fixed primary light source. However, some of the secondary photons still carry information about the point where the primary photons hit the WOM as pointed out in Section~\ref{Sec:Introduction}, namely when the secondary photon travels directly towards the SiPM array, without spiralling around the WOM symmetry axis.

Based on the work described in~\cite{Ernst,Vagts,Eckardt}, we demonstrate here that the distribution of registered photoelectrons in the eight SiPM groups allows to estimate on a statistical basis the original azimuth angles of the primary photon hit positions on the WOM. Hence, the granularity of the SiPM array is advantageous compared to a single large-area photomultiplier simply covering the complete WOM exit surface.

\begin{figure}[ht]
    \centering
    \includegraphics[width = 0.5\textwidth]{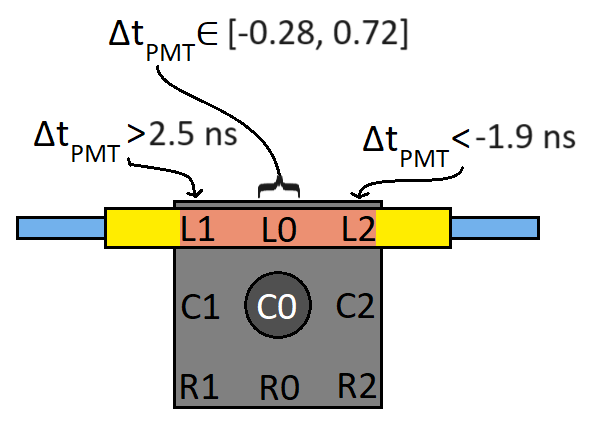}
    \caption{Positions of muons crossing the LS detector cell selected by the arrival time difference between the two PMT signals at each of the two long plastic scintillators in the small-scale cosmics test setup.}
    \label{fig:alpha}
  \end{figure}

To do so, we have used the cosmics test setup to select muons traversing the LS box and in the subsequent analysis restricting them to a defined range of crossing points and incidence angles. For this, both long scintillators were placed on top of each other in three positions called left (L), central (C), and right (R). In addition, the difference of signal arrival times of both PMTs in each long scintillator, $\Delta t_{PMT}$, were constrained to small intervals such that the muon incidence angle on the LS box was restricted to a small range around \SI{0}{\degree} and the muon crossing points to a small area. The corresponding muon crossing areas are shown in Fig.~\ref{fig:alpha}.

Each crossing area corresponds to an azimuth angle with respect to the WOM symmetry axis, which we call $\alpha$, with $\alpha=0$ referring to the R2 position. We then reconstruct this azimuth angle using the collected integrated yield in each of the eight SiPM groups. For this, we use the average $\alpha$ position of each SiPM group $\alpha_i$ ($i=0, ..., 7$), knowing the orientation of the SiPM ring array on the WOM. The event-wise reconstructed azimuth angle is called $\phi_{ew}$. For the calculation of $\phi_{ew}$ in each event, we calculate a vector sum of $x_i$ and $y_i$ values from the $\alpha_i$ values weighted by the collected integrated yield $Q_i$ in the corresponding SiPM group: $x_i = \cos(\alpha_i) \cdot Q_i$, $y_i = \sin(\alpha_i) \cdot Q_i$, $X = \sum_{i=0}^7 x_i$, $Y = \sum_{i=0}^7 y_i$. From $X$ and $Y$, the value of $\phi_{ew}$ is calculated as: $\phi_{ew} = \arctan(Y/X)$.
The reconstruction of $\phi_{ew}$ is possibly biased if there is an intrinsic non-uniformity in the integrated yield in the eight SiPM groups, which can be caused by different sources: e.g. differences in SiPM efficiencies, differences in optical couplings, or shift of the SiPM array with respect to the WOM position. We estimate this non-uniformity by measuring the $\phi_{ew}$ distribution for the C0 position, for which one expects approximately a uniform integrated yield in the absence of bias sources as mentioned above.

\begin{figure}[ht]
    \centering
    \includegraphics[width = 0.49\textwidth]{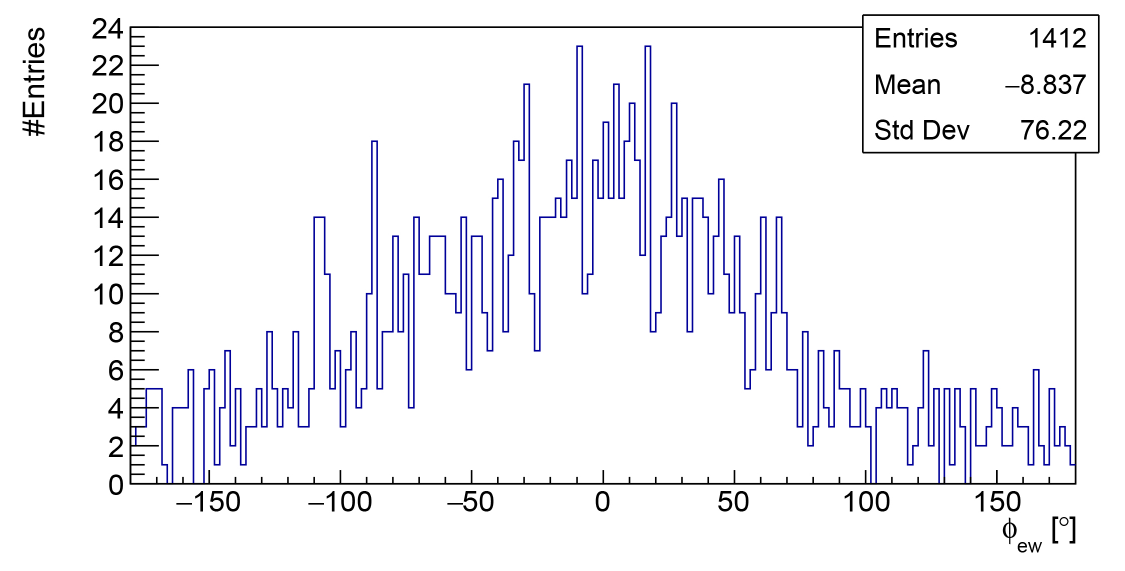}
    \includegraphics[width = 0.49\textwidth]{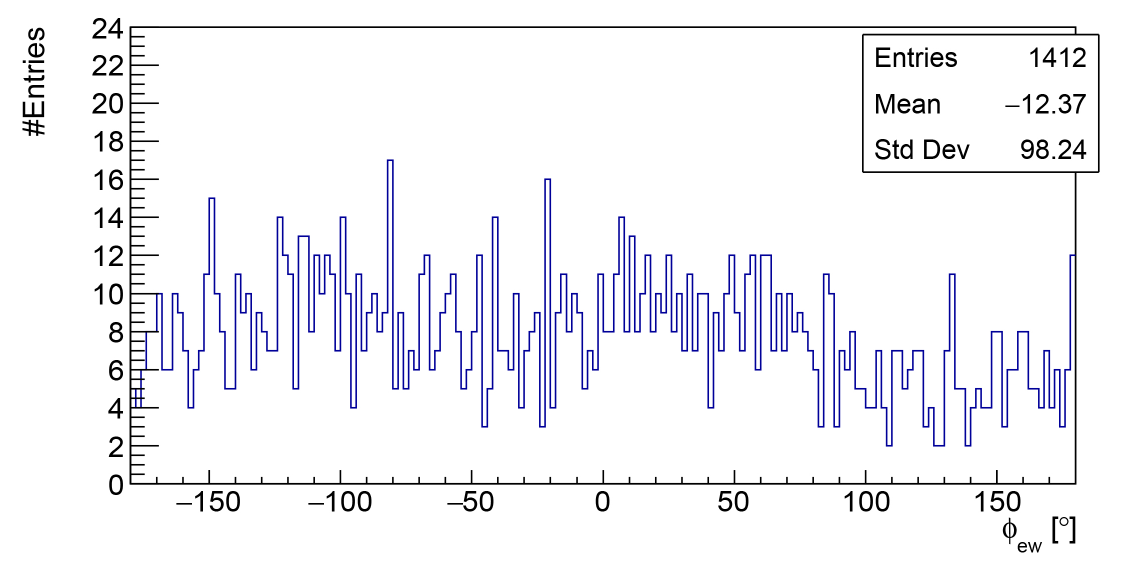}
    \caption{ $\phi_{ew}$ distribution measured at position C0. \textit{Left:} Before correction for non-uniformity. \textit{Right:} After correction for non-uniformity.}
    \label{fig:NonUniformity}
\end{figure}

Fig.~\ref{fig:NonUniformity} (\textit{left}) shows the $\phi_{ew}$ distribution for the C0 position, which clearly deviates from a uniform distribution. From the measurement at C0, we extract for each SiPM group $i$ the average integrated yield $Q_{i,av}^{\text{C0}}$ and calculate the following correction factor $\overline{Q_{i,av}^{\text{C0}}} = Q_{i,av}^{\text{C0}}/\sum_{k=0}^7 Q_{k,av}^{\text{C0}}$, which is then used as a multiplicative correction factor for each measured integrated yield $Q_i$.
As a cross-check that the correction procedure works, we show in Fig.~\ref{fig:NonUniformity} (\textit{right}) the $\phi_{ew}$ distribution for the C0 position after applying the correction. The $\phi_{ew}$ distribution is indeed much more uniform. The deviation from non-uniformity is likely caused by the fact that the definition of the C0 area is not perfectly symmetric with respect to the WOM symmetry axis.
\begin{figure}[ht]
    \centering
    \includegraphics[width = 0.49\textwidth]{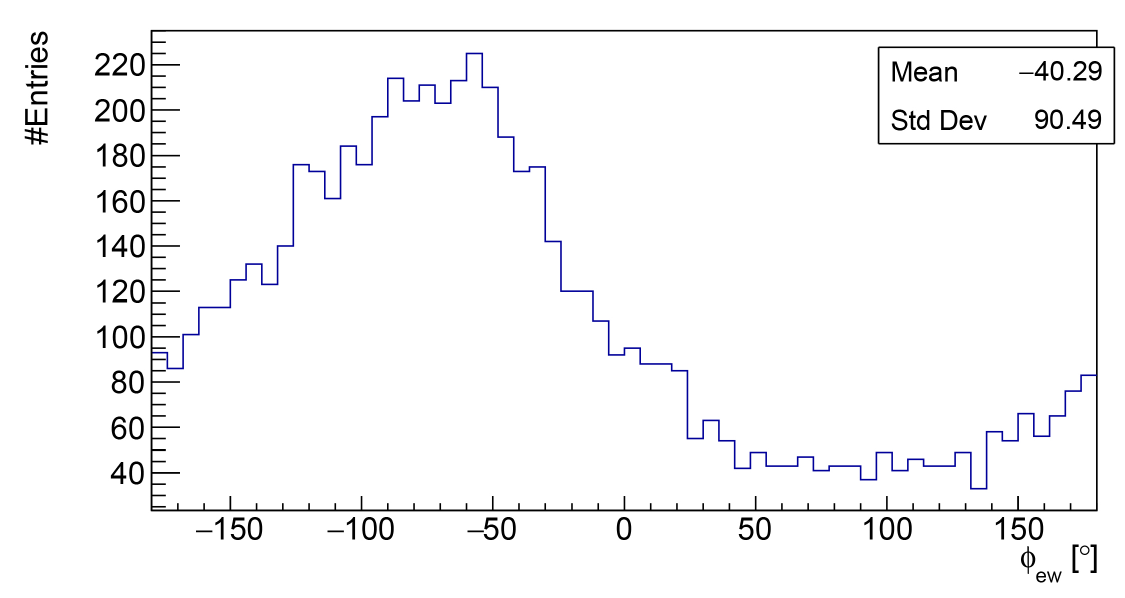}
    \includegraphics[width = 0.49\textwidth]{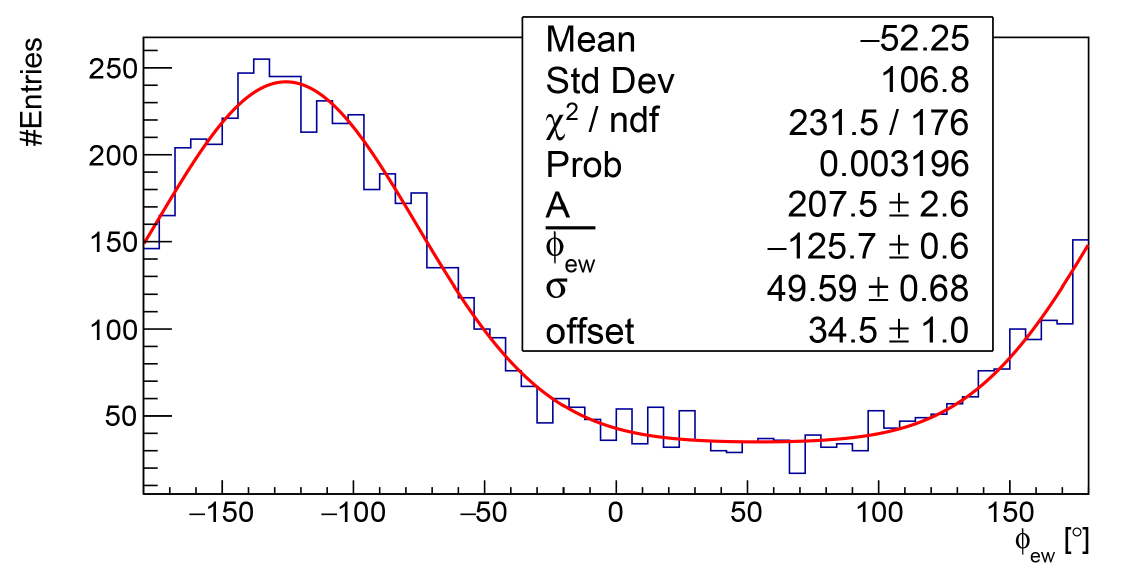}
    \caption{ $\phi_{ew}$ distribution measured at position C1. \textit{Left:} Before correction for non-uniformity. \textit{Right:} After correction for non-uniformity, incl. a fit to the distribution with a Gaussian (periodic in $\phi_{ew}$, plus a constant background term).}
    \label{fig:phi_ew_example_C1}
\end{figure}

Fig.~\ref{fig:phi_ew_example_C1} shows the $\phi_{ew}$ distribution as an example for the C1 position before (\textit{left}) and after (\textit{right}) applying the correction obtained from the measurement at the C0 position.

\begin{figure}[ht]
    \centering
    \includegraphics[width = 0.8\textwidth]{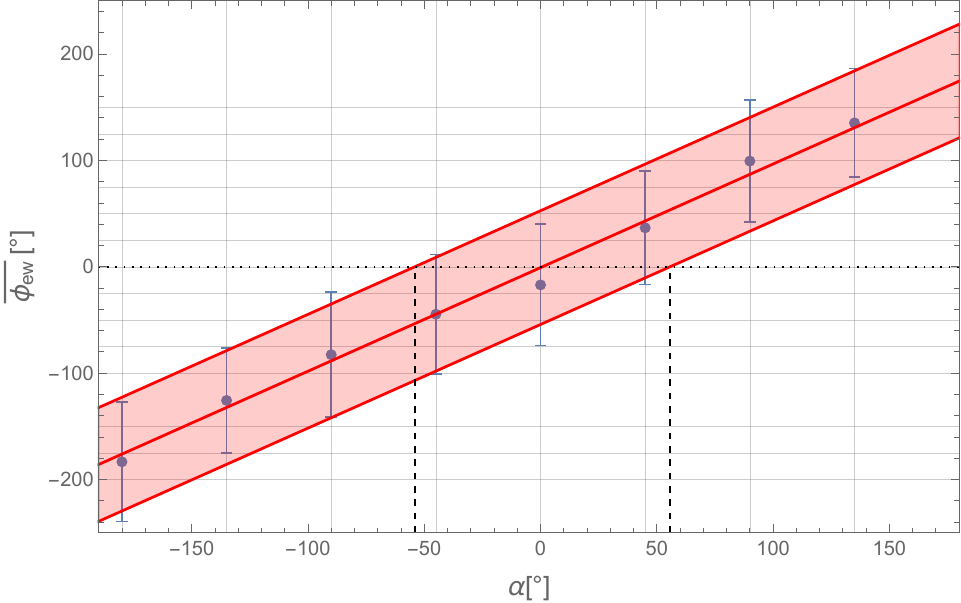}
    \caption{Average $\phi_{ew}$ value and corresponding standard deviation of the $\phi_{ew}$ distribution as a function of $\alpha$, after non-uniformity correction using the measurement at the C0 position.}
    \label{fig:phiew_alpha}
\end{figure}

Fig.~\ref{fig:phiew_alpha} shows the average over all events of the corrected $\phi_{ew}$ values, $\overline{\phi_{ew}}$, as a function of the particle azimuth angle $\alpha$ defined by the corresponding particle crossing area. The error bar quantifies the standard deviation $\sigma$ of the reconstructed $\phi_{ew}$ distribution and is about \SI{54.9}{\degree} on average. The quoted mean, $\overline{\phi_{ew}}$, and the standard deviation $\sigma$ for each  $\alpha$ are obtained from a fit, using as a fit function a Gaussian, which is periodic in $\phi_{ew}$, plus a constant background term, the result of which is shown for the C1 position in Fig.~\ref{fig:phi_ew_example_C1} (\textit{right}). To a very good approximation, the $\overline{\phi_{ew}}$ values show a linear dependence on $\alpha$, which allows to estimate $\alpha$ in a single event with an uncertainty of about \SI{55}{\degree}. The quoted precision depends on various parameters: 1) the distance of the light source to the WOM as well as the size of the box and the reflectivity of its walls, since the detection of primary photons being reflected on the cell walls before hitting the WOM will dilute the angular information; 2) the length of the WOM tube, since the  fraction of photons that spiral inside the WOM before being detected in the SiPM array decreases with tube length.

Final remark: The non-uniformity in the integrated yield in the SiPM groups could be also estimated by adding a LED to the SiPM array that sends UV-light pulses into the scintillator and then detecting the integrated yields in the eight SiPM groups. One could even avoid the determination of the non-uniformity, if one is able to measure the integrated yield in the SiPM groups for known particle crossing points, either using test beam data or using in the actual experiment information from other detectors that allow the determination of the particle crossing points.

\section{Simulation of photon transport inside a WOM and experimental validation}
\label{Sec:WOMTubeCharacterization}
The design of the WOM aims to capture the scintillation photons created in the LS volume and then guide the secondary photons towards the SiPMs at the end of the tube. As discussed in Section~\ref{Sec:SiPMDetectionEfficiency}, a lot of effort was made to increase the detection efficiency of the scintillation light. In addition to the discussed measures, one has to validate the WOM design and whether the secondary photons are efficiently guided towards the SiPMs.
\begin{figure}[ht]
    \centering
    \includegraphics[width=0.8\textwidth]{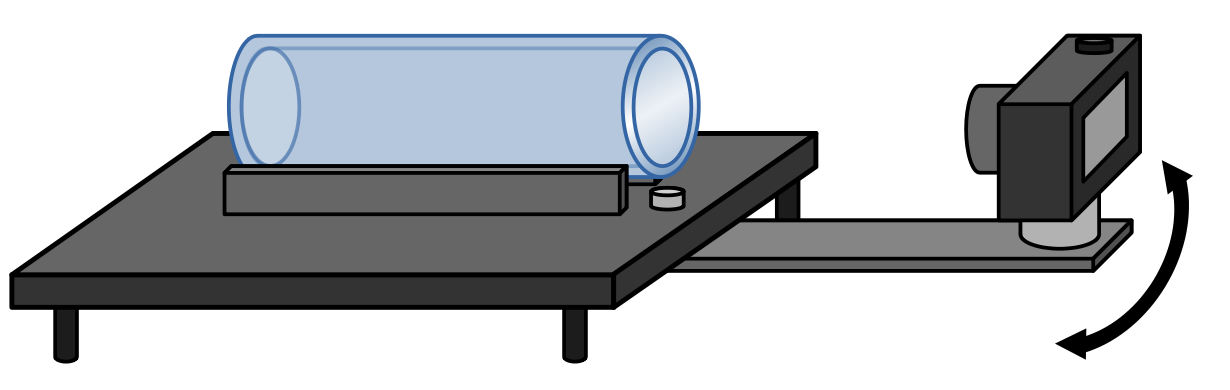}
    \caption{Sketch of the WOM test stand. The camera is fixed to a lever arm and can be rotated to change the observed photon exit angles. The centre of the WOM (which is sitting on two aluminium rods) is at the same height as the centre of the camera sensor.}
    \label{fig:WOMTeststandSketch}
\end{figure}

To better understand the characteristics of the photon transport inside the WOM, a setup was developed to test the properties of different tubes and compare the results to a Geant4 simulation. This setup, as sketched in Fig.~\ref{fig:WOMTeststandSketch}, is centred around a WOM and a digital single-lens reflex camera. The WOM sits on two aluminium rods, painted matte black to reduce reflection of stray light, to ensure minimal contact with the holding structure altering the outer surface of the tube. Primary photons are generated by a LED with a peak emission wavelength of \SI{375}{\nano\meter}~\cite{Nichia_LED} close to the peak emission wavelength \SI{360}{\nano\meter} of the LS (Section~\ref{Sec:LiquidScintillatorPurification}). The camera is mounted on a lever arm and can be rotated to capture photos of the WOM end surface. The rotation axis is aligned with the end of the WOM such that by rotating the camera and taking photos at different angles, we can capture secondary photons with various exit angles from the WOM end. This angular information is not accessible when taking data with SiPMs optically coupled to the WOM, since they integrate over all photon angles. Conversely, the camera is not optically coupled to the WOM, meaning that it can only capture photons transmitted through the front surface of the WOM. It can therefore not collect any photons above the angle for internal total reflection, as those photons cannot leave the WOM.
\begin{figure}[ht]
    \centering
    \includegraphics[width=0.325\textwidth]{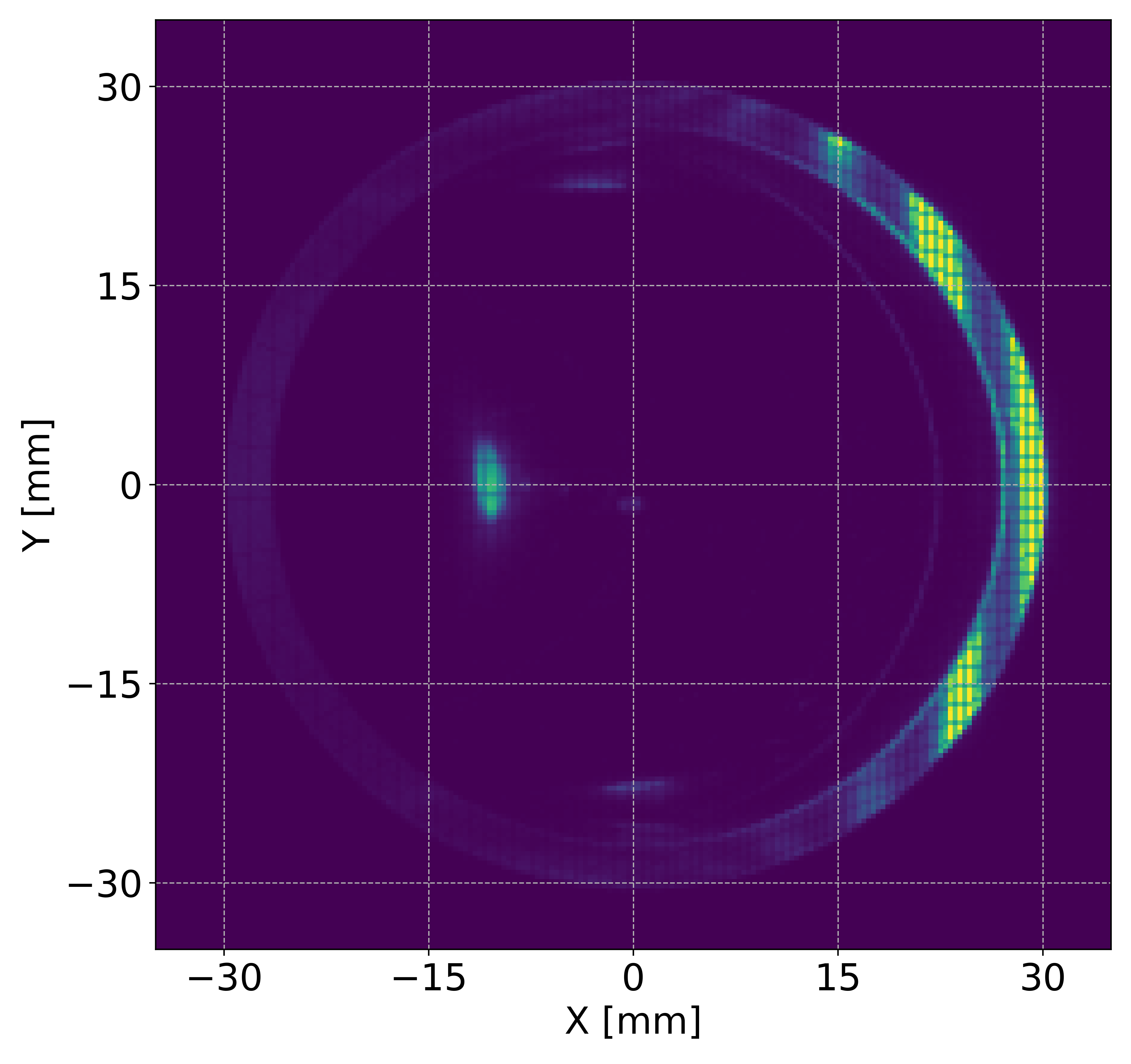}
    \includegraphics[width=0.325\textwidth]{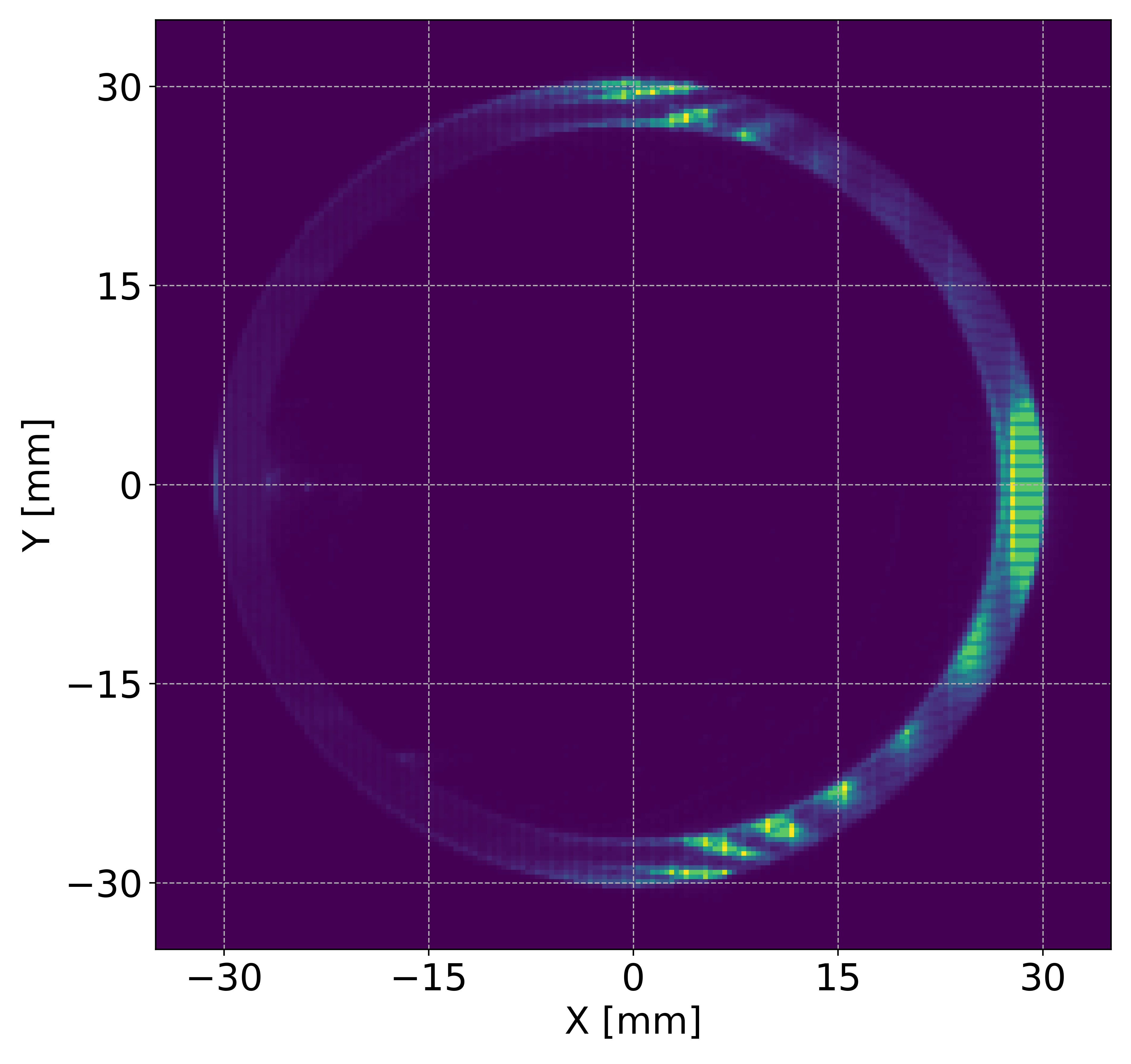}
    \includegraphics[width=0.325\textwidth]{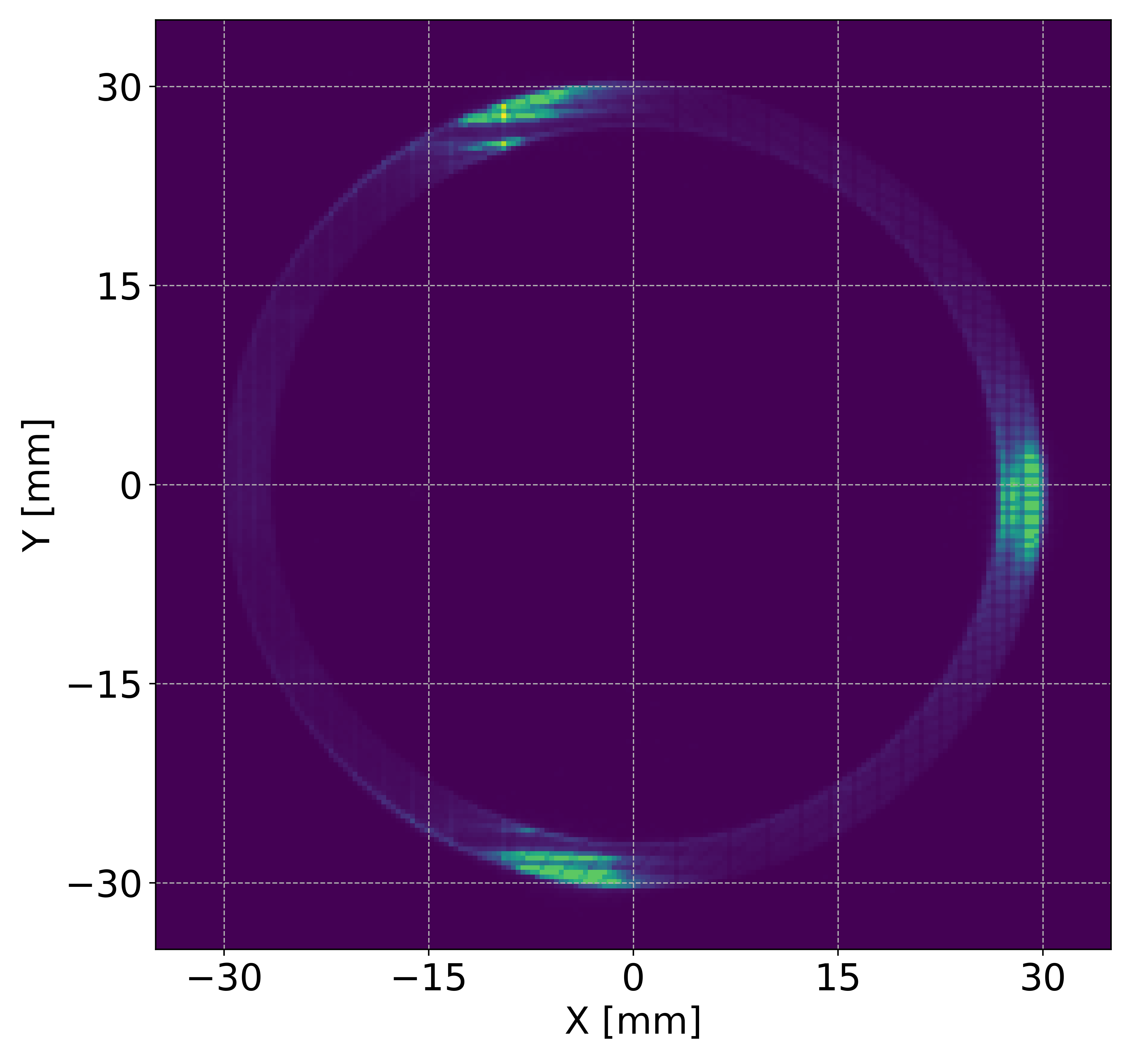}
    \includegraphics[width=0.325\textwidth]{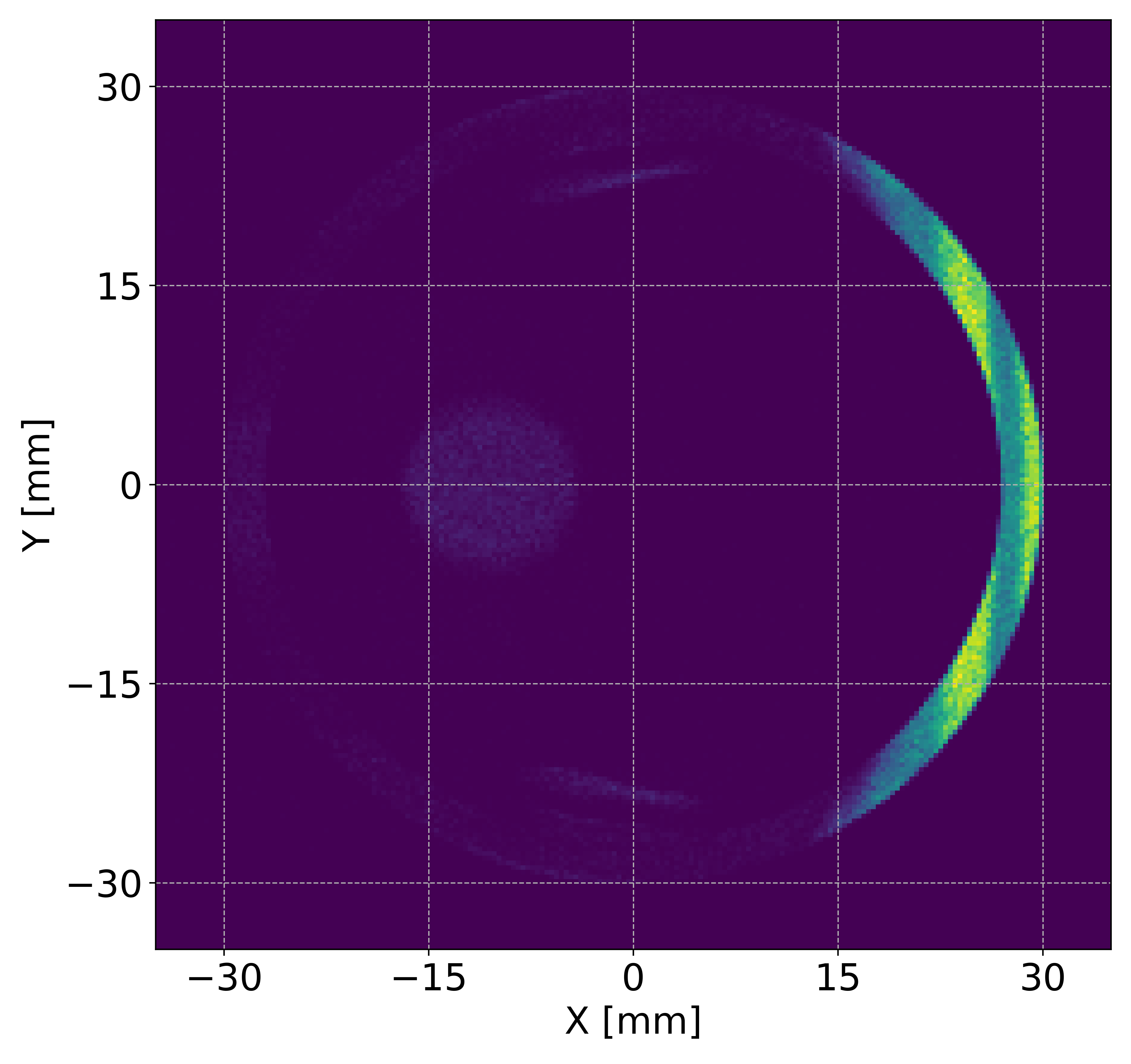}
    \includegraphics[width=0.325\textwidth]{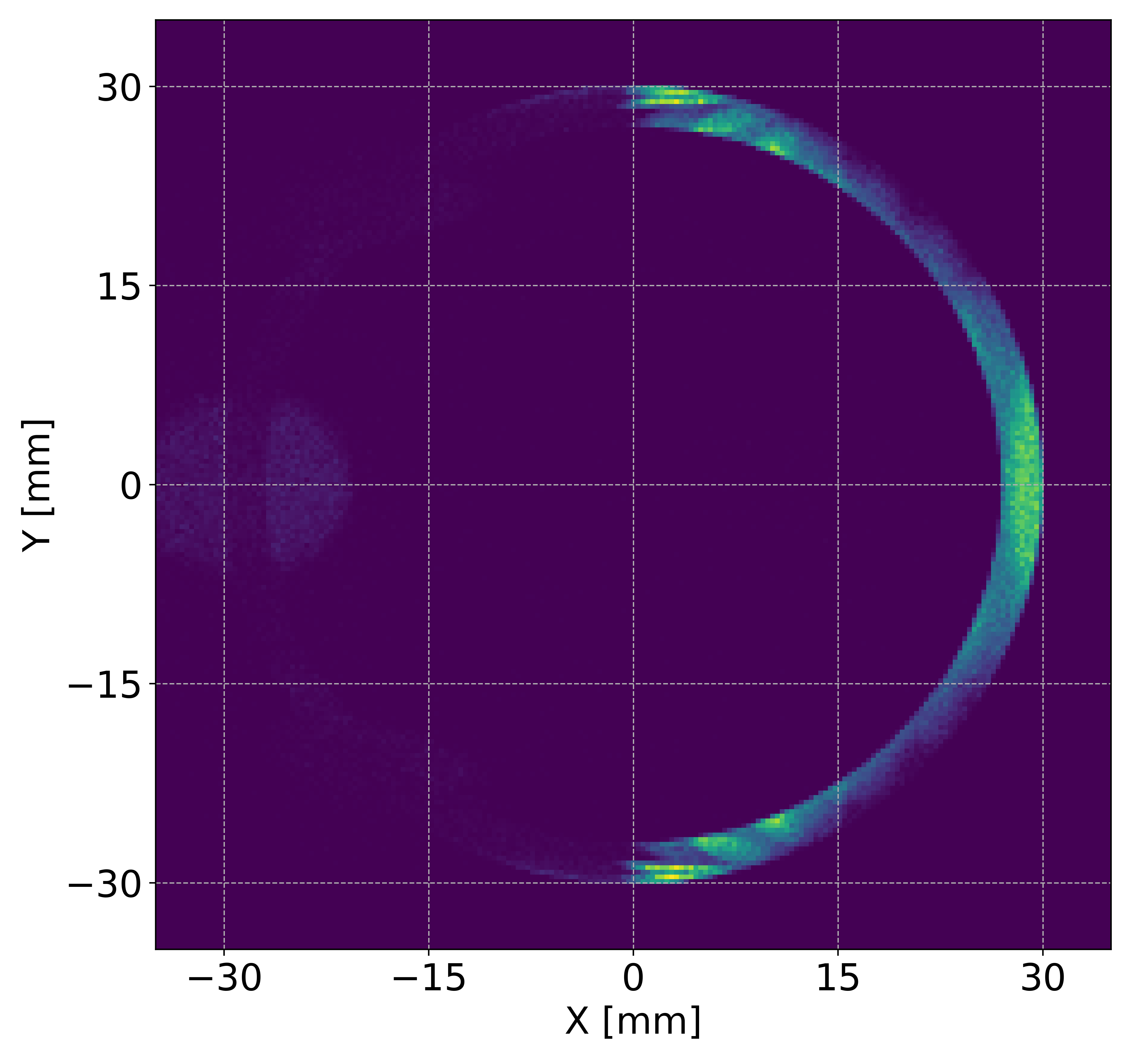}
    \includegraphics[width=0.325\textwidth]{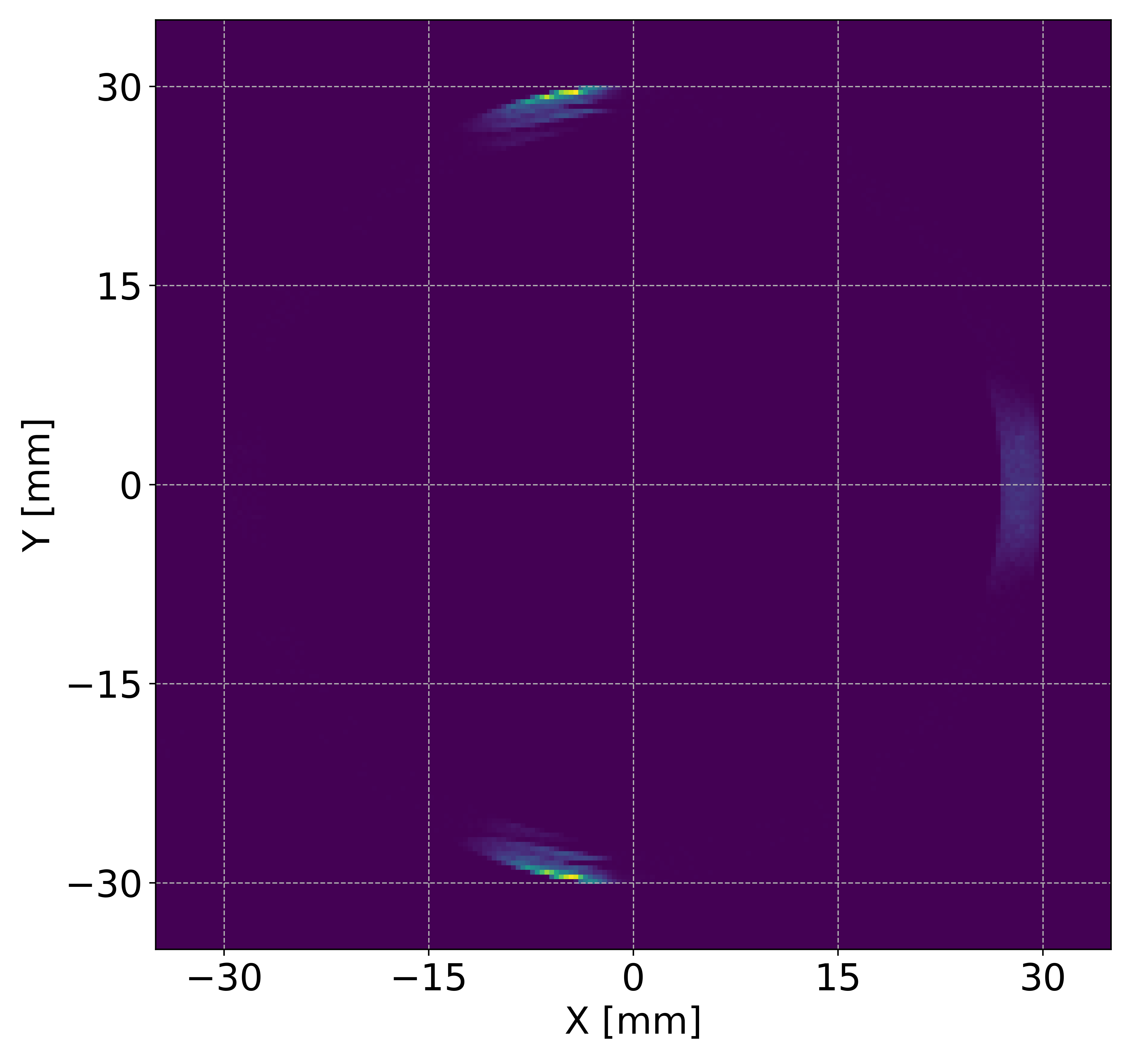}
    \caption{\textit{Top:} Photos of the front surface of a WOM, taken at the test stand. The LED is located on the right side of the WOM tube and was placed at half length and height of the tube. The photos were taken at different camera angles of \SI{20}{\degree} (\textit{left}), \SI{30}{\degree} (\textit{middle}), and \SI{40}{\degree} (\textit{right}), with the camera being rotated towards the WOM side opposite of the LED. \textit{Bottom:} Images with the same setup and camera angles simulated in Geant4.}
    \label{fig:WOMSurfacePhotos}
\end{figure}

In a first step, this setup was used to validate the simulation of the photon transport inside the WOM. It has shown that the secondary photons form distinct patterns on the front surface of the tube. Those patterns change with the position of the primary light source and the camera angle, i.e. the exit angle of the secondary photons. Fig.~\ref{fig:WOMSurfacePhotos} shows exemplary photos of a WOM and the corresponding simulated photos. The real photos underwent processing to compare them to the simulation. Firstly, only the blue colour channel of the raw images was selected. The red and green channels are dominated by noise as they are insensitive to the blue secondary photons. Secondly, a coordinate transformation is applied to the pixels on the camera sensor. By using the size of the camera sensor, the focal length of the camera lens, and the distance and angle of the camera relative to the WOM, this transformation can reconstruct the original geometry of the surface of the tube. 
While we are working on a calibration of the LED brightness and the camera detection efficiency, we cannot calculate the absolute brightness from the readout of the camera sensor yet. The colour scale of each photo is adjusted to its maximum brightness, which means they only display relative values. Therefore, the comparison of the photos remains mostly qualitative at this stage of the analysis. The photos in Fig.~\ref{fig:WOMSurfacePhotos} show that the simulation can reproduce the patterns on the front surface of the WOM. The positions of those nontrivial structures clearly coincide, while the relative heights of the different peaks in brightness can differ and need further investigation.
Nevertheless, qualitatively the simulation seems to properly simulate the photon-transport inside the WOM, as the patterns of brightness on the WOM surface and therefore the most probable photon paths can be reproduced. Additional comparisons have been conducted with varying LED positions along the length of the tube. Similarly to the displayed photos the brightness patterns can be reproduced qualitatively, therefore further validating the proper simulation of photon transport inside the tube.

\section{Mechanical design and construction of the full-size detector prototype}
\label{Sec:LiquidScintillatorDetectorCell}

\begin{figure}[htb]
    \centering
    \includegraphics[width = 1.0\textwidth]{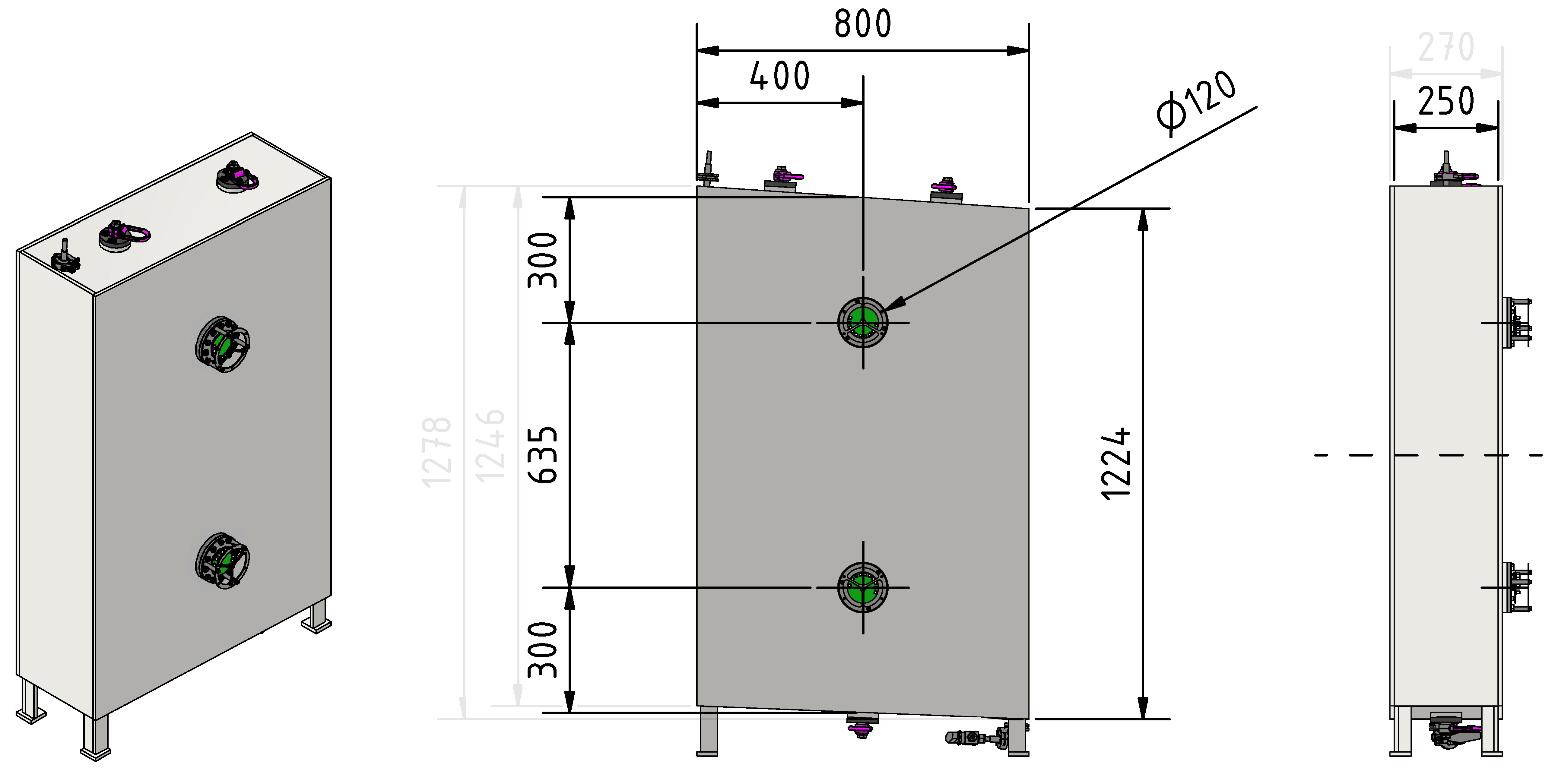}
    \caption{CAD drawing of the full-size LS-SBT detector prototype cell. All dimensions are given in mm, the thickness of the corten steel walls is 10\,mm.}
    \label{fig:cell}
\end{figure}

The goal of this work was to characterise the performance of a full-size LS-SBT prototype detector cell in a dedicated test beam campaign: The realised LS cell (see Fig.~\ref{fig:cell}) is representative for a vertical side segment of the SBT. It has an outer width of 800\,mm, while its side height varies from 1224\,mm to 1246\,mm (corresponding to the shape of the widening decay vessel, with the detector cell's upper corners of the longer side being 1278\,mm above the lower corners of the shorter side). The outer depth of the cell is 270\,mm, thus giving a LS thickness of 250\,mm. To save material, the cell was built from corten steel (S355 J2W 1.8965) sheets of 10\,mm thicknes, resulting in a total weight of the filled detector prototype of $\sim$550~kg.

Six large corten steel sheets were welded together forming the detector cell, with all possible welding seams inside this box. Two circular openings of 70\,mm diameter were cut into the cell's front and equipped with corten steel flanges to accommodate the WOM PMMA vessels. These are horizontally centred on the front sheet at a distance of 400\,mm to the cell sides, and vertically separated by 635\,mm at a distance of 300\,mm from the upper / lower cell sheets. Two crane hooks welded to the top and and two crane hooks at the bottom allow for transport. Short corten steel standoffs welded to the bottom support the cell in an upright position, ensuring the vertical sides being orthogonal to the floor. The box is furthermore equipped with two short KF16 stainless steel nipples at the highest and lowest points intended for filling and emptying of the LS cell. The bottom opening can be closed with a KF16 ball valve, while the top opening is connected to an expansion vessel (made of a standard ISO-K~160 stainless steel full nipple of 256\,mm length and two flanges). 

After welding, the inside of the detector cell was cleaned with acetone and spray-coated via the WOM openings with two layers of primer followed by two layers of reflective coating (see Section~\ref{Sec:CellWallReflectivity}).

\bigskip
The construction of the final SHiP LS-SBT detector will require about 4000 PMMA vessels housing the same number of WOMs. Selecting an appropriate manufacturing method is thus essential to control the detector expenses -- potential options comprise injection moulding, adhesive bonding, or welding. The current PMMA vessel design uses machined, semi-finished pipes and sheet material that are joined by ultrasonic welding: This process provides the flexibility to adjust the dimensions of the PMMA vessels at any time to meet changing requirements of R\&D, while also remaining cost efficient.

\begin{figure}[htb]
    \centering
    \includegraphics[width = 1.0\textwidth]{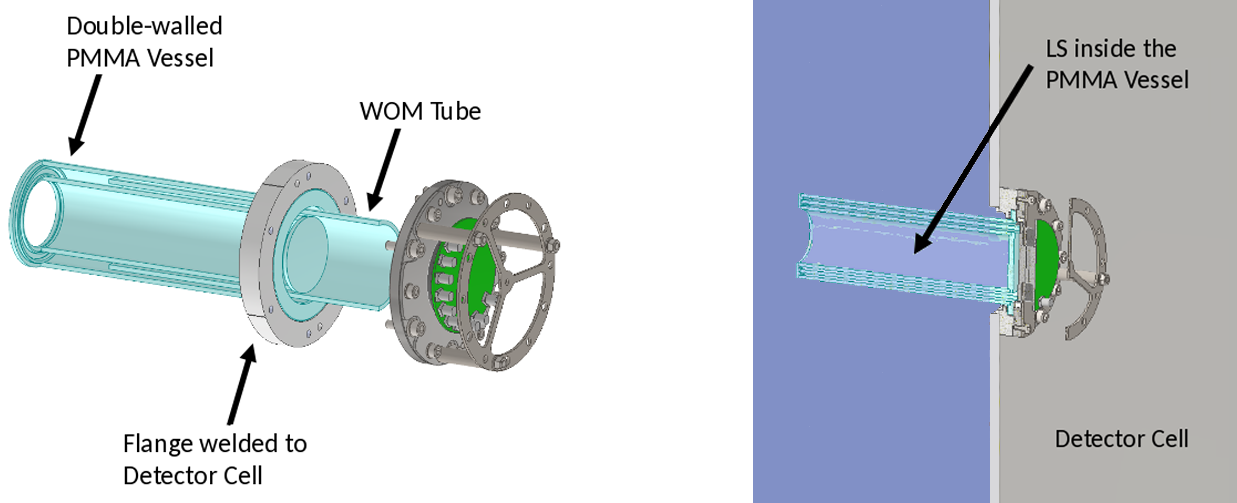}
    \caption{\textit{Left:} CAD sketch of the PMMA WOM vessel illustrating the integration of WOM tube and eMUSIC readout into the LS detector cell. \textit{Right:} The double-walled structure of the PMMA vessel design increases the active detection volume by allowing LS within the WOM tube.}
    \label{fig:WOM_vesselsketch}
\end{figure}

The WOM vessels consist of two concentric tubes of extruded PMMA with 3\,mm thickness, a length of 223\,mm, and outer radii of 50\,mm and 70\,mm, respectively. On the "far" side, the ends of the vessel tubes are connected and joined in a leak-tight way by a PMMA ring of 70\,mm outer diameter and 23\,mm width. At the "readout" side, the cavity created between the vessel tubes remains open for WOM insertion and readout. A circular groove in the PMMA ring at the "far" side of the vessel secures the position of the WOM within. At the "readout" side, the inner vessel tube is sealed with a PMMA disk, while a PMMA ring welded to the outer tube allows fixture to the detector box and liquid sealing. This double-walled vessel geometry provides an air layer around the WOM (required for total reflection of the WLS photons, see Section~\ref{sec:DetectorCellConcept}), but also permits the LS to surround the WOM from both the inside and the outside, thus reducing the active detector volume by less than $0.5\,\% $ (see Fig.~\ref{fig:WOM_vesselsketch}).

The ultrasonic PMMA welding used a HiQ sonotrode with 70\,mm diameter and integrated energy directors with a height of $0.5$\,mm that was developed by Hermann Ultraschall. Preliminary tests revealed the need for a dedicated fixture holding the tubes during the welding process: This ensures reproducible placement and alignment of the components, and also prevents the PMMA tube from rotating, thus maintaining the precision of the assembly. 

In order to reduce stress in the welds, the PMMA vessels were afterwards annealed at \SI{80}{\degree}C for several hours and then slowly cooled. All WOM vessels were tested for leakage using helium at 3\,bar overpressure.

\section{DESY test beam exposure of the full-size detector prototype}
\label{Sec:DESYtestbeam}
In order to characterise and understand its performance, a full-scale liquid scintillator SBT cell prototype was tested in October 2022 at the DESY II test beam area 24~\cite{Diener:2018qap}. 
Even though the response of the LS-SBT to the used positron beam of various energies is different from MIPs, the study is well-suited for an initial detector characterization.

\subsection{Detector setup}
\label{Sec:DetectorSetup}
The detector cell was mounted in the test beam area on a rotating platform, fastened to a movable table. It was mounted upright with the face containing the WOM vessels and electronics positioned away from the beam. The coordinate system was defined such that the horizontal axis on the detector cell is labelled as $x$, the vertical axis $y$ and the beam direction $z$. In this coordinate system, the movable table allowed for translation of the cell along $x$ and $y$, and the rotating platform allowed for rotation around the $y$-axis of the cell in steps of \SI{15}{\degree}. The movable table was approximately 3\,m away from the beam exit window. The beam height in the test beam area is approximately 1.7\,m and a 2\,mm~$\times$~2\,mm secondary collimator was used to focus the beam. The momentum spread in the beam was measured to be $(158\pm6)$\,MeV/c over the full momentum range. The DESY test beam telescope, consisting of four small plastic scintillators coupled to PMTs, was located between the collimator and the LS prototype detector, in line with the beam. A detailed description of the beam and telescope can be found in~\cite{Diener:2018qap}. The coincidence signal of these four PMTs was used to trigger the data acquisition with the WaveCatcher digitiser.

In the specified coordinate system, the central point of the cell is the origin of the $x$ and $y$ axes. The angular orientation is defined such that when the wall of the detector that houses the WOM vessels and electronics is orthogonal to the beam, the angle is \SI{0}{\degree}. Clockwise rotation of the cell on the platform is defined as a positive angle. For the orientation of the detector prototype in the upright position as shown in Fig. \ref{fig:cell}, the WOM located in the upper place is called "WOM up" in the following, the WOM located in the lower place "WOM down" accordingly.

\subsection{Readout electronics and data acquisition}
\label{Sec:ElectronicsAndDAQ}
For the readout of photons collected by the two WOM, a novel and adaptable readout electronics system has been devised to accommodate the considerable flexibility required during the developmental phases of detector R\&D~\cite{Alt_2023}. Each readout board is equipped with a total of 40 Hamamatsu S14160-3050HS SiPMs~\cite{Hamamatsu-S14160} to convert the photons delivered by the WOM into electrical charge signals and is equipped with a complement of 40 Hamamatsu S14160-3050HS SiPMs~\cite{Hamamatsu-S14160}. A shared voltage source powers each SiPM, with a 10\,nF capacitor placed at the cathode to provide additional input charge to the cathode during the break down process.The anodes of the SiPMs are individually connected to a 100-pin SAMTEC LSHM-150-XX.X-X-DV-A-N high density connector by a 0\,$\Omega$ resistor. All signal lines are meticulously isolated from adjacent ones through pins that establish ground connections, thus averting unwanted cross-talk. The inclusion or exclusion of specific SiPMs in the readout can be managed by the selective removal of the 0\,$\Omega$  resistor. Moreover, in cases where a voltage-sensitive amplifier is preferred over a transimpedance amplifier, an additional resistor can be utilised to link the SiPM anodes to ground, thereby offering the desired electrical configuration matching the voltage sensitive amplifier.

For the amplification of the SiPM signals, a fitting amplification board was developed, utilising the eMUSIC ASIC provided by Scientifica~\cite{MUSIC-ASIC}. This board serves as the housing platform for the eMUSIC chip itself and an ATmega328P-AU microprocessor responsible for configuring the MUSIC ASIC during detector initialisation.
The interconnection between the amplification board and the SiPM board is facilitated through the same SAMTEC connector mentioned above. Between the high-density connector and the input to the eMUSIC chip the SiPMs signal lines are re-organised into eight groups, with each group comprising five neighboured SiPMs. These eight groups are subsequently linked to the eight input channels of the eMUSIC chip.
This chip features an array of configurable parameters accessible through the ATmega micro-controller by the Serial Peripheral Interface protocol. It offers two distinct gain configurations, enhancing versatility in signal amplification. Moreover, the ASIC embedded within the eMUSIC chip incorporates tuneable Pole-Zero cancellation, resulting in output signals characterised by a Full Width at Half Maximum of less than 10 nanoseconds. In cases where signal shaping is unnecessary, the Pole-Zero cancellation can be bypassed. The board also includes SMA connectors designed for seamless interfacing with the digitisation unit, thereby ensuring the efficient transmission of amplified signals for further analysis and processing.

The 16 analog signals from two boards housing the eMUSIC amplifiers are connected by  3\,m long RG174 cables to a single WaveCatcher module~\cite{WaveCatcher}. The WaveCatcher electronics comprises a family of electronic boards based on the SAMLONG switched capacitor array technology, designed as an alternative to conventional ADC-based digitisers or oscilloscopes. In the setup at DESY, a 16-channel WaveCatcher board has been used. It allows for the accurate acquisition and processing of signals spanning from -1.25 V to +1.25 V with up to 4096 discrete level steps. This high resolution is critical for capturing and characterising subtle variations within the input signals. The input bandwidth is limited to 500     MHz while the sampling rates, offer a variable range between 0.4 and 3.2 GS/s (Giga Samples per second). The combination of their high bandwidth and adjustable sampling rates renders the WaveCatcher boards particularly suitable for the precise acquisition of high-speed signals, such as very short pulses. This capability is pivotal for applications necessitating time-domain analysis and pulse characterisation.

\subsection{Collected data}
\label{Sec:CollectedData}
\begin{figure}[ht]
    \centering   
    \begin{subfigure}[b]{0.62\textwidth}
        \centering
        \includegraphics[width =1.0\textwidth]{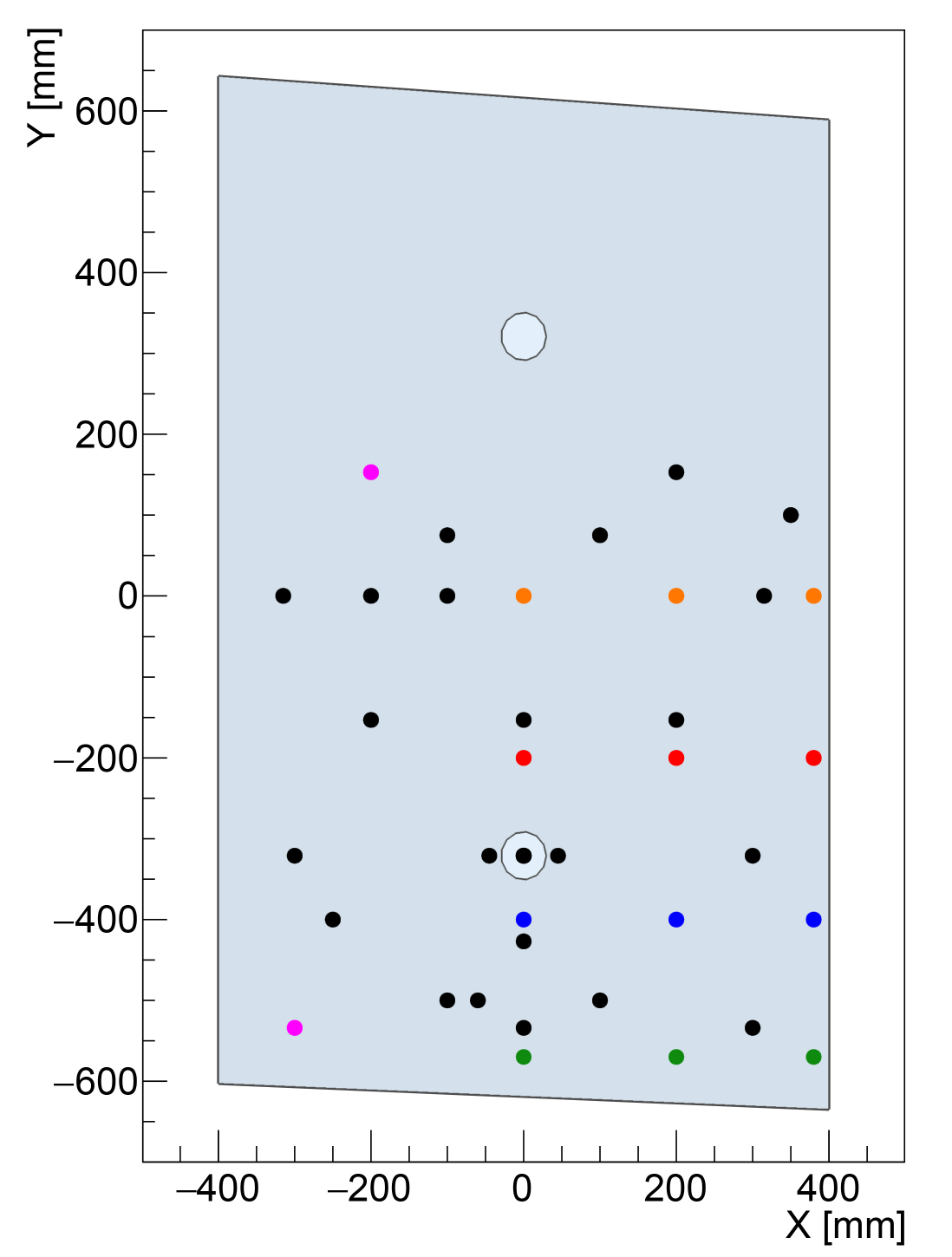}
        \caption*{} \label{fig:deglocations_0_75}
    \end{subfigure}
    \hspace{5mm}
    \begin{subfigure}[b]{0.31\textwidth}
         \centering
         \includegraphics[width = \textwidth]{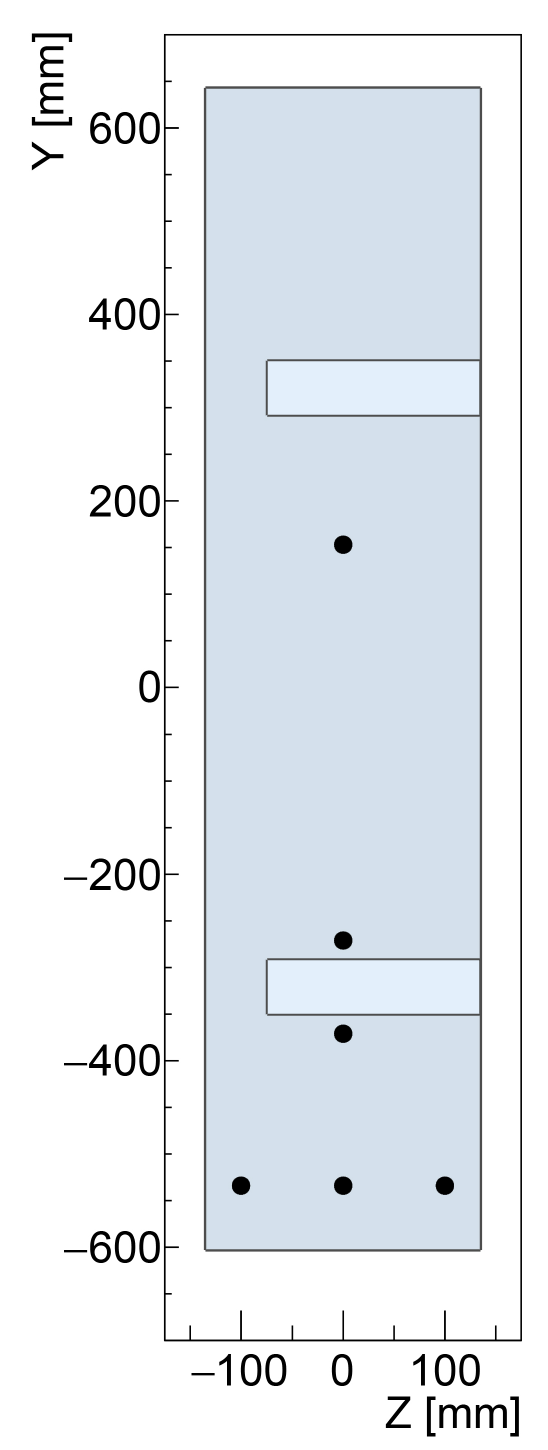}
         \caption*{} \label{fig:deglocations_90}
    \end{subfigure}
    \caption{Locations of all particle crossing points on the detector cell. \textit{Left:} Points mark the location where the particle crosses the central plane of the detector for incident beam angles between 0 and \SI{75}{\degree}. \textit{Right:} Points mark the location where the particle enters the detector from the side for incident beam angles of \SI{90}{\degree}. Colours indicate points that have been grouped together in subsequent analysis. The outer walls of the detector cell and the locations of the WOMs are shown by outlines.}
    \label{fig:deglocations}
\end{figure}

Several particle crossing points within the detector cell were measured, along with various angles between the beam and cell. All measured particle crossing points on the cell are shown in Fig.~\ref{fig:deglocations}.

For each incident beam angle, only a subset of these positions were recorded; measured particle crossing points at each beam angle can be found in Table~\ref{tab:beampoints}.

\begin{table}[ht]
\begin{minipage}{0.45\textwidth}
    \footnotesize
    \setlength{\tabcolsep}{2pt}
    \begin{tabular}{r r r rrr rrrrrrr}
    \toprule
        \textbf{Beam X} && \textbf{Beam Y} &&&& \multicolumn{7}{r}{\textbf{Angle}} \\
        \textbf{[mm]}   && \textbf{[mm]}   &&&& \multicolumn{7}{r}{\textbf{[\SI{}{\degree}]}} \\
    \midrule
           0 &&  153 &&&&   &    &    &    &    &    & 90 \\
        -200 &&  153 &&&& 0 & 15 & 30 & 45 & 60 & 75 &    \\
         200 &&  153 &&&&   & 15 & 30 & 45 & 60 & 75 &    \\ 
    \hline
         350 &&  100 &&&&   &    &    &    &    & 75 &    \\
    \hline
        -100 &&   75 &&&& 0 &    & 30 &    & 60 &    &    \\
         100 &&   75 &&&& 0 &    & 30 &    & 60 &    &    \\
    \hline      
           0 &&    0 &&&& 0 & 15 & 30 & 45 & 60 & 75 &    \\
        -100 &&    0 &&&& 0 &    &    &    &    &    &    \\
        -200 &&    0 &&&& 0 &    &    &    & 60 & 75 &    \\
         200 &&    0 &&&& 0 &    & 30 &    & 60 & 75 &    \\
        -315 &&    0 &&&&   & 15 & 30 & 45 &    &    &    \\
         315 &&    0 &&&&   & 15 & 30 & 45 &    &    &    \\
         380 &&    0 &&&& 0 &    &    &    &    &    &    \\  
    \hline
           0 && -153 &&&& 0 &    & 30 &    & 60 &    &    \\
        -200 && -153 &&&& 0 & 15 & 30 & 45 & 60 & 75 &    \\
         200 && -153 &&&&   & 15 & 30 & 45 & 60 & 75 &    \\
    \hline
           0 && -200 &&&& 0 &    &    &    &    &    &    \\
         200 && -200 &&&& 0 &    &    &    &    &    &    \\
         380 && -200 &&&& 0 &    &    &    &    &    &    \\
    \hline      
           0 && -271 &&&&   &    &    &    &    &    & 90 \\
    \hline    
           0 && -371 &&&&   &    &    &    &    &    & 90 \\
\bottomrule  
    \end{tabular}
\end{minipage}
\begin{minipage}{0.45\textwidth}
    \centering
    \footnotesize
    \setlength{\tabcolsep}{2pt}
    \begin{tabular}{r r r rrr rrrrrrr}
    \toprule
        \textbf{Beam X} && \textbf{Beam Y} &&&& \multicolumn{7}{r}{\textbf{Angle}} \\
        \textbf{[mm]}   && \textbf{[mm]}   &&&& \multicolumn{7}{r}{\textbf{[\SI{}{\degree}]}} \\
    \midrule
           0 && -321 &&&&   & 15 & 30 & 45 & 60 & 75 &    \\
         -45 && -321 &&&&   & 15 & 30 & 45 & 60 & 75 &    \\
          45 && -321 &&&&   & 15 & 30 & 45 & 60 & 75 &    \\
        -300 && -321 &&&&   & 15 & 30 & 45 & 60 & 75 &    \\
         300 && -321 &&&&   & 15 & 30 & 45 & 60 & 75 &    \\
    \hline
           0 && -400 &&&& 0 &    &    &    &    &    &    \\
         200 && -400 &&&& 0 &    &    &    &    &    &    \\
        -250 && -400 &&&&   & 15 &    &    &    &    &    \\
         380 && -400 &&&& 0 &    &    &    &    &    &    \\
    \hline    
           0 && -427 &&&& 0 &    & 30 &    & 60 &    &    \\
    \hline   
         -60 && -500 &&&&   &    & 30 &    &    &    &    \\
        -100 && -500 &&&& 0 &    & 30 &    & 60 & 75 &    \\
         100 && -500 &&&& 0 &    & 30 &    & 60 & 75 &    \\
    \hline   
           0 && -534 &&&&   & 15 & 30 & 45 & 60 & 75 & 90 \\
        -100 && -534 &&&&   &    &    &    &    &    & 90 \\
         100 && -534 &&&&   &    &    &    &    &    & 90 \\
        -300 && -534 &&&& 0 & 15 & 30 & 45 & 60 &    &    \\
         300 && -534 &&&&   & 15 & 30 & 45 & 60 &    &    \\
    \hline     
           0 && -570 &&&& 0 &    &    &    &    &    &    \\
         200 && -570 &&&& 0 &    &    &    &    &    &    \\
         380 && -570 &&&& 0 &    &    &    &    &    &    \\
    \bottomrule
    \end{tabular}
\end{minipage}
\caption{All particle crossing points on the detector cell for each angle of detector rotation (compare Fig.~\ref{fig:deglocations}). The positive angles indicate a clockwise rotation of the cell around the $Y$ axis.}
\label{tab:beampoints}
\end{table}

For each beam position and angle, data was collected at five beam energies: 1.4\,GeV, 2.4\,GeV, 3.4\,GeV, 4.4\,GeV, and 5.4\,GeV. In each run, 10000 events were recorded.

\subsection{Monte Carlo simulation of the LS detector cell}
\label{Sec:MonteCarloSimulation}

A Geant4 simulation of the full detector cell was developed including the exact cell dimensions and corresponding wall thicknesses, properties of the LS, and the relevant properties for subsequent light transport. The LAB (Hyblene-113 from SASOL company) chemical composition and density are taken from the datasheet provided by the manufacturer. Refractive indices of the LS are taken from~\cite{Tseung} and the light yield of 10830 photons per MeV deposition is from~\cite{Anderson}. The emission spectrum of PPO dissolved in LAB is taken from~\cite{Marrodan}. The slow and fast decay time components are 4.3\,ns and 13.4\,ns, respectively, as measured in~\cite{Onken}. The absorption length of the LS both before and after purification are taken from measurements performed in Mainz (Section~\ref{Sec:LiquidScintillatorPurification}). The reflective coating of the inner cell wall is implemented in the simulation with reflectivity measurements performed in Mainz (Section~\ref{Sec:CellWallReflectivity}). The refractive indices of PMMA are taken from~\cite{Sultanova}. The attenuation lengths of the PMMA making up the WOMs and the vessel housing them are both taken from~\cite{Bodmer}. The chemical composition of the WLS WOM coating is implemented in the simulation. The attenuation length of the WLS, as well as an estimated WLS layer thickness of 20~$\mu$m are based on measurements performed in Berlin
for glass slides, (Section~\ref{Sec:WOMAbsorptionProbability}), assuming that the layer thickness on the WOM tubes is at least as large as the minimal values achieved on glass slides. The SiPM window in the simulation is in direct contact with the WOM surface. The refractive index of the SiPM window and SiPM detection efficiency depending on wavelength are taken from the manufacturer's datasheet. All 40 SiPMs per WOM are included in the simulation but for analysis purposes they are considered in groups of five to mimic the detector readout.

To produce simulated events, a particle gun of defined energy, position, and direction is specified, simulating electromagnetic showers within the detector cell, scintillation and Cerenkov processes, and optical photon transportation within the LS and PMMA components until absorption in the SiPM material. The part of photon propagation within the WOM was already discussed in Section~\ref{Sec:WOMTubeCharacterization}. Extractable quantities from the simulation include the total energy deposition and number of scintillation photons produced inside the detector cell, as well as the number and time distribution of detected photons in each SiPM group. 

\begin{figure}
    \centering
    \includegraphics[width = 0.5\textwidth]{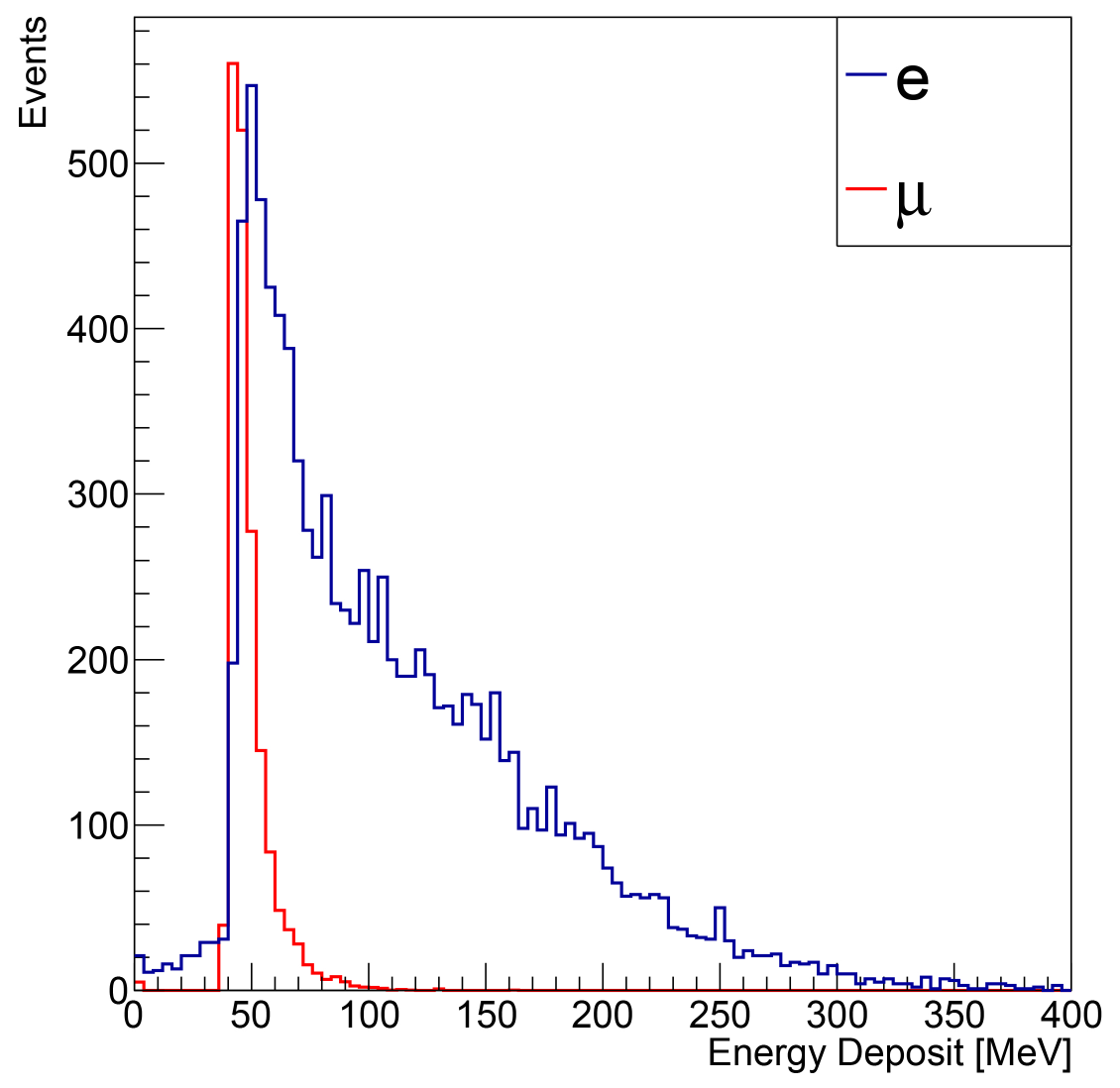}
    \caption{Energy deposit inside the LS in simulation for a positron (\textit{blue}) and muon (\textit{red}) hitting the centre of the cell with a beam angle of \SI{0}{\degree} and an energy of 1.4\,GeV. The mean (or most probable) energy deposition for a positron is 113 MeV (or 50 MeV) and the standard deviation of the energy deposition distribution is 62 MeV. The mean (or most probable) energy deposit for a muon is 49 MeV (or 44 MeV), and the standard deviation of the energy deposition distribution is 9 MeV. \hbox{10000 events} were simulated for both samples. (The distribution for the muon has been re-scaled for visualisation purposes to have similar maximum amplitudes in the positron and muon distributions.)}
    \label{fig:simulation_edep}
\end{figure}

Fig.~\ref{fig:simulation_edep} shows the distribution of energy deposited inside the LS for a beam angle of \SI{0}{\degree} and energy of 1.4\,GeV. The distribution peaks at 50 MeV and has a mean of about 113 MeV and a standard deviation of about 62 MeV. As a comparison, the most probable value of the energy loss of a minimum ionising particle seen in the liquid scintillator is around 44 MeV.

\begin{figure}
    \centering
    \includegraphics[width = 1.0\textwidth]{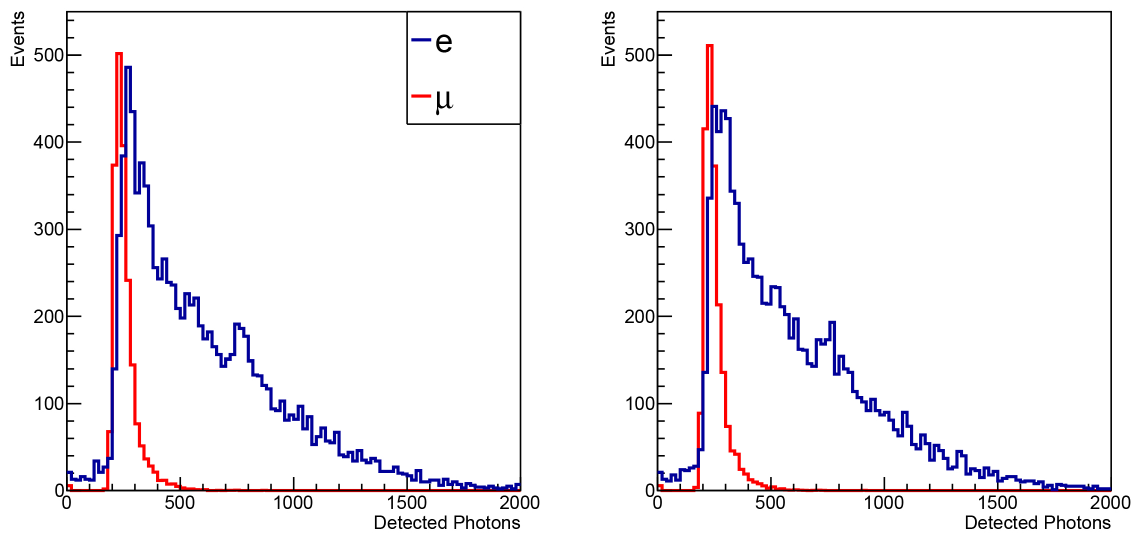}
    \caption{Example distribution of detected photons in simulation for a positron (\textit{blue}) and \hbox{muon (\textit{red})} hitting the centre of the cell. The beam angle is \SI{0}{\degree} and energy is 1.4\,GeV. The reflectivity of the inner cell walls is \SI{65}{\percent} of the measured values. 10000 events were simulated for both samples. (The distribution for the muon has been re-scaled for visualisation purposes to have similar maximum amplitudes in the positron and muon distributions.) \textit{Left:} WOM down, \textit{right:} WOM up.}
    \label{fig:simulation_example}
\end{figure}

As an example, the distributions of detected photons in each WOM for a beam in the centre of the cell at \SI{0}{\degree} and 1.4\,GeV are shown in Fig.~\ref{fig:simulation_example}
using a reflectivity of the inner cell walls of \SI{65}{\percent} of the measured values as shown in Section~\ref{Sec:CellWallReflectivity}.

There are some notable differences between the simulation and physical setup, namely, the fact that in the simulation the reflective coating of the inner cell walls is completely uniform, while the coating was observed to vary widely over different regions within the detector cell due to the observed rust stains as described in Section~\ref{Sec:CellWallReflectivity}. There is also direct contact between the WOMs and SiPMs in the simulation, while a silicon gel pad was used in the detector cell. The DAQ electronics are not present in the simulation, when photons reach the SiPM detection surface they are counted according to the implemented SiPM detection efficiency. In this case, effects of the electronic components such as SiPM dark counts, crosstalk, or afterpulsing, as well as electronic noise, saturation, or delays due to cable length are not implemented in the simulation. Of these differences the discrepancy in reflective coating between the simulation and detector cell, along with effects of the electronics are expected to have the largest impact.

\subsection{Efficiency of the full-size detector prototype}
\label{Sec:Efficiency}
\begin{figure}
    \centering
    \includegraphics[width=0.8\textwidth]{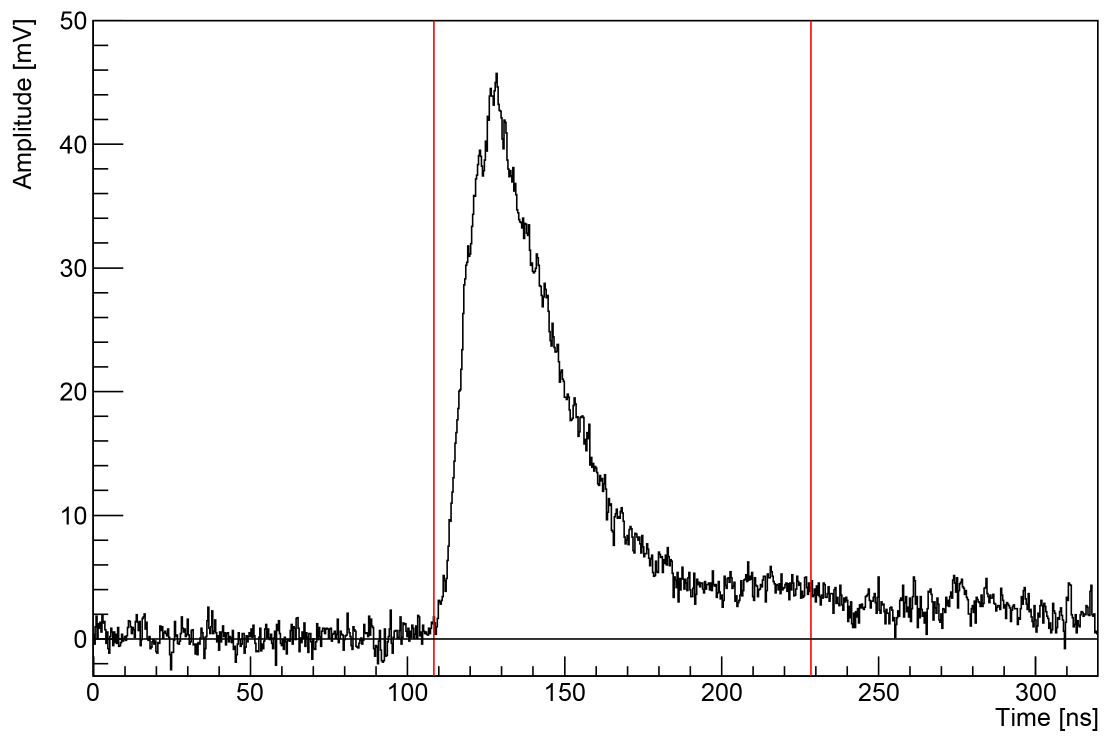}
    \caption{Example waveform of one SiPM group for one event with the incoming beam of 1.4\,GeV energy and \SI{0}{\degree} incident angle located at the central point of the detector cell, after baseline correction. The \textit{red} vertical lines show the integration window used for analysis.}
    \label{fig:wf_example}
\end{figure}

The collected light measured in each event, or integrated yield, is quantified by integrating the waveforms of all SiPM channels over time, yielding units of mV$\times$ns. The integration time of each waveform is taken to be 120\,ns, 20\,ns before the peak amplitude and 100\,ns after the peak, in order to contain as much of the waveform as possible while minimising the effect of dark counts, crosstalk, and afterpulsing. The baseline of the waveform is corrected using the minimum mean value of the waveform within a sliding window of 50 bins width inside a 50 ns time interval located well before the start of the signal. Fig.~\ref{fig:wf_example} shows an example signal waveform for a single SiPM group after baseline correction. No gain measurement of the SiPMs was obtained due to noise levels obscuring the single photoelectron peak, so we do not have a measurement of the waveform measured in mV as function of time measured in ns is created by a single photoelectron. In this case, we do not know precisely how many photoelectrons are detected for a given integrated waveform in mV$\times$ns. Work is ongoing to characterise the SiPM signals but for this work we simply use the integrated yield as a metric for the amount of light collected, without quantifying the amount of photoelectrons observed.
 
\begin{figure}[ht]
	\centering
	\includegraphics[width=1.0\textwidth]{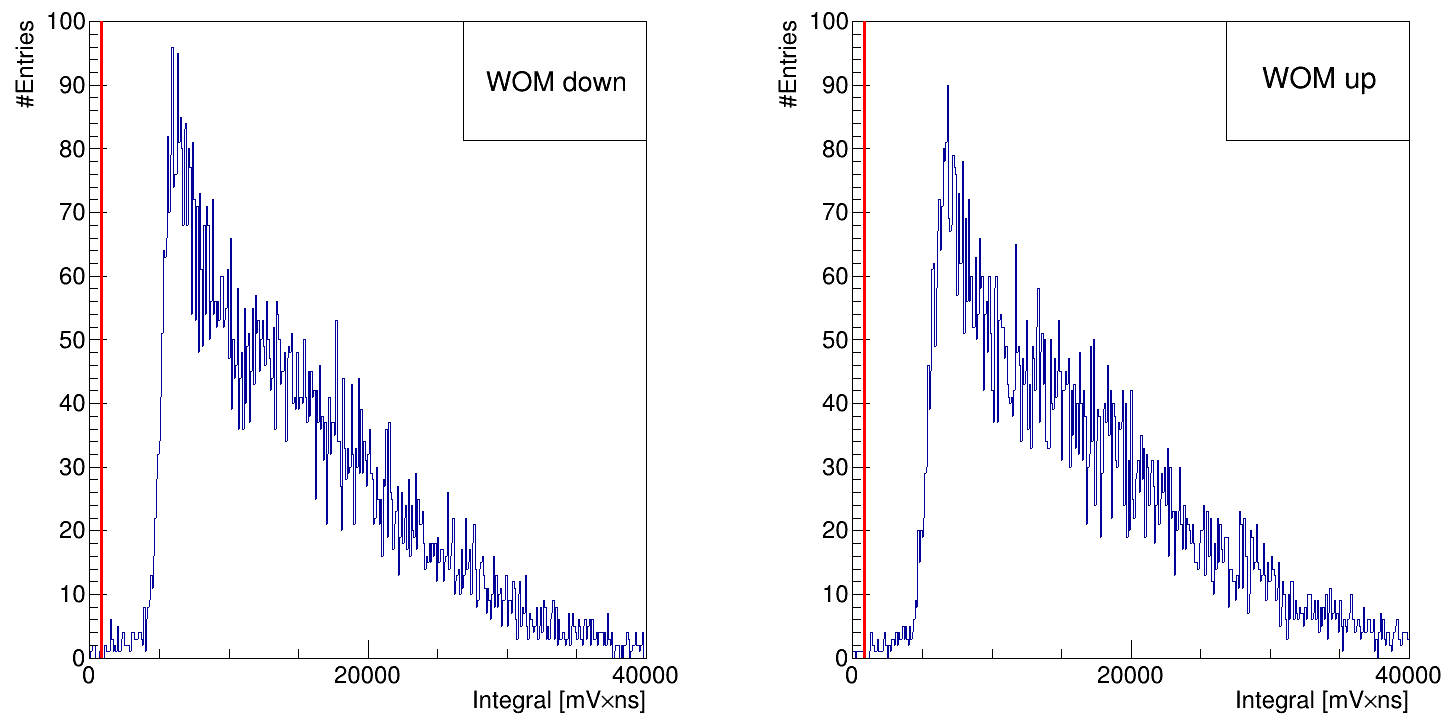}
	\caption{Distribution of the signal integrals (in 120\,ns) for an example measurement, summed over all SiPMs in each WOM, with the incoming beam of 1.4\,GeV energy and \SI{0}{\degree} incident angle located at the central point of the detector cell. The \textit{red} line indicates the threshold used to reject dark-count events. \textit{Left:} WOM down, \textit{right:} WOM up.}	
	\label{fig:DataSignalIntegral_WOM}
\end{figure}

Fig.~\ref{fig:DataSignalIntegral_WOM} shows the distribution of the signal integrals (in 120\,ns) for an example measurement (beam energy: 1.4\,GeV, \SI{0}{\degree} rotation angle) summed over all SiPMs in each WOM, including the threshold value used in the efficiency determination. The shape of the distributions look similar to distribution of detected photons in the simulation (see Fig.~\ref{fig:simulation_example}). In particular, in data as well as in the simulation there is a small fraction of events with a low integrated yield. In the simulation, this can be traced back to rare cases where the positron is back-scattered from the entrance steel wall producing only very small energy depositions in the liquid scintillator from follow-up processes.  

 In order to define the threshold at which to reject the events, a measurement was performed by triggering on one of the beam telescope PMTs while the beam was switched off. For each of these events, the signal from the eight SiPM groups from a WOM were summed up, after applying a baseline correction. 
 \begin{figure}[ht]
	\centering
	\includegraphics[width=1.0\textwidth]{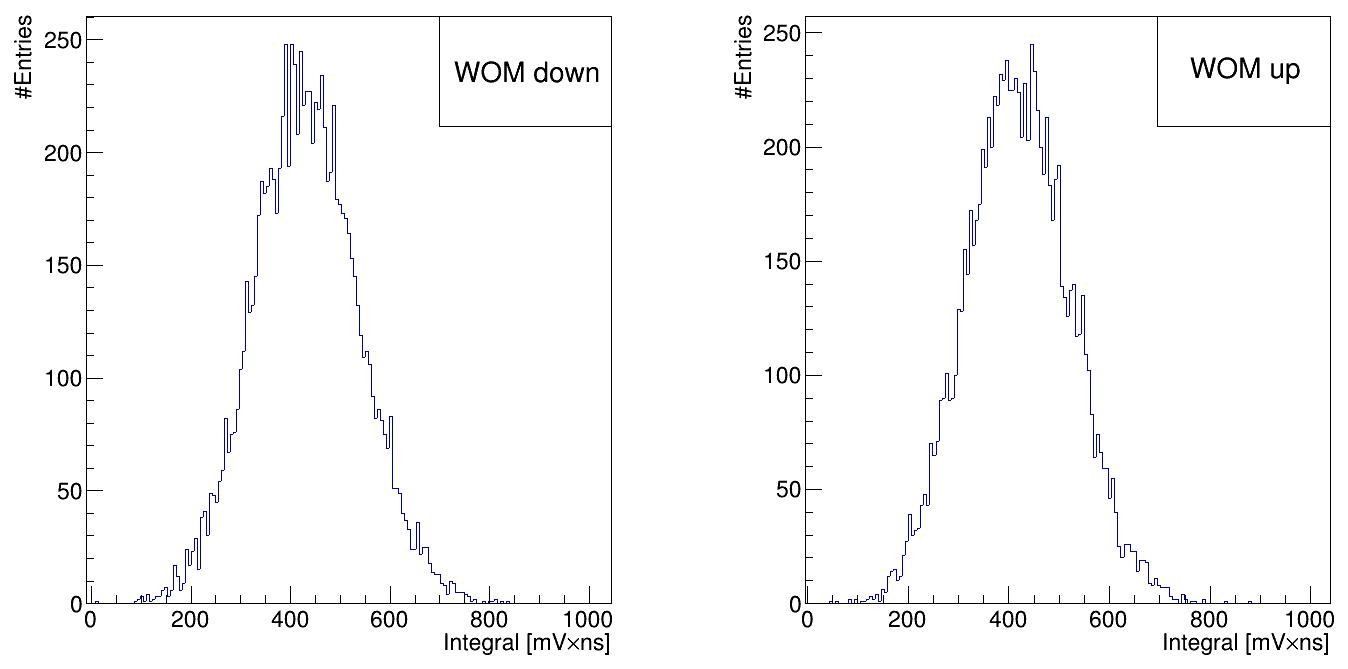}
	\caption{Distribution of the signal integrals (in 120\,ns), summed over all SiPMs for the beam-off measurement to estimate the expected integrated yield from dark-counts as described in the text. \textit{Left:} WOM down, \textit{right:} WOM up.}	
	\label{fig:DC_WOM}
\end{figure}
The integrated yield collected by a WOM in such a measurement was calculated as the time-integral of the summed signal over a window of 120\,ns. Given this time-window size, the probability
to detect photons in the SiPMs
generated either by cosmics crossing the detector or by natural radioactivity resulting in scintillation light in the detector is rather small. Hence, it is expected that any photoelectron
yield seen in the SiPM groups is dominated by SiPM dark counts due to the typical dark-count rate per SiPM in the MHz range.
From the histogram of charge calculated for 10000 of such "dark-count" events (Fig.~\ref{fig:DC_WOM}), the maximum value for the integrated yield of a dark-count event for each WOM was estimated. This maximum was chosen as the threshold for possible dark-count events. The events were rejected if the sum signal integral over SiPMs in a WOM was lower than 800~$\si{ns}\times\si{mV}$. This threshold is also shown in Fig. \ref{fig:DataSignalIntegral_WOM} as a red vertical line.

From the measurements with \SI{0}{\degree} rotation angle at different beam energies, the efficiencies of the detector, $\varepsilon$, and for each WOM, $\varepsilon_{d}$, $\varepsilon_{u}$, were estimated. The efficiency has been calculated as the ratio of accepted events to total recorded events:
 \begin{equation}
 	\varepsilon_{d/u} = \frac{N_{tot}-N_{rejected}}{N_{tot}} = \frac{N_{(integral_{d/u}>800 ~\si{mV}\times\si{ns})}}{N_{tot}}.
 \end{equation} 
 The uncertainties have been calculated at $\SI{68}{\percent}$ confidence interval with the Clopper-Pearson method \cite{Clopper-Pearson}. 

\begin{figure}[ht]
	\centering
    \includegraphics[width=1.0\textwidth]{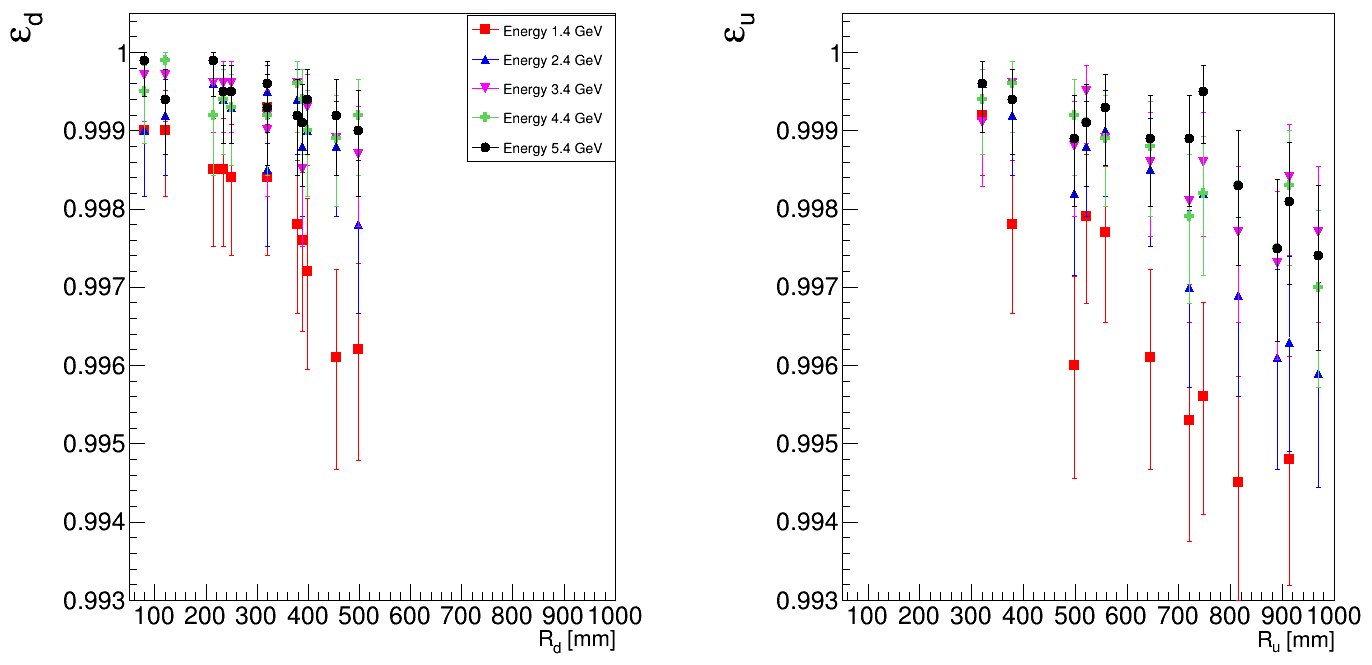}
	\caption{Efficiency results for the two WOMs as a function of distance between particle crossing point and centre of the WOM under study.}	
	\label{fig:eff_WOM}
\end{figure}

Fig.~\ref{fig:eff_WOM} shows the efficiency of the two WOMs as a function of distance between the particle crossing point and the centre of the WOM under study, $R_d$ and $R_u$ respectively. As expected, the efficiency decreases with decreasing beam energy and increasing distance between the particle crossing point and the given WOM. For the largest possible distance and lowest beam energy, the efficiency is larger than \SI{99.3}{\percent} at \SI{68}{\percent} confidence level, which is significantly above the efficiency values quoted for a previous test detector at similar electron/positron energies and distances between the beam position and the WOM tube under study~\cite{SHIP:2021tpn}.

We consider the quoted efficiency values as lower limits, because there was no beam telescope counter installed downstream of the detector cell so that it is not guaranteed that the positron passes the detector cell. As already discussed above, a small fraction of events are observed in the simulation in which the positron is back-scattered from the steel entrance wall of the vessel, leading to a very small or even close-to zero energy deposition in the liquid scintillator. Moreover, in the data it is not excluded that the PMTs of the beam telescope have triggered because the incoming positron generated an electromagnetic shower in the upstream collimator while no or only very low-energetic shower particle(s) hit the detector cell. In future test beam measurements, an additional beam counter will be placed downstream of the detector to guarantee that triggered events are induced by particles traversing the complete cell.

\begin{figure}[ht]
	\centering
	\includegraphics[width=0.8\textwidth]{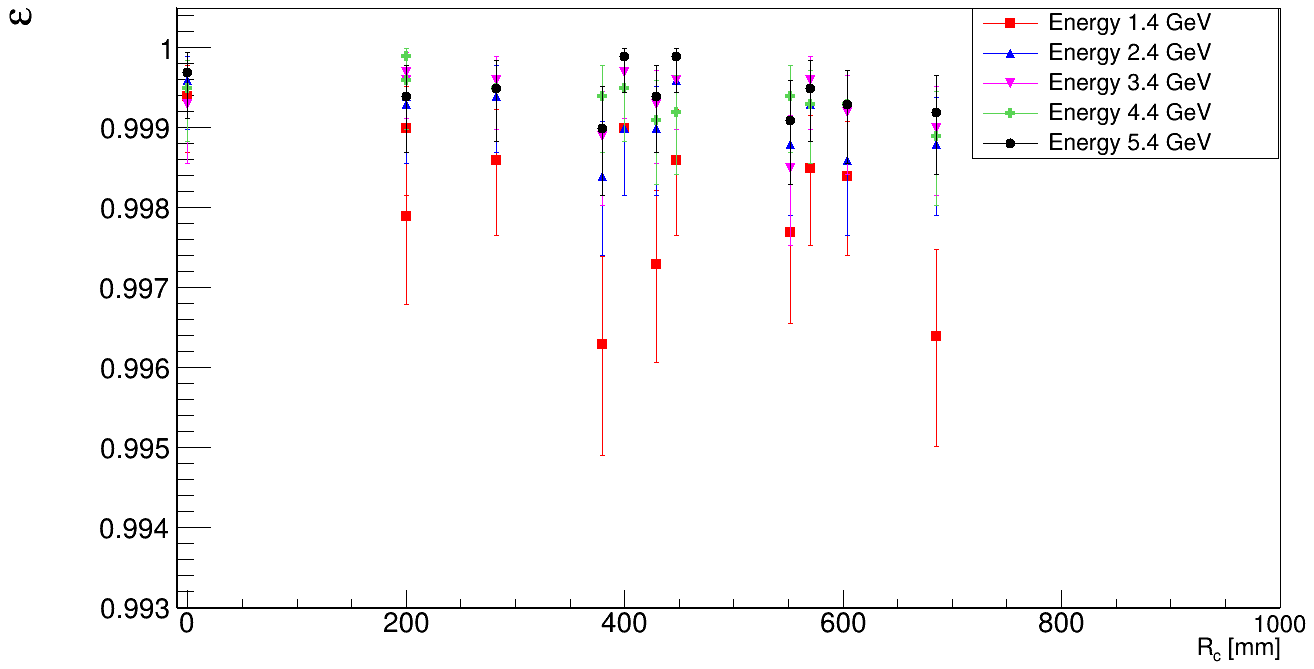}
	\caption{Total efficiency for the detector as a function of distance $R_c$ between the particle crossing point and the centre of the detector cell, requiring that at least one of the summed WOM signals has passed the threshold of $800~\si{mV}\times\si{ns}$.}	
	\label{fig:eff_tot}
\end{figure}

To calculate the total efficiency of the detector, an OR condition is required: 
 \begin{equation}
    \varepsilon = \frac{N_{((integral_{d}>800~\si{mV}\times\si{ns})\vee (integral_{u}>800~\si{mV}\times\si{ns}))}}{N_{tot}}.
 \end{equation}
Fig.~\ref{fig:eff_tot} shows this combined efficiency as a function of the distance between the particle crossing point and the centre of the detector, called $R_c$. As expected, the combined efficiency is higher than the single-WOM efficiencies with at least \SI{99.5}{\percent} at \SI{68}{\percent} confidence level and increases with beam energy.

From the comparison of Fig.~\ref{fig:DC_WOM} and Fig.~\ref{fig:simulation_example}, we infer that the efficiency values measured for 1.4 GeV positrons can likely be achieved as well in the case of minimum ionizing particles.

\subsection{Light collection and detector response uniformity}
\label{Sec:LightCollectionandDetectorResponse}
As mentioned in Section~\ref{Sec:Efficiency} we do not have a measurement of a single photoelectron signal in our SiPMs, which mean that we cannot convert the number of photons detected in simulation to the measured integrated yields, or vice-versa. Here we compare simulated results to measurement in a qualitative way. The simulation described in Section~\ref{Sec:MonteCarloSimulation} was used to produce a dataset with the same incoming beam angles and energies as those measured at the test beam. As discussed in Section~\ref{Sec:CellWallReflectivity}, the detector's inner walls showed rust stains. Hence, the reflectivity of the detector cell was assumed to be below that of the measured samples. To account for this, simulated datasets with reflectivities of \SI{100}{\percent}, \SI{90}{\percent}, \SI{80}{\percent}, \SI{70}{\percent}, \SI{65}{\percent}, \SI{60}{\percent}, and \SI{50}{\percent} of the measured reflectivity were produced for all particle crossing points at \SI{0}{\degree} and compared to data. All simulations assume that the reflectivity within the detector cell is uniform for all surfaces, which may not be the case, but in the absence of measurements from the detector cell we make this approximation. 
To find the reflectivity value
that brings the simulation in best agreement with the data,
we define a $\chi^2$ function
which compares the average
photon yield in the simulation ($simulation_i$) and the measured average integrated yield ($data_i$) for all particle crossing points ($i=1, ..., N$).
In order to directly compare the simulation to data, a scale parameter $\lambda$ was applied to the simulated photon yields, depending on the reflectivity $\alpha$. The $\chi^2$ between these distributions was then summed over all particle crossing points $i$ for a given reflectivity, and this number is minimised:
\begin{equation}
    \chi_{Tot}^2\left(\lambda(\alpha), \alpha\right) = \sum_i\frac{(\lambda(\alpha)\cdot simulation_i(\alpha) - data_i)^2}{\sigma_i^2},
\end{equation}
where $\sigma_i$ is the standard deviation of the simulated photon yield distribution divided by $\sqrt{N_{events}}$, and $\lambda$ is a variable fit parameter. It was found that the high-reflectivity samples relatively overestimated the amount of light collection for beam positions in the corners of the box, while lower reflectivities produced distributions closer to observation. The simulation with \SI{65}{\percent} of the measured reflectivity showed the lowest $\chi^2_{Tot}$ so was chosen as the optimal simulation for comparison to data. No other simulation parameters were altered. 

\begin{figure}[ht]
    \centering
    \includegraphics[width = 0.8\textwidth]{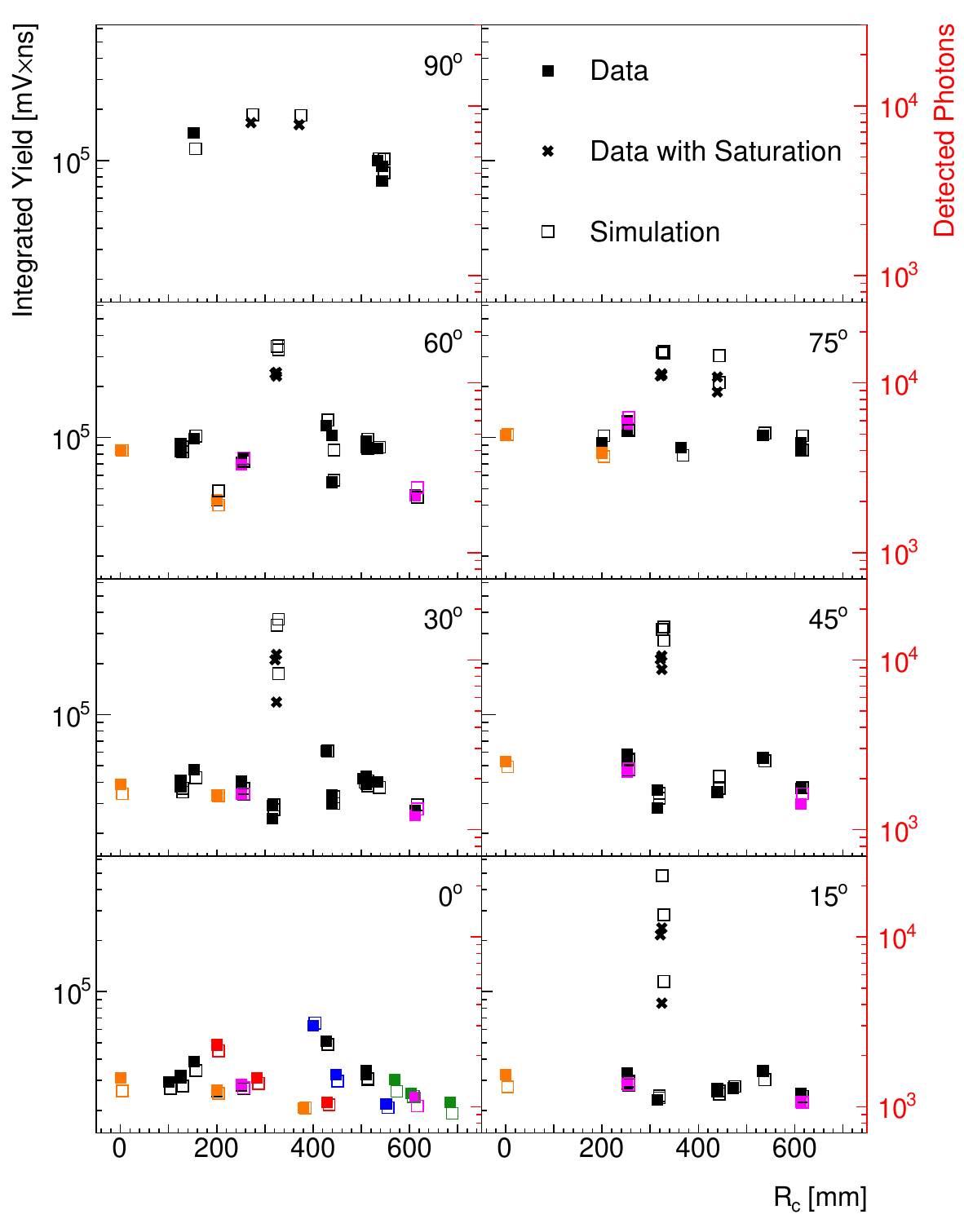}
    \caption{Average integrated yield for various beam angles and particle crossing points as a function of distance from the origin of the cell, compared with the average number of detected photons in the simulation. Raw data and simulation are shown without scaling or normalisation. Rather, they are superimposed with the axis in black (\textit{right}) corresponding to measured data and the axis in red (\textit{left}) corresponding to simulation. The axis range of detected photons has been matched to that of integrated yield for visual purposes using a scaling determined by the $\chi^2$ minimisation presented above. Beam energy for all cases is 1.4\,GeV. Cross markers indicate data where the electronics showed saturation for some or all waveforms. The cell wall reflectivity in simulation is \SI{65}{\percent} of the measured values.}
    \label{fig:chargealldeg}
\end{figure}

A comparison between the measured average integrated yields and the average photon yields in the simulation with a cell wall reflectivity reduced to \SI{65}{\percent} of measured values can be seen in Fig.~\ref{fig:chargealldeg}. 
In data, a general increase in integrated yields is observed with increasing angle between the beam and the detector, since with increasing angle the incoming particle traverses a longer path length in the liquid scintillator and thus deposits more energy. For all angles, the integrated yield is lowest for particle crossing points furthest away from the centre of the detector cell. This is expected since the scintillation photons produced further away from the WOMs have a higher chance of being lost due to absorption in the LS or poor reflectivity in the cell walls before reaching a WOM. The integrated yield is highest at all angles for certain points between in the $R_c$ range 200-400\,mm, corresponding to beam positions very close to a WOM. Some points have such a high integrated yield that the electronics became saturated, which are denoted by a different marker. We observe that the detector response is not uniform over the cell, when incoming particles enter very close to a WOM where the integrated yield is highest. The response could likely be made more uniform by increasing the reflectivity of the inner walls, increasing the light collection for incoming particles close to the edges of the detector cell. In the absence of a gain measurement for the SiPM groups so we cannot directly compare simulation and data with each other, but they have been visualised in the same plot with corresponding axes. Simulated distributions for all angles show very similar behaviour to the measured data, indicating that the scaling between integrated yield and photons in simulation is linear. Despite many unknown factors that are approximated in the simulation it appears to model the detector quite accurately. One notable difference between data and simulation is the saturated data points, which all lie below the corresponding simulation. This is expected because the simulation does not exhibit any saturation, whereas the integrated yields for these points are cutoff. It is likely that including electronic signatures such as variable gain for each SiPM channel, cross talk and after-pulsing in the SiPMs, as well as dark counts into the simulation could yield an even closer agreement with data. Work is ongoing to simulate the waveform signal produced by the SiPM groups for direct comparison between simulation and data.

\begin{figure}[ht]
    \centering
    \includegraphics[width = \textwidth]{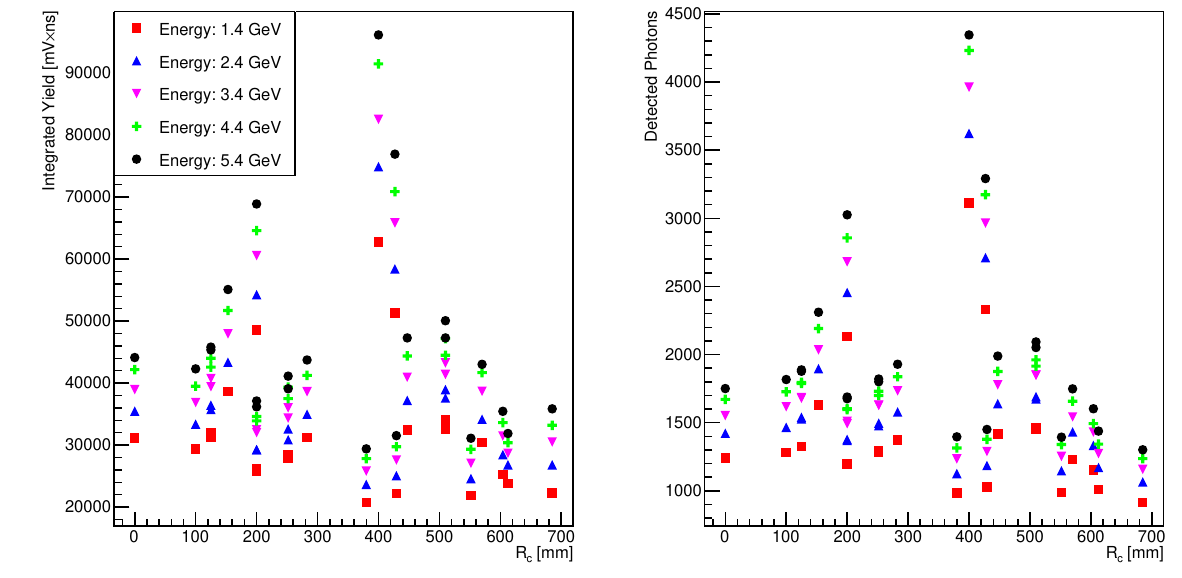}
    \caption{\textit{Left:} Average integrated yield for various beam energies and particle crossing points as a function of distance from the centre of the cell for a beam angle of \SI{0}{\degree}. \textit{Right:} Average number of detected photons in simulation for the same particle crossing points and beam energies. The cell wall reflectivity in simulation is \SI{65}{\percent} of the measured values.}
    \label{fig:charge0degenergies}
\end{figure}

The comparison between integrated yields for different particle crossing points and beam energies at an angle of \SI{0}{\degree} is shown in Fig.~\ref{fig:charge0degenergies}, along with the same for simulation.
As expected, the amount of light collected increases with beam energy for all particle crossing points. The shape of the integrated yield distributions for the various energies show consistent behaviour with each other. This is a promising result, indicating that the integrated yield at a given particle crossing point strictly increases with energy deposition. The same behaviour is observed in the simulation. This comparison is not possible for data at other beam angles since nearly all data points at beam energies higher than 1.4\,GeV are affected by electronic saturation.

\begin{figure}[ht]
    \centering
    \includegraphics[width = 0.5\textwidth]{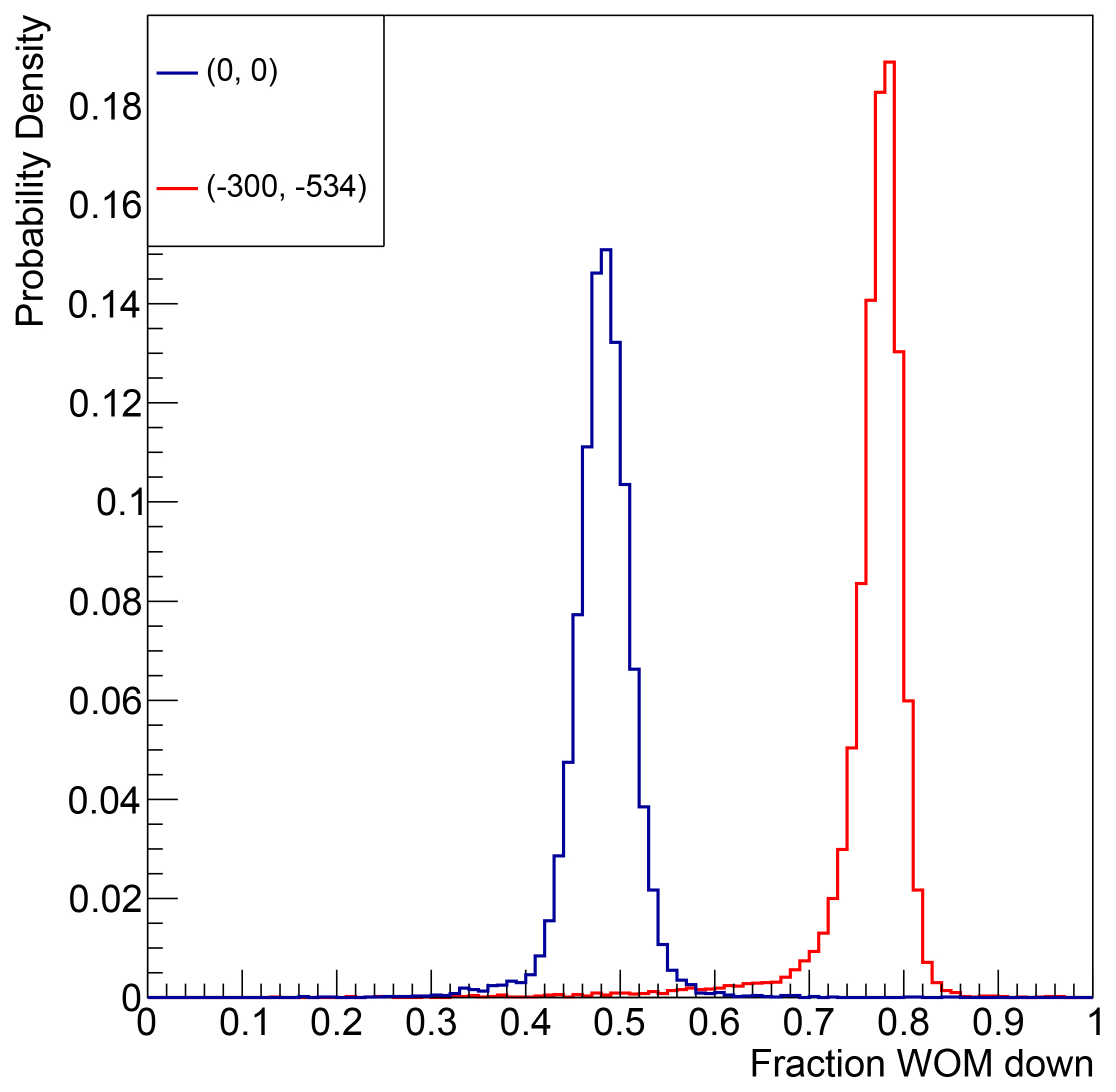}
    \caption{Probability distribution of the fractional integrated yield detected in WOM down for two particle crossing points. The particle crossing point in the lower left corner of the detector has a much higher probability of a larger fractional integrated yield in WOM down, as expected. This suggests that the likelihood reconstruction should be able to properly reconstruct such events.}
    \label{fig:womfrac}
\end{figure}

\begin{figure}
     \centering
     \begin{subfigure}[b]{0.49\textwidth}
         \centering
         \includegraphics[width=\textwidth]{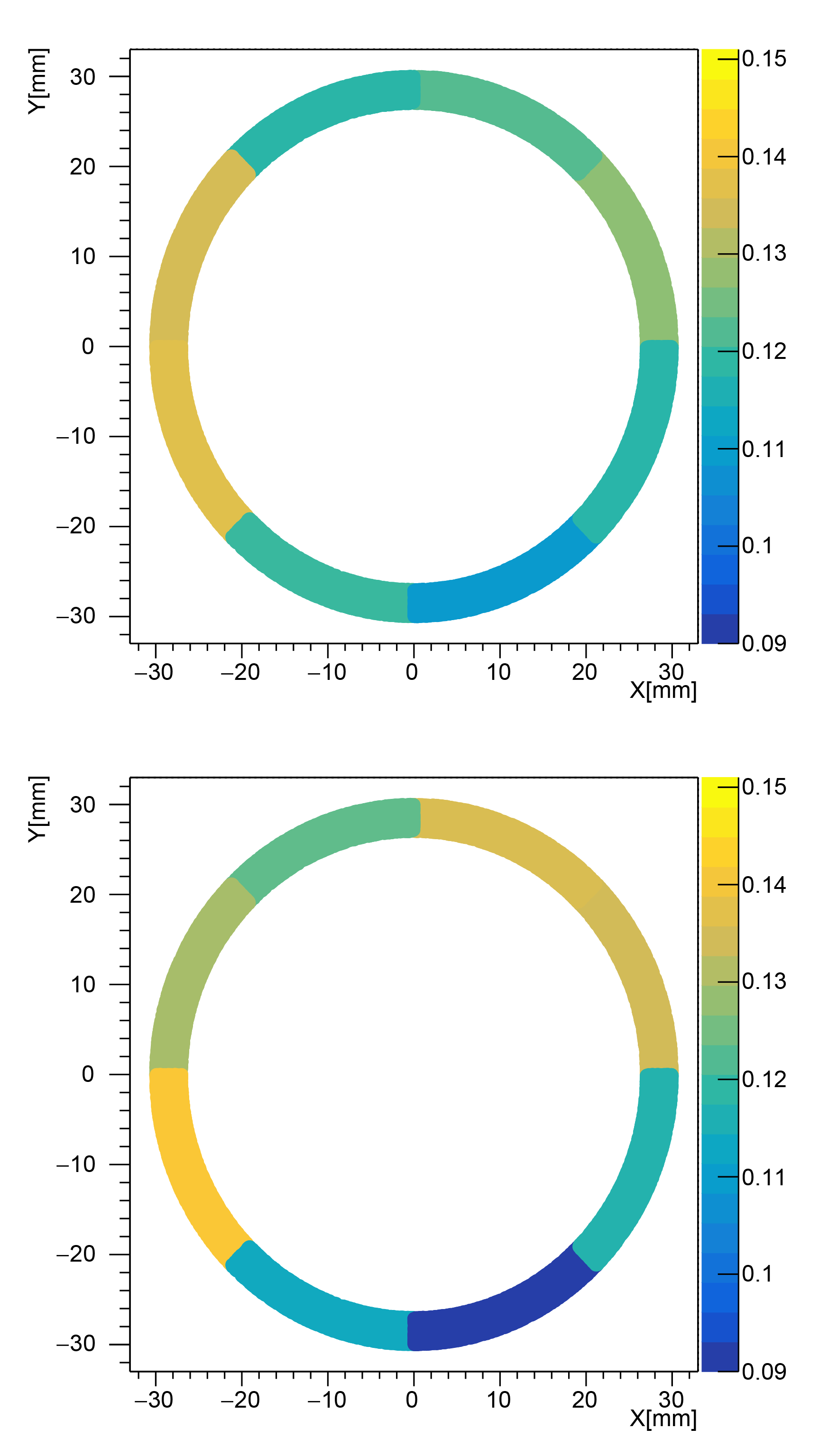}
         \caption{}\label{fig:phi_prob_a}
     \end{subfigure}
     \hfill
     \begin{subfigure}[b]{0.49\textwidth}
         \centering
         \includegraphics[width=\textwidth]{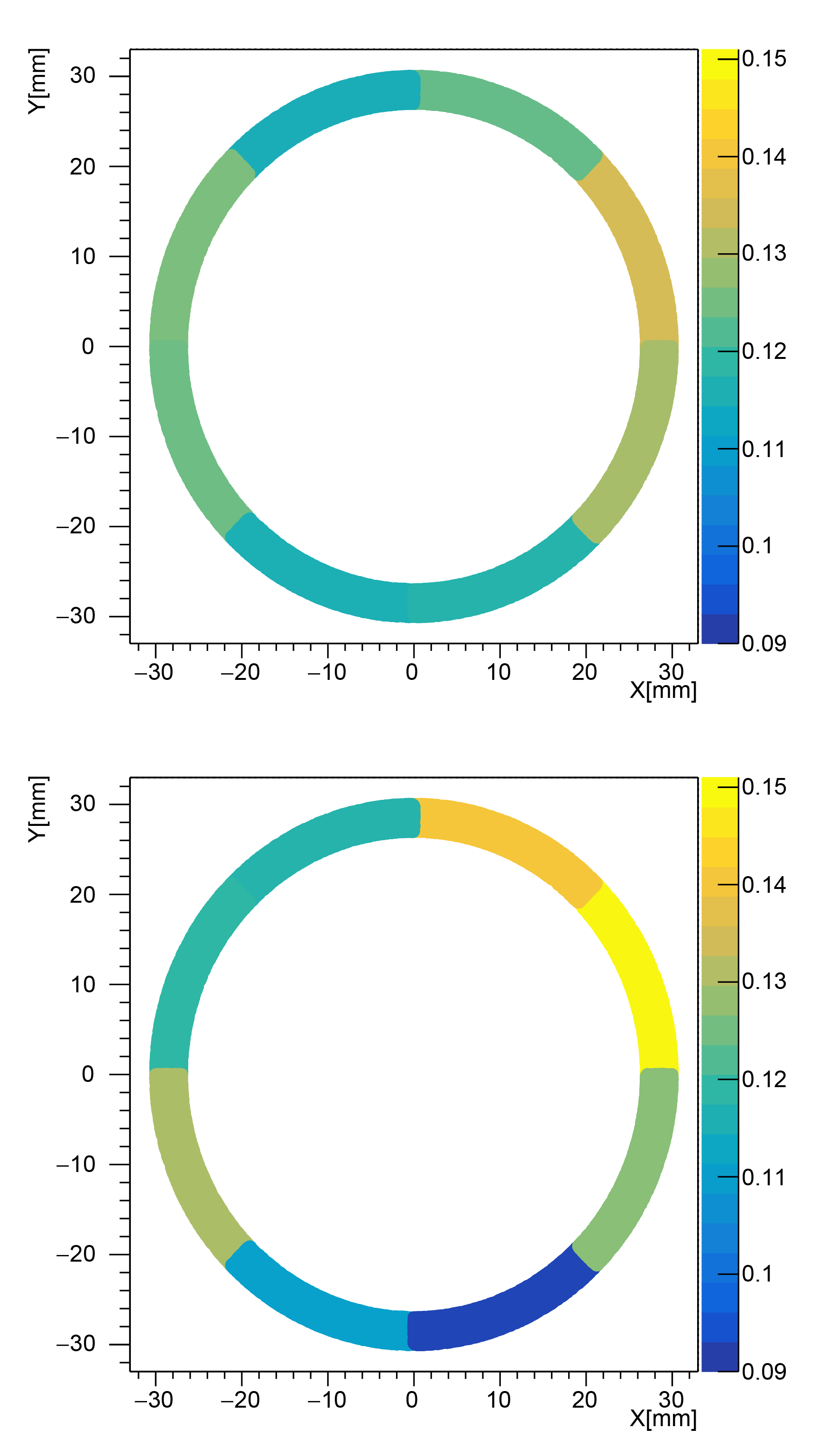}
         \caption{}\label{fig:phi_prob_b}
     \end{subfigure}
        \caption{Most probable value of integrated yield per SiPM channel divided by total integrated yield of respective WOM in WOM up (\textit{top}) and WOM down (\textit{bottom}) for particle crossing point (-200, 0) (\textit{left}) and (200, 0) (\textit{right}). When the particle crosses to the left (or right) of the WOMs, we see a higher portion of signal in the left (or right)-most channels, so these distributions can give us information on the $x$ value of an incoming particle.}
        \label{fig:phi_prob}
\end{figure}

\begin{figure}[ht]
    \centering
    \includegraphics[width = 0.8\textwidth]{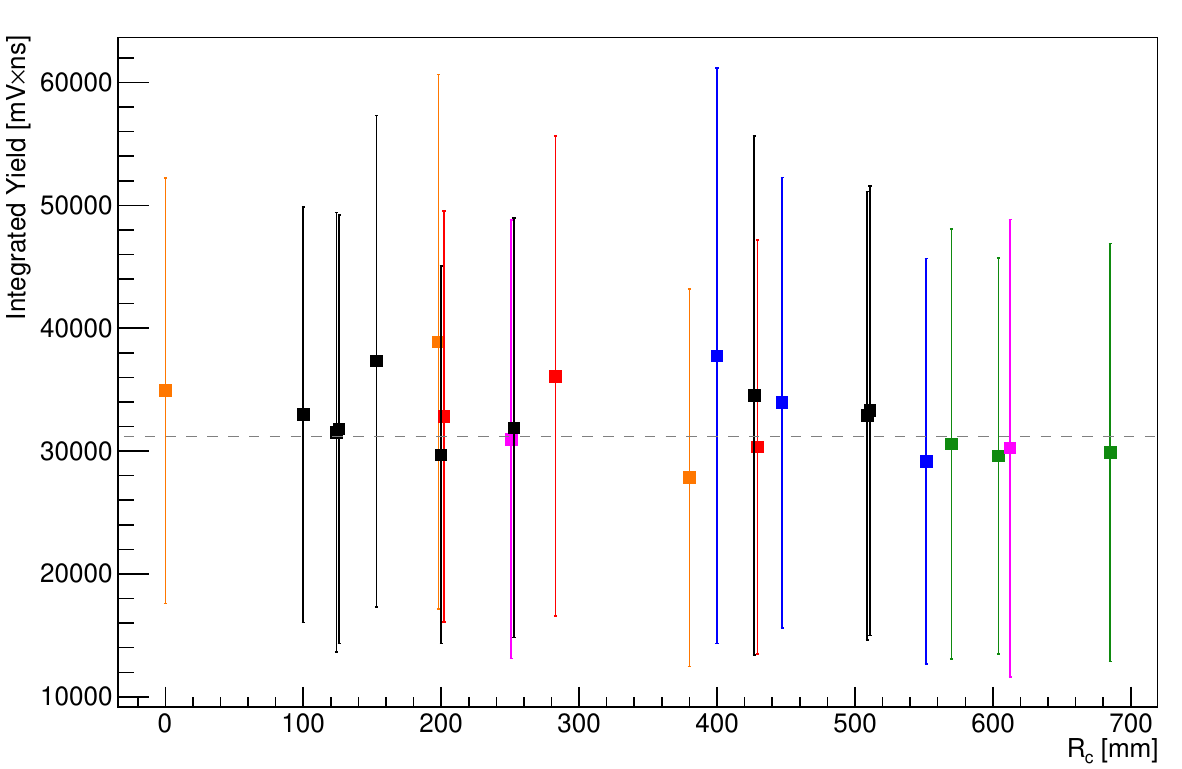}
    \caption{Integrated yield after likelihood-based correction with beam incidence angle of \SI{0}{\degree} and energy of 1.4\,GeV. The measured fractions of integrated yield in each WOM and in each SiPM group of each WOM are used to reconstruct the most likely particle crossing point of the respective event. Error bars are showing the standard deviations of the distributions, the relative sizes of which are only slightly increased compared to the relative size of standard deviation of the energy deposition observed in the MC simulation (compare Fig. \ref{fig:simulation_edep}). The \textit{dashed} line is the reference value used for the correction.}
    \label{fig:likelihood_pe}
\end{figure}

\begin{figure}[ht]
    \centering
    \includegraphics[width = 0.8\textwidth]{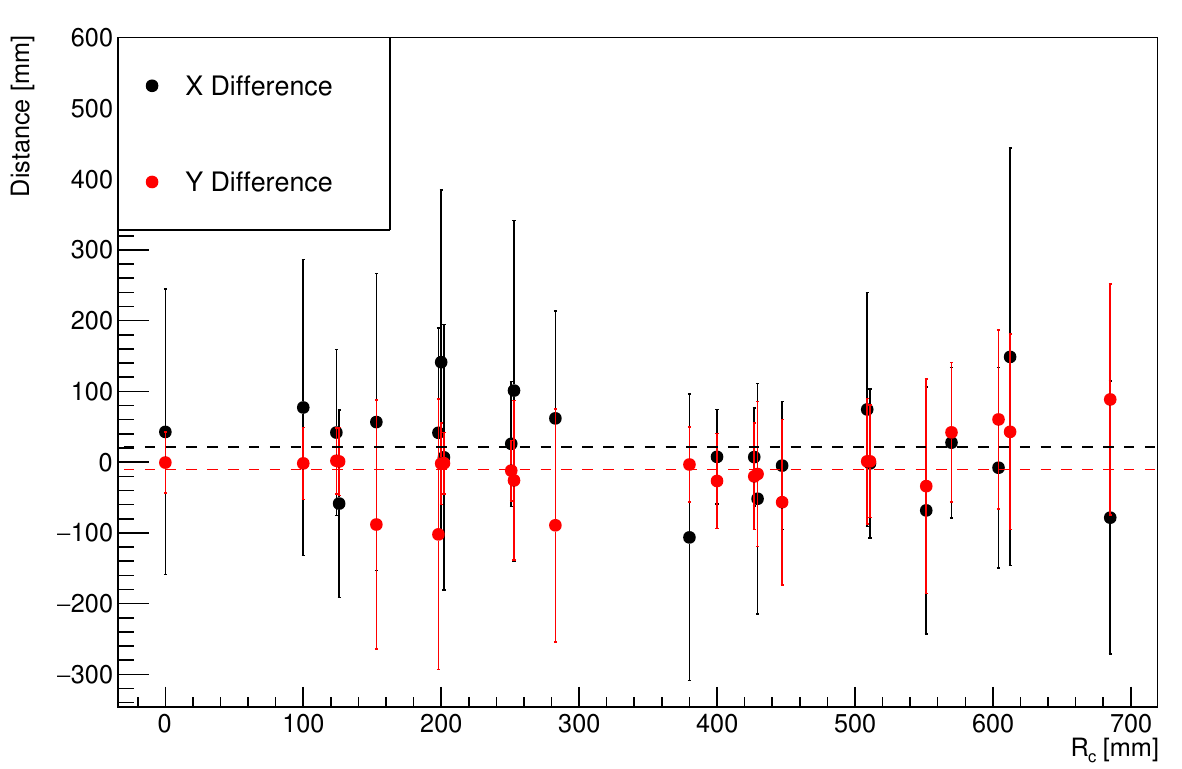}
    \caption{Mean difference between the $x$ and $y$ values of each point and their reconstructed values, with corresponding standard deviations resulting from the likelihood-based correction procedure for the integrated yield. The \textit{dashed} lines represent the mean of each distribution averaged over all particle crossing points.}
    \label{fig:likelihood_pe_residuals}
\end{figure}

\begin{figure}[ht]
    \centering
    \includegraphics[width = 0.8\textwidth]{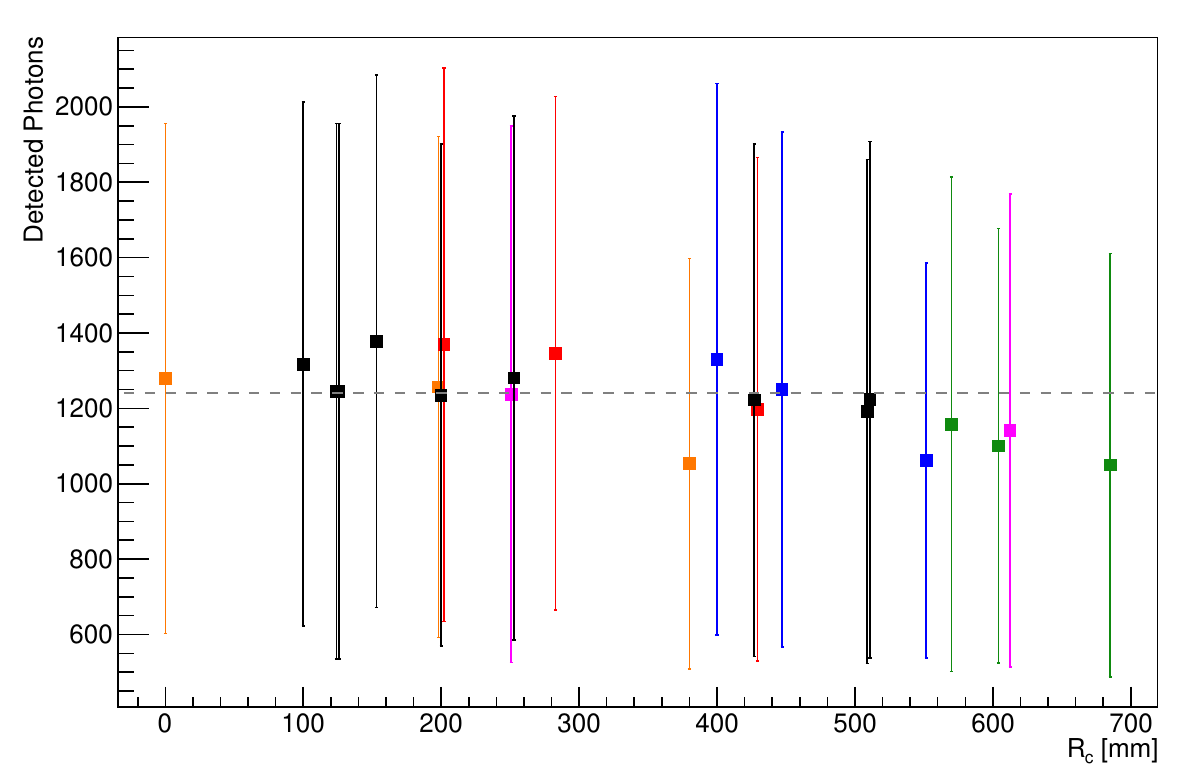}
    \caption{Detected photons in simulation after likelihood-based correction. The fractions of photons in each WOM and in each SiPM group of each WOM are used to reconstruct the most likely particle crossing point for the respective event. Error bars indicate the standard deviations of the distributions. The \textit{dashed} line is the reference value used for the correction. The cell wall reflectivity in simulation is \SI{65}{\percent} of the measured values.}
    \label{fig:likelihood_photon_sim}
\end{figure}

In the SHiP experiment, the location of a particle entering the detector is a priori unknown. To reduce the uncertainty on the measured energy deposition due to the observed non-uniformity in the integrated yield, which varies by $\sim^{+\SI{100}{\percent}}_{-\SI{35}{\percent}}$ with respect to the central particle crossing point, a likelihood-based correction procedure was developed. All data used in this study is taken with a beam angle of \SI{0}{\degree} and energy of 1.4\,GeV. The centre point of the cell is chosen as the reference value for the correction of the integrated yield. The variables chosen for the correction are the fractional integrated yields in each WOM and each SiPM group, which give information on the particle $y$ position and $x$ position, respectively. For each particle crossing point, the probability density distribution of each of these variables is determined from the measurements and the ratio between the total integrated yield measured in the central point and each other particle crossing point is calculated. Then for each event, the likelihood is calculated that the measured data are in agreement with a given position. The point with the highest likelihood is chosen to apply the previously calculated ratio of integrated yields for the chosen particle crossing point as a correction factor. In this way, the corresponding correction will be exact if the point is properly reconstructed for each event. Sample probability distributions of the ratio of integrated yields in WOM down for the central crossing point and a corner crossing point are shown in Fig.~\ref{fig:womfrac}.
Sample most probable values of the fractional integrated yield per SiPM channel are shown in Fig.~\ref{fig:phi_prob}.
Results of the integrated yield correction can be found in Fig.~\ref{fig:likelihood_pe}.
The correction works particularly well for particle crossing points whose pre-correction integrated yield differed significantly from the integrated yield at the central position. Before correction, we see a variation in integrated yields of $\sim^{+\SI{100}{\percent}}_{-\SI{35}{\percent}}$, with respect to the central point. After correction this is reduced to $\sim^{+\SI{25}{\percent}}_{-\SI{10}{\percent}}$. The presented correction procedure also provides an estimate for the particle crossing point. The mean difference in $x$ and $y$ between each point and its reconstructed value and the corresponding standard deviation are shown in Fig.~\ref{fig:likelihood_pe_residuals}.
The standard deviation in the mean $x$ difference is larger than in the $y$ difference, which is expected since the WOMs are separated along the $y$ direction, hence this topology provides more precision in the $y$ coordinate. This likelihood-based approach is able to estimate the position of an incoming particle with a resolution of at most 30 cm along $x$ and 20 cm along $y$ and biases of at most 14 cm along $x$ and 10 cm along $y$. The likelihood-based correction method was applied as well to the simulation, with results shown in Fig.~\ref{fig:likelihood_photon_sim}.
It performs similarly to that of data, with variations before correction ranging from $\sim^{+\SI{150}{\percent}}_{-\SI{40}{\percent}}$ with respect to the central point. After correction this range is reduced to $\sim^{+\SI{10}{\percent}}_{-\SI{20}{\percent}}$. 
 
With a more sophisticated reconstruction technique using machine learning, the precision in the particle crossing point can be significantly improved as demonstrated in Section \ref{Sec:SpatialResolution}. Therefore, we expect that such a machine-learning technique has also the potential to improve the uniformity in the integrated yield determination, which will be further studied in the future with a further prototype detector consisting of several cells. We expect that it will also significantly improve the correction to simulation. Considering the promising agreement between data and simulation, it is possible that in the future a machine-learning approach could be taken for particle reconstruction, by training with simulation and applying it to data.

\subsection{Time response of the full-size detector prototype}
\label{Sec:TimeResolution}

In this section, we present the time response of the LS-SBT prototype detector for a selected set of particle crossing points on the detector cell as sketched in Fig.~\ref{fig:deglocations}. If not stated otherwise, the beam energy was always chosen to be 1.4\,GeV and the incidence angle \SI{0}{\degree}.

\begin{figure}[ht]
 	\centering
 	\includegraphics[width =0.8\textwidth]{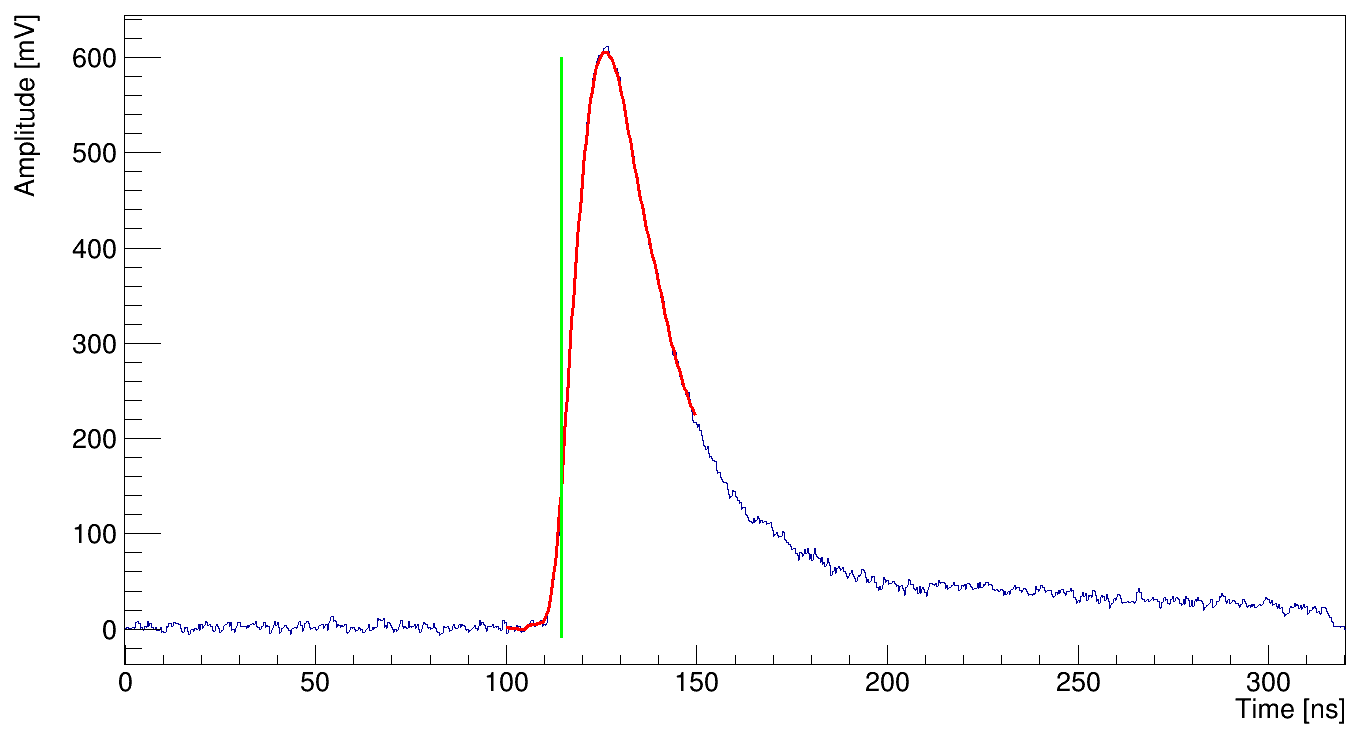}
 	\caption{Sum of the signals from the eight SiPM channels of one WOM for one event, after baseline correction. The \textit{red} line is the result of the smoothing procedure performed on the signal. The \textit{green} vertical line indicates the signal time at $\SI{25}{\percent}$ of the maximum of the summed signals.}	
 	\label{fig:smoothing}
 \end{figure}

\begin{figure}[ht]
	\centering
	\includegraphics[width = \textwidth]{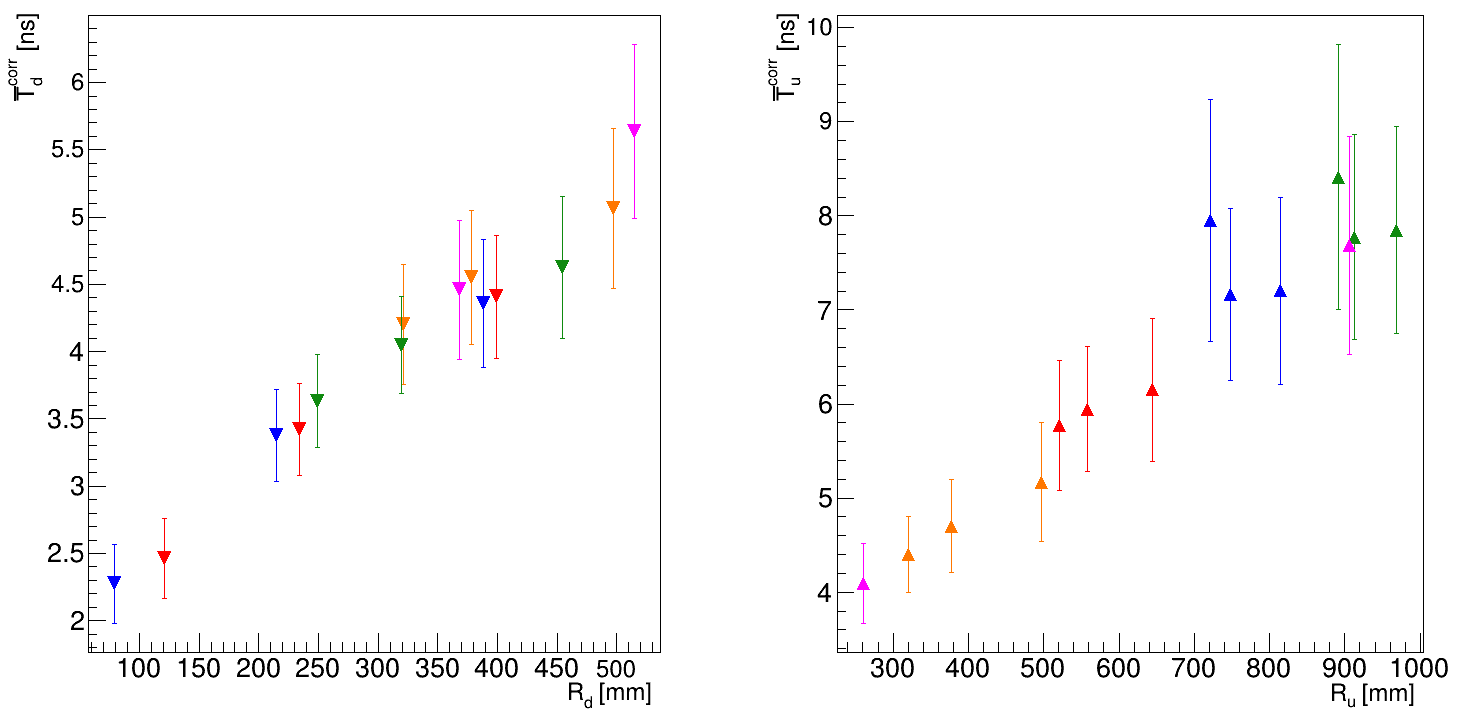}
    \caption{Signal arrival time at the WOMs as a function of the distance between particle crossing point and the centre of the respective WOM. \textit{Left:} WOM down, \textit{right:} WOM up.}	
	\label{fig:time_WOM}
\end{figure}

In order to study the time response of the detector, the signal arrival time for both WOMs was estimated using a constant fraction discrimination (CFD) method with a threshold value of $\SI{25}{\percent}$ of the maximum amplitude of each signal. For each event, the waveforms from the eight channels of each WOM were summed to reduce the fluctuation due to electronic noise. Afterwards, a smoothing was applied and the time at which the summed signal reaches the $\SI{25}{\percent}$ of its maximum was determined by interpolation (Fig.~\ref{fig:smoothing}). 
 \begin{figure}[ht]
 	\centering
 	\includegraphics[width = 0.5\textwidth]{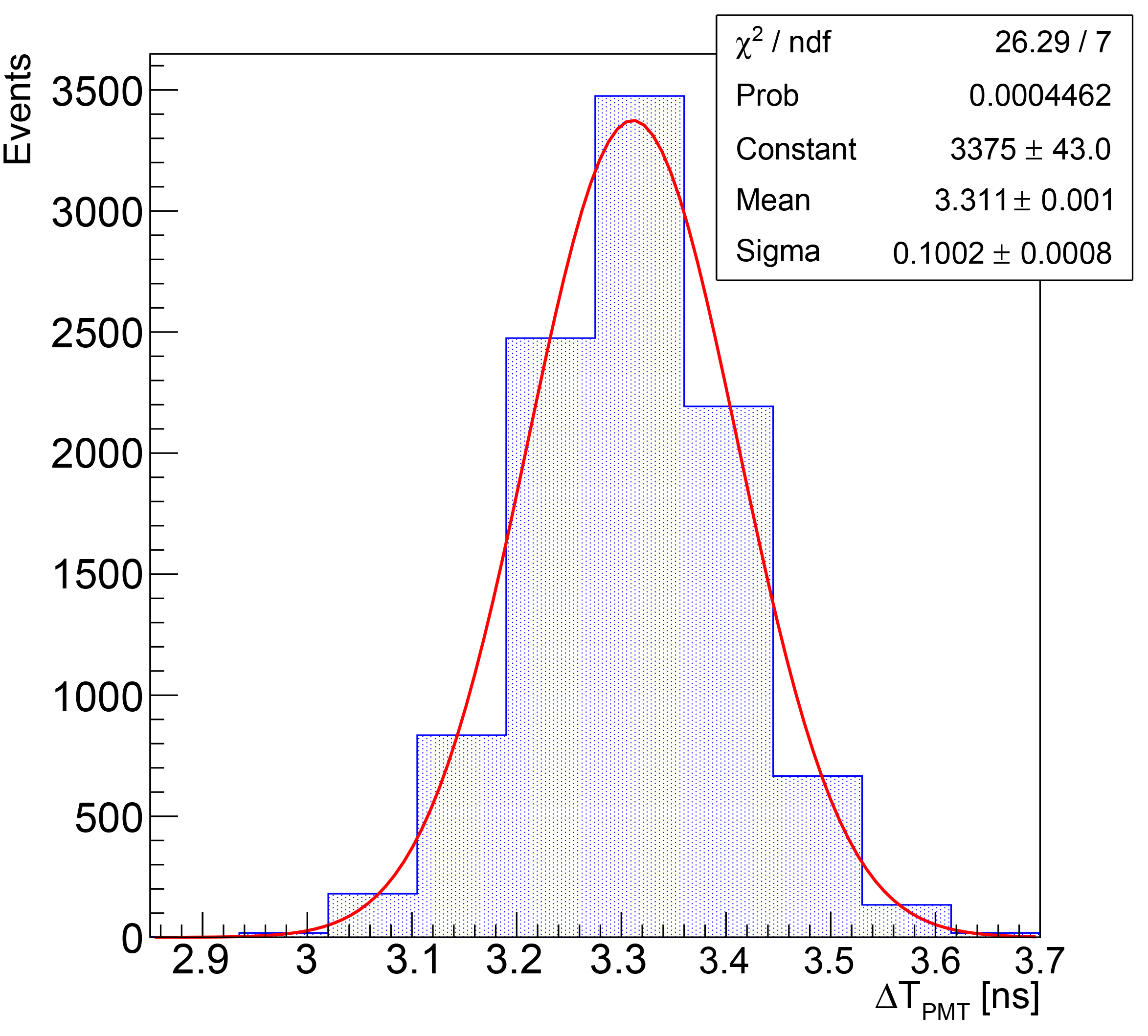}
 	\caption{Distribution of $\Delta T_{PMT}$ for the four test beam telescope scintillators used for
    triggering the data acquisition of the WaveCatcher digitiser.}	
 	\label{fig:varepsilonres}
 \end{figure}
From the resulting times, called $\bar{T}_{d}$ and $\bar{T}_{u}$, the average arrival time  of the four trigger PMT signals, $T_{PMT_i}$, from the beam telescope, $\bar{T}_{trigger}=\sum_i T_{PMT_i}/4$, was subtracted, also using for each PMT signal the CFD threshold of $\SI{25}{\percent}$. Any variation in the time offset from event to event, $T_0$, cancels in this difference: $\bar{T}_{d}^{corr} = \bar{T}_{d}-\bar{T}_{trigger}$ and $\bar{T}_{u}^{corr} = \bar{T}_{u}-\bar{T}_{trigger}$. The intrinsic time uncertainty of $\bar{T}_{trigger}$ channels can be deduced from the distribution of $\Delta T_{PMT}=\big(\frac{T_{PMT_1}+T_{PMT_2}}{2}-\frac{T_{PMT_3}+T_{PMT_4}}{2}\big)/2$, which is plotted in Fig.~\ref{fig:varepsilonres}. From a Gaussian fit, the standard deviation of $\Delta T_{PMT}$ is found to be 0.1\,ns. Fig.~\ref{fig:time_WOM} shows for each particle crossing point the mean value of the corresponding $\bar{T}_{u}^{corr}$ and $\bar{T}_{d}^{corr}$ distributions as a function of the distance between the particle crossing point and the centre of the corresponding WOM, $R_u$ and $R_d$, and as an error bar the corresponding standard deviation of the distribution as an estimate of the single-event time resolution.

\begin{figure}[ht]
	\centering
	\includegraphics[width =0.8\textwidth]{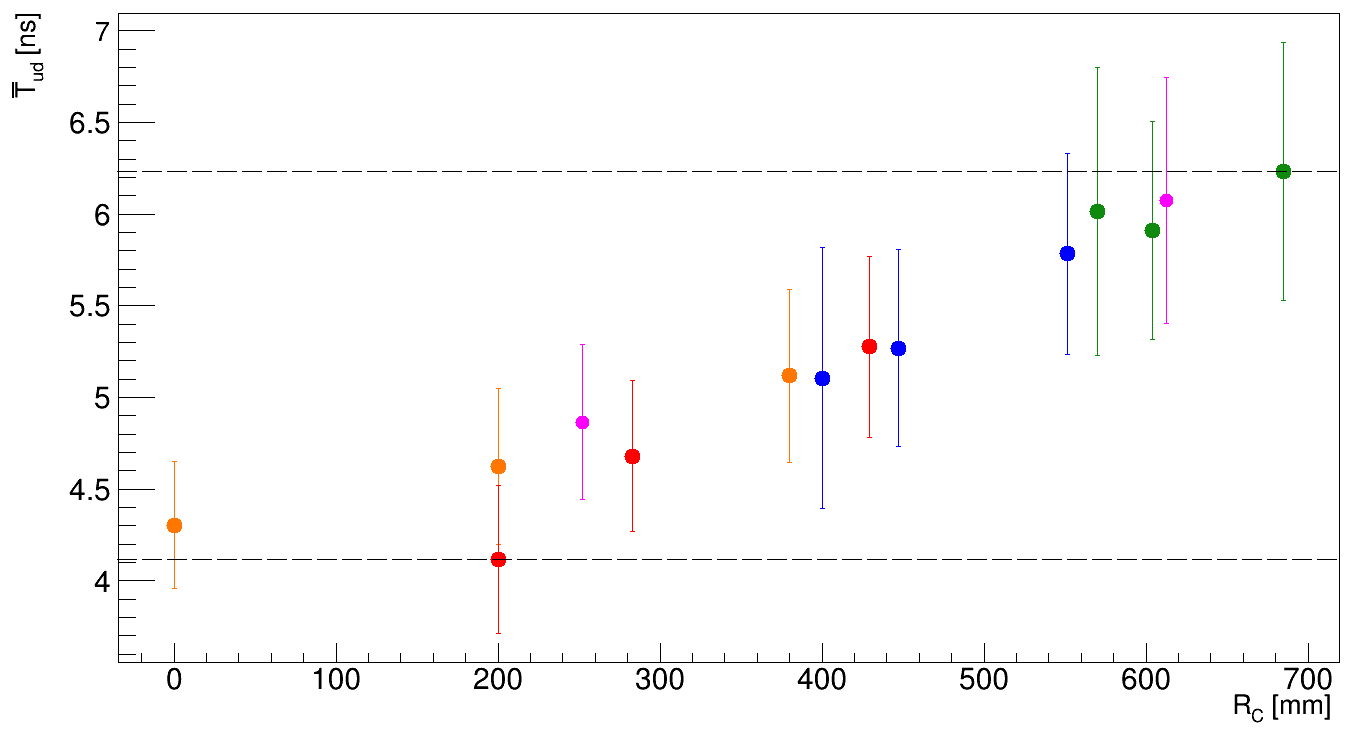}
	\caption{Estimated particle arrival time at the detector $\bar{T}_{ud}$ as a function of distance between particle crossing point and centre of the cell. The \textit{dashed} lines depict the time variation over the whole box.}	
	\label{fig:time_tot}
\end{figure}

For each particle crossing point, the particle arrival time at the detector cell was estimated for each event by calculating the average between the two times: $\bar{T}_{ud} =(T_{u}^{corr} + T_{d}^{corr})/2$.
\begin{figure}[ht]
	\centering
	\includegraphics[width =1.0\textwidth]{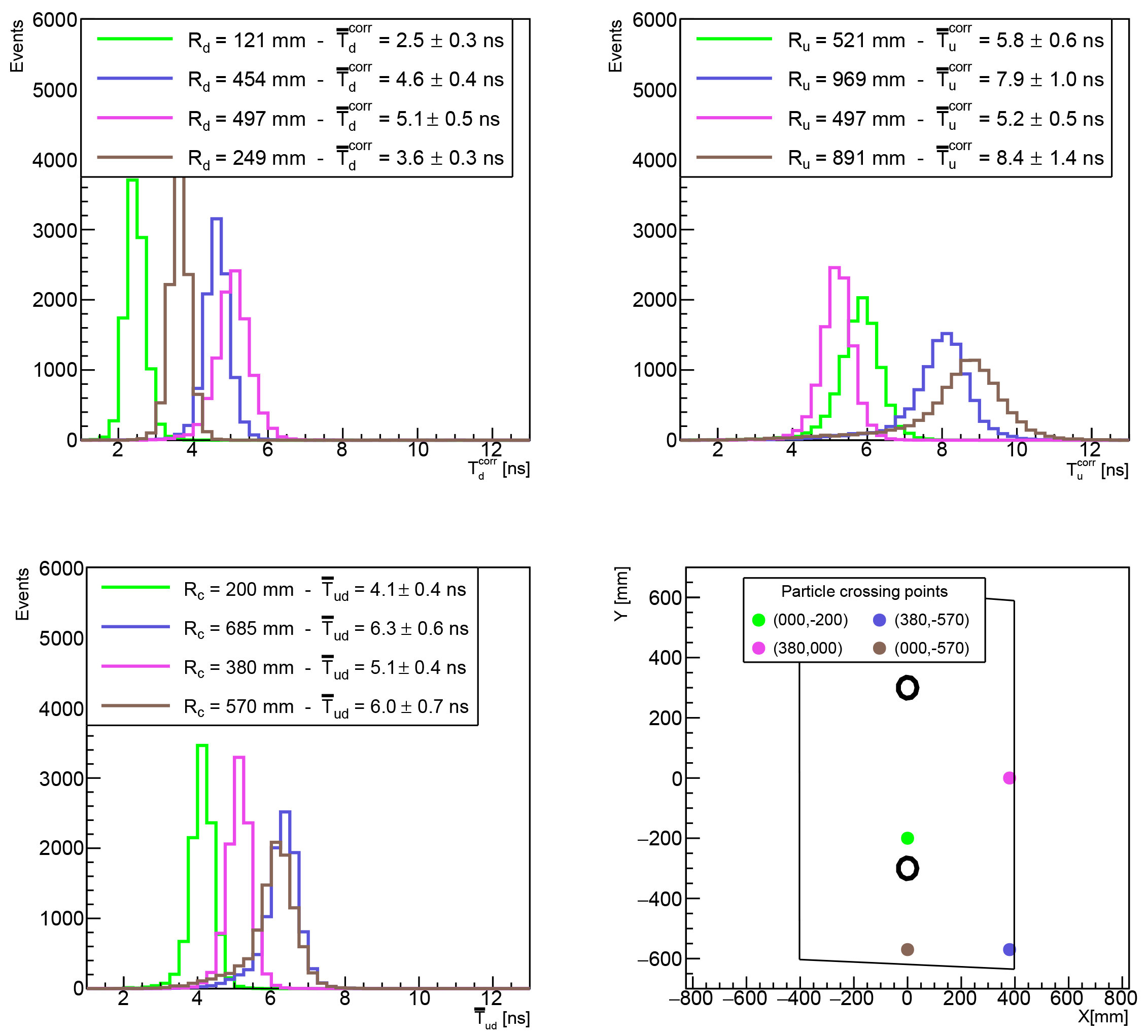}
    \caption{Exemplary distributions of $T_{d}^{corr}$, $T_{u}^{corr}$, and $\bar{T}_{ud}$ for four selected particle crossing points, as shown in \textit{bottom right}. The distribution for each position contains 10000 events.}	
	\label{fig:time_res}
\end{figure}
The mean and the standard deviation of $\bar{T}_{ud}$ for different particle crossing points are shown in Fig.~\ref{fig:time_tot}. $\bar{T}_{ud}$ distributions are shown in Fig.~\ref{fig:time_res} for selected points. As expected, the $\bar{T}_{ud}$ value depends much less on the particle crossing point than the individual times $\bar{T}_{u}^{corr}$ and $\bar{T}_{d}^{corr}$. Over all tested points, $\bar{T}_{ud}$ varies within a 2\,ns time window.

For different positions, the standard deviation of the $\bar{T}_{ud}$ distributions varies between 0.4\,ns and 0.8\,ns. These time uncertainties are significantly larger than the uncertainty on $\bar{T}_{trigger}$ of 0.1\,ns. As a result, we do not correct the standard deviation values of the $\bar{T}_{ud}$ distributions for the uncertainty on $\bar{T}_{trigger}$.

Without any further correction, the intrinsic time resolution of the detector is $\pm{1}$\,ns, due to the time variation in $\bar{T}_{ud}$ if the particle crossing point is unknown, plus a statistical uncertainty of at most 0.8\,ns, which could be improved in the future by increasing the light collection through a higher inner cell wall reflectivity.

\begin{figure}[ht]
	\centering
	\includegraphics[width =0.8\textwidth]{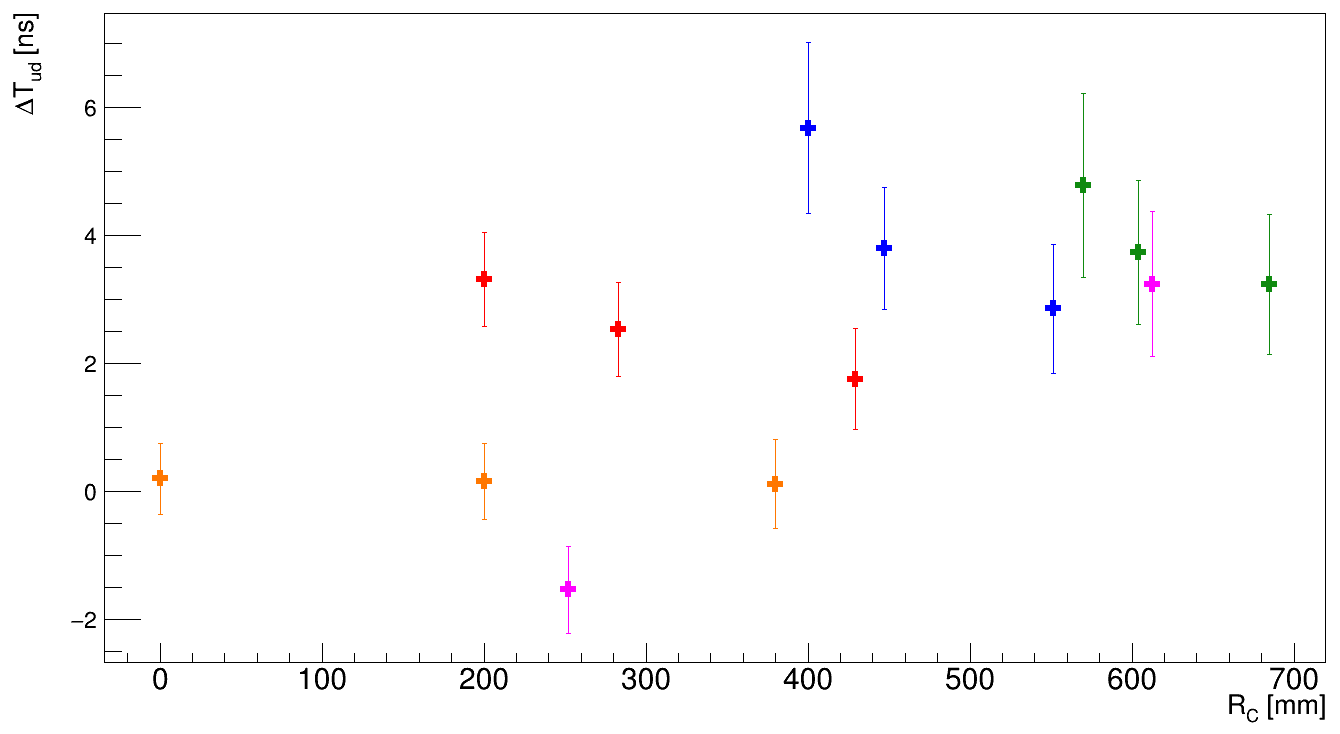}
    \caption{$\Delta\bar{T}_{ud}$ as a function of distance between particle crossing point and centre of the cell.}	
	\label{fig:DeltaT}
\end{figure}

To study the possibility to reduce the intrinsic time variation of about $\pm{1}$\,ns, the difference in arrival time between the two WOMs has been calculated: $\Delta\bar{T}_{ud} =\bar{T}_{u}^{corr} - \bar{T}_{d}^{corr}$. The mean values and standard deviations of $\Delta\bar{T}_{ud}$ for the same particle beam positions considered in Fig.~\ref{fig:time_tot} are shown in Fig.~\ref{fig:DeltaT}.
As expected, the sign of $\Delta\bar{T}_{ud}$ allows one to distinguish easily whether the particle crossing point is in the upper ($y>0$) or in the lower half ($y<0$) of the detector cell. However, there is no easy distinction possible between different $x$ values within the upper or lower half of the cell. A simple correction of $\bar{T}_{ud}$ using only the measured $\Delta\bar{T}_{ud}$ to decrease the $\pm{1}$\,ns variation over the detector cell is not straightforward. Therefore, we study a similar correction strategy as already used in Section \ref{Sec:LightCollectionandDetectorResponse}.

\begin{figure}[ht]
    \centering
    \includegraphics[width = 0.5\textwidth]{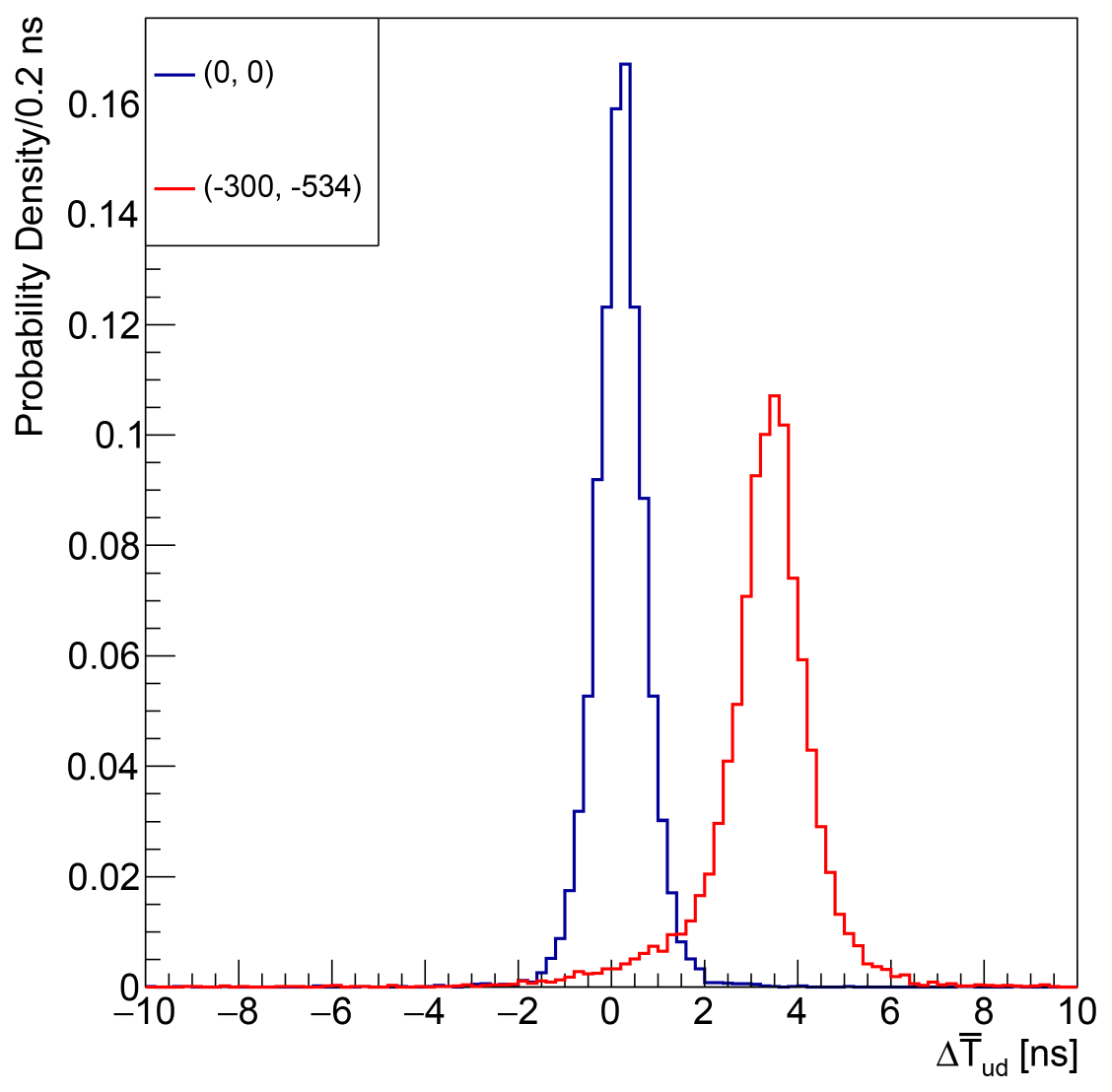}
    \caption{$\Delta\bar{T}_{ud}$ probability distributions for the central particle crossing point and a crossing point in the lower left corner of the detector. $\Delta\bar{T}_{ud}$ is most probable around 0\,ns for the central crossing point, while in the corner crossing point, the highest probabilities are closer to 4\,ns. This adds additional discrimination capability to the likelihood-based position reconstruction.}
    \label{fig:timediff}
\end{figure}

\begin{figure}[ht]
    \centering
    \includegraphics[width = 0.8\textwidth]{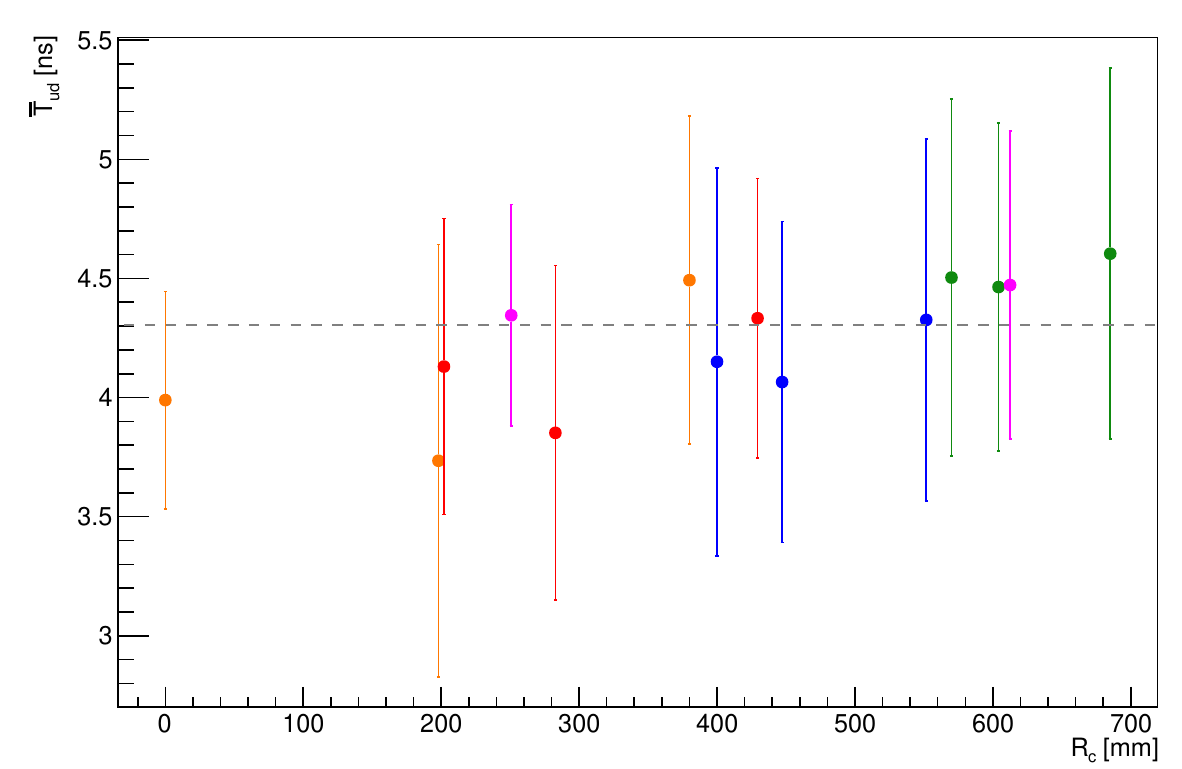}
    \caption{$\bar{T}_{ud}$ distributions after application of the likelihood-based correction for a beam incidence angle of \SI{0}{\degree} and beam energy of 1.4\,GeV. The measured fractions of integrated yield in each WOM and in each SiPM group of each WOM, as well as $\Delta\bar{T}_{ud}$, are used to reconstruct the most likely particle crossing point for the respective event. The error bars show the standard deviations of the distributions. The \textit{dashed} line indicates the reference value used for the correction.}
    \label{fig:likelihood_timeavg}
\end{figure}

\begin{figure}[ht]
    \centering
    \includegraphics[width = 0.5\textwidth]{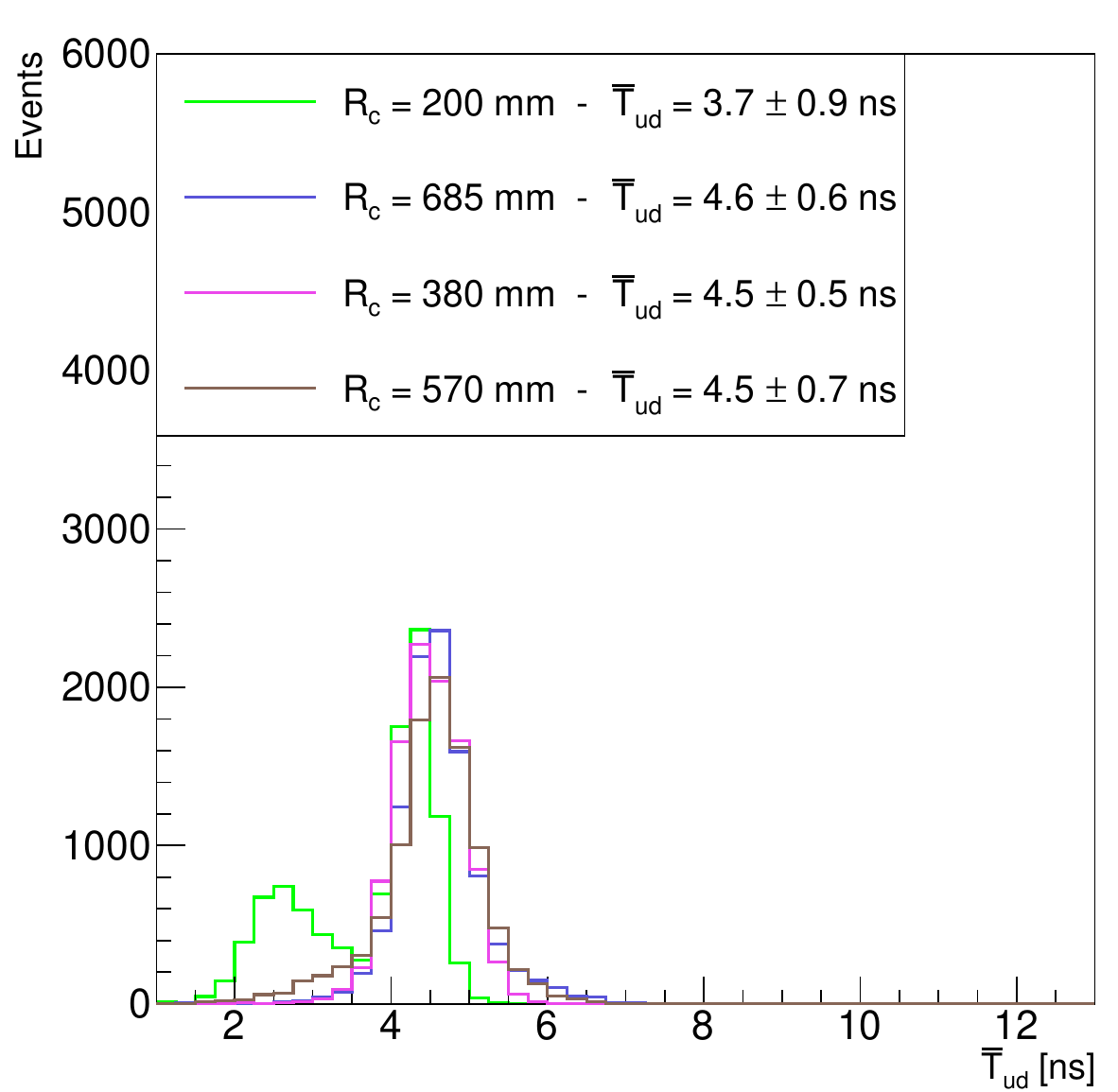}
    \caption{$\bar{T}_{ud}$ distributions for selected points after likelihood-based correction. Colours refer to particle crossing points as indicated in Fig.~\ref{fig:time_res}.}
    \label{fig:likelihood_timeavghist}
\end{figure}

As with the integrated yield in Section~\ref{Sec:LightCollectionandDetectorResponse}, $\bar{T}_{ud}$ was corrected using an event likelihood-based method. The distributions used to perform the correction are the same as before, with the addition of using also the $\Delta\bar{T}_{ud}$ probability densities. Sample probability distributions of $\Delta\bar{T}_{ud}$ for the central crossing point and a crossing point in the corner of the detector are shown in Fig.~\ref{fig:timediff}.
Results of the correction can be seen in Fig.~\ref{fig:likelihood_timeavg}.
The distributions shown in Fig.~\ref{fig:time_res} after the likelihood-based correction are shown in Fig.~\ref{fig:likelihood_timeavghist}.
The corrected distributions lie much closer to $\bar{T}_{ud}$ of the central point, while standard deviations are comparable to their pre-correction values. The variation between $\bar{T}_{ud}$ for the centre of the cell and the farthest away particle crossing point is reduced from about $2$\,ns before correction to $0.4$\,ns. Several points are over corrected such that for a much closer point the largest variation observed is about $-0.7$\,ns in $\bar{T}_{ud}$. One particular example of such a particle crossing point has a $R_c$ value of 200\,mm and is located on the connection line between both WOM tubes (green colour code in Figs.~\ref{fig:time_res}). In about 2/3 of the events, the correction works very well, whereas in 1/3 of the events the correction shifts by about $-1.8$\,ns on average. As with the integrated yield correction, this is expected to improve
significantly with increased integrated yield and with a machine-learning algorithm for reconstruction.

The simulation does not yet include waveforms, so no comparison between data and simulation with respect to timing is possible. Studies are ongoing to implement a waveform into the simulation, which would allow for a more in-depth comparison to data as is currently possible with the light collection response.

\subsection{Reconstruction of the particle crossing point}
\label{Sec:SpatialResolution}

If the WOMs in the LS detector cell register light produced by a charged particle crossing the cell somewhere, the best estimate of the particle crossing point without any further information is the centre of the cell assuming a uniform distribution of the particle crossing point coordinates in $x$ and $y$. As an uncertainty one could then assume the standard deviation of a uniform distribution, which is the cell size in $x$ and $y$ divided by $\sqrt{12}$. For the cell dimension under study, this would correspond to $80/\sqrt{12} \approx23$\,cm for the $x$ coordinate and to $120/\sqrt{12} \approx 35$\,cm. As a result of the positioning of the two WOMs at the same $x$ coordinates but at different $y$ coordinates, the collected light share and the difference in signal arrival time between the two WOMs allows one to easily deduce whether the particle crossing point has a positive or a negative $y$ value with respect to the centre of the cell. Therefore, one would shift the best estimate for $y$ to the $y$ position of the corresponding WOM, and the standard deviation for the $y$ coordinate would be divided by a factor of $2$ resulting in about 17\,cm. As already discussed and shown in Section \ref{Sec:LightCollectionandDetectorResponse}, the light yield share between the two WOMs together with the measured photoelectron distribution over the eight SiPM groups caries much more information about the particle crossing coordinates using event likelihoods.

We present in the following a neural-network (NN) based analysis that uses the individual integrated yields and signal arrival times in the eight SiPM groups of each WOM to estimate the $x$\ and $y$\ positions of the positron beam hitting the cell. The method significantly improves the bias in the reconstructed position and its estimated uncertainty beyond the likelihood method applied in Section \ref{Sec:LightCollectionandDetectorResponse}.

The neural network utilises integrated yields of each SiPM group and timing information as the input layer~\cite{Jadidi_2023}. Each basic neural network consists of an input layer with 32 neurons, followed by three hidden layers, each with Rectified Linear Unit (ReLU) activation functions, batch normalisation layers, and dropout procedures to prevent overfitting \cite{Mehta_2019}. The architecture embraces a deep ensemble model strategy \cite{Ganaie_2022}, where three identical basic networks are trained, and their outputs are concatenated and averaged to reduce bias induced by random initiating. A four-fold cross-validation approach~\cite{scikit-learn} is employed to enhance model robustness and minimise overfitting, which partitions the dataset into training, validation, and test sets. A dataset with 10,000 events when the beam was set to 5.4\,GeV has been used to train the network. The test set contains $20\,\%$ of the data and is excluded from training the network entirely. The rest of the dataset is used for training and validating the model. Also, other datasets with different beam energies can play the test set role to examine the model performance. In total, 12 basic models are trained through this scheme. Hyperparameter tuning, vital for model optimisation, is carried out using the random search method, seeking the best configuration to minimise the Mean Absolute Error (MAE), which serves as the loss function for training the neural network. One note is that there are two different models; one is trained and tested with MC simulation data, and the other is trained with the test beam data. In this project, Keras, which is an open-source Python package, has been used. Keras wraps the functionality of another package, TensorFlow \cite{abadi2016tensorflow}.

\begin{figure}
    \centering
    \includegraphics[width=0.45\textwidth]{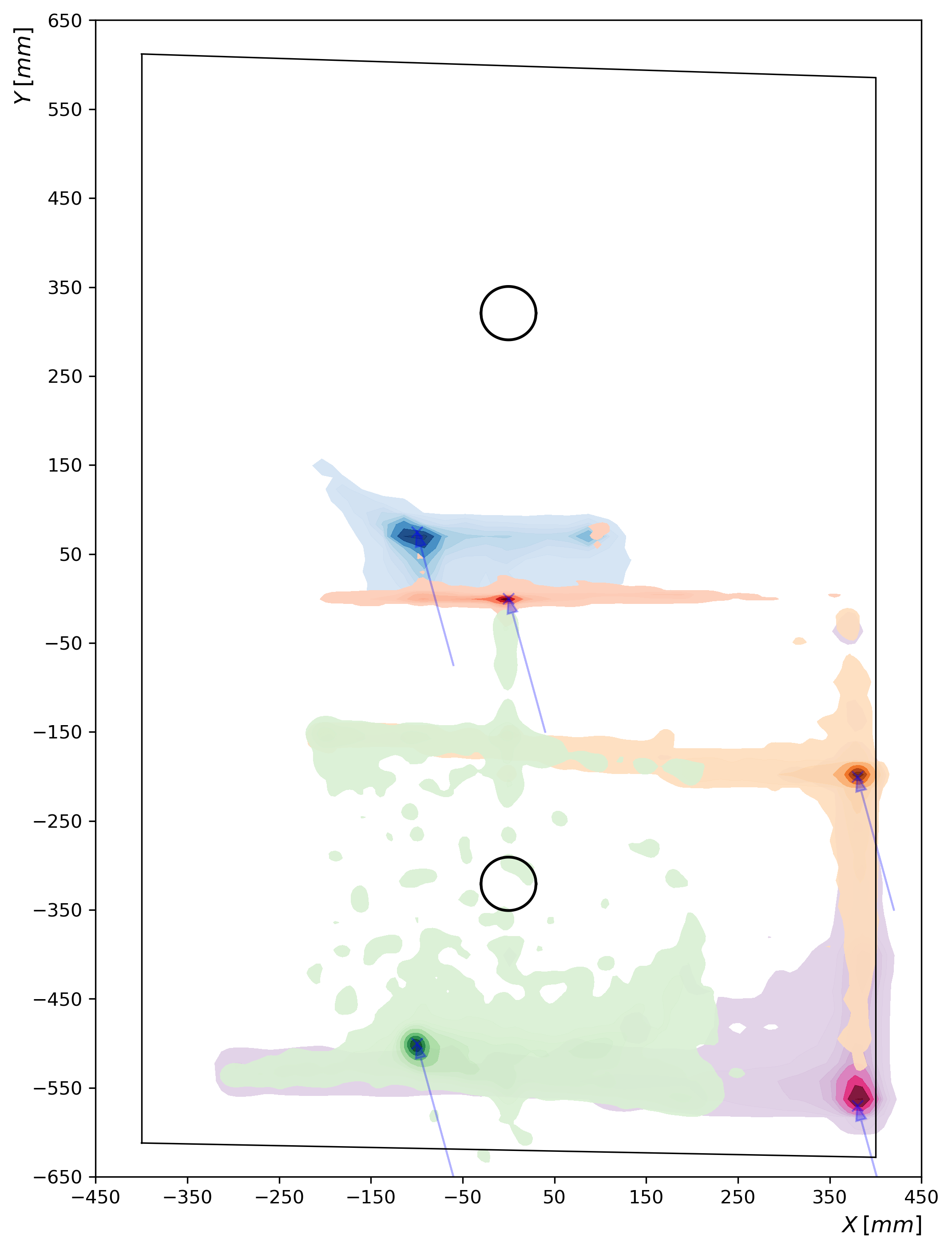}
    \hspace{5mm}
    \includegraphics[width=0.45\textwidth]{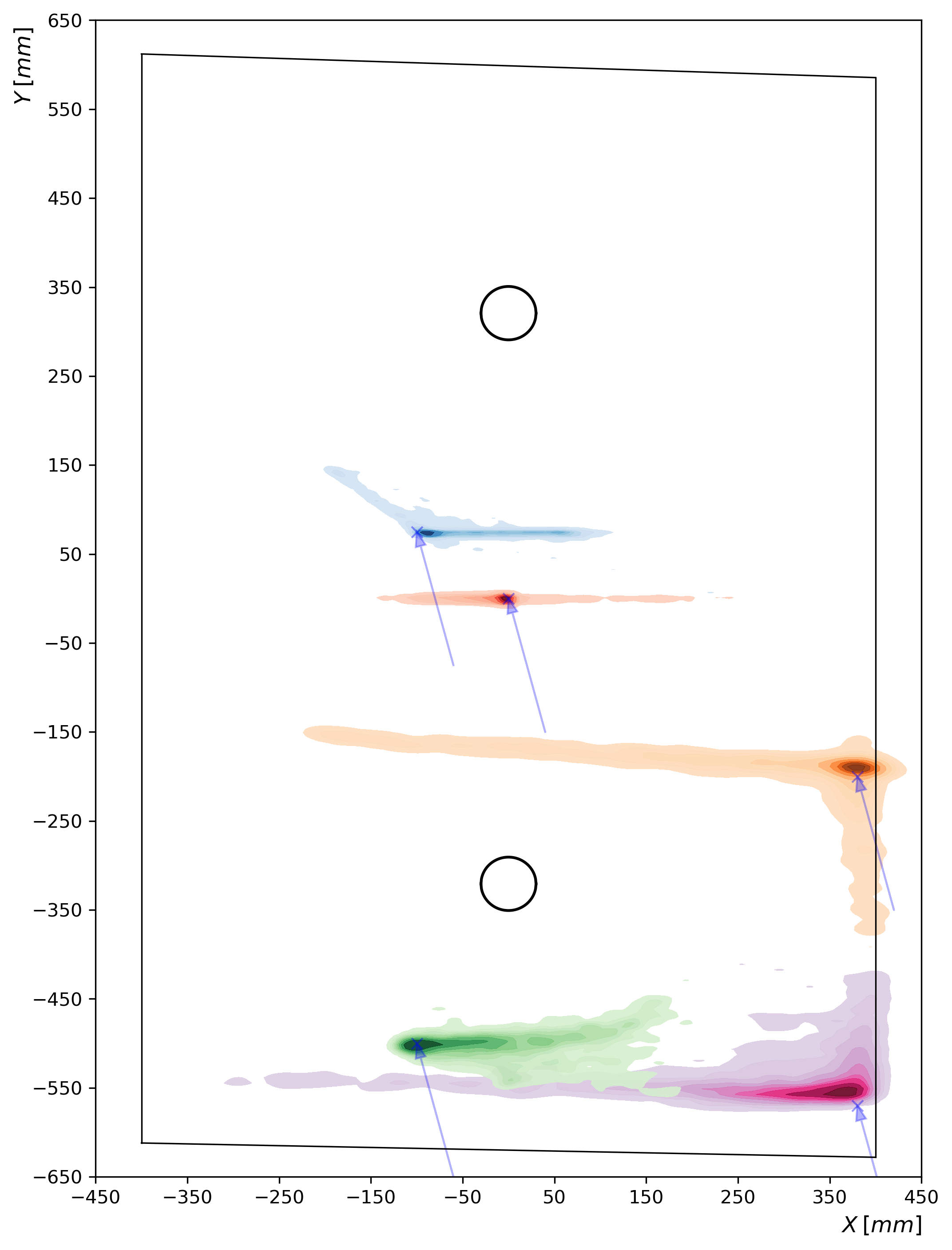}\\
    \vspace{2mm}
    \includegraphics[width=0.45\textwidth]{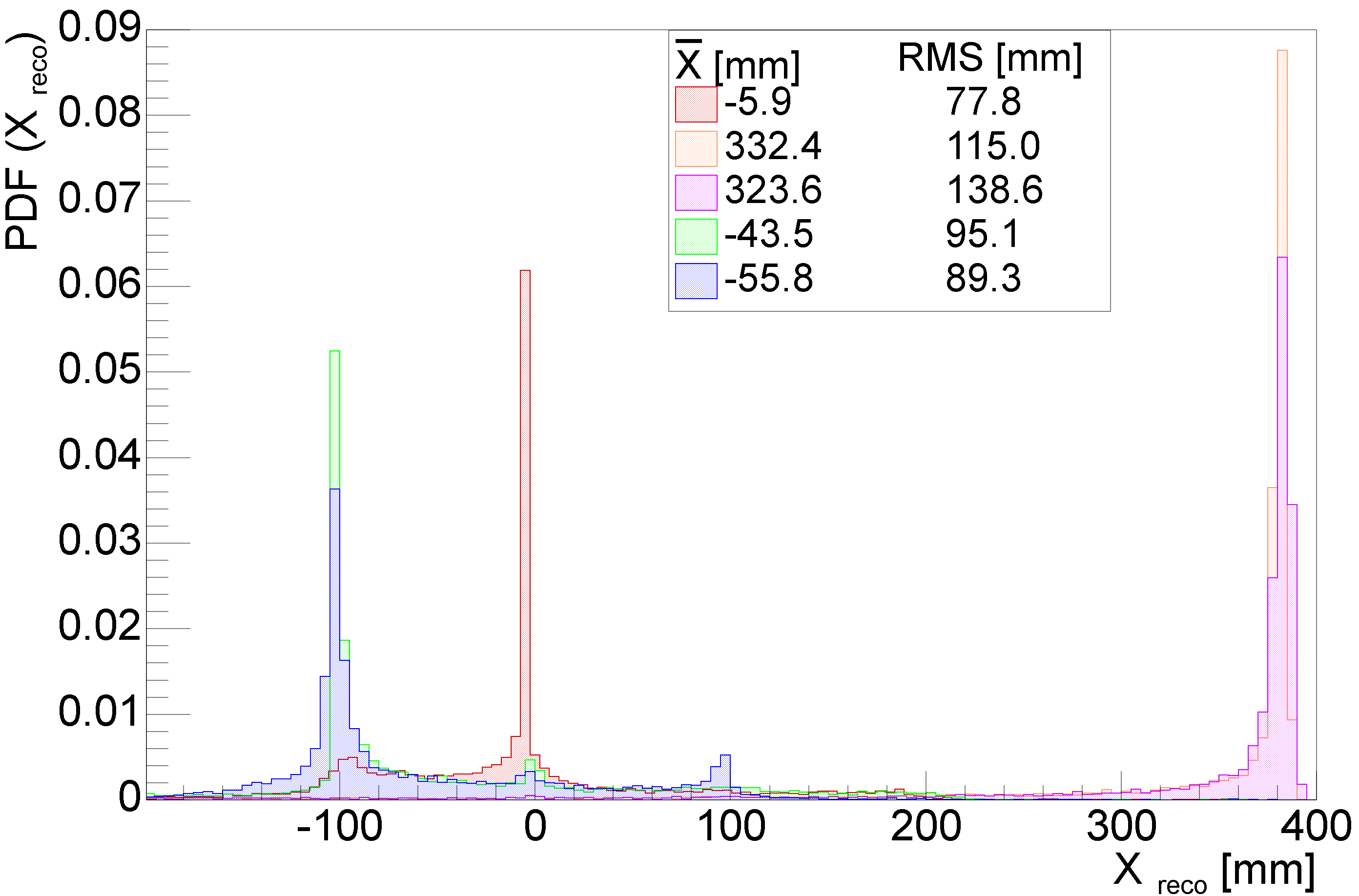}
    \hspace{5mm}
    \includegraphics[width=0.45\textwidth]{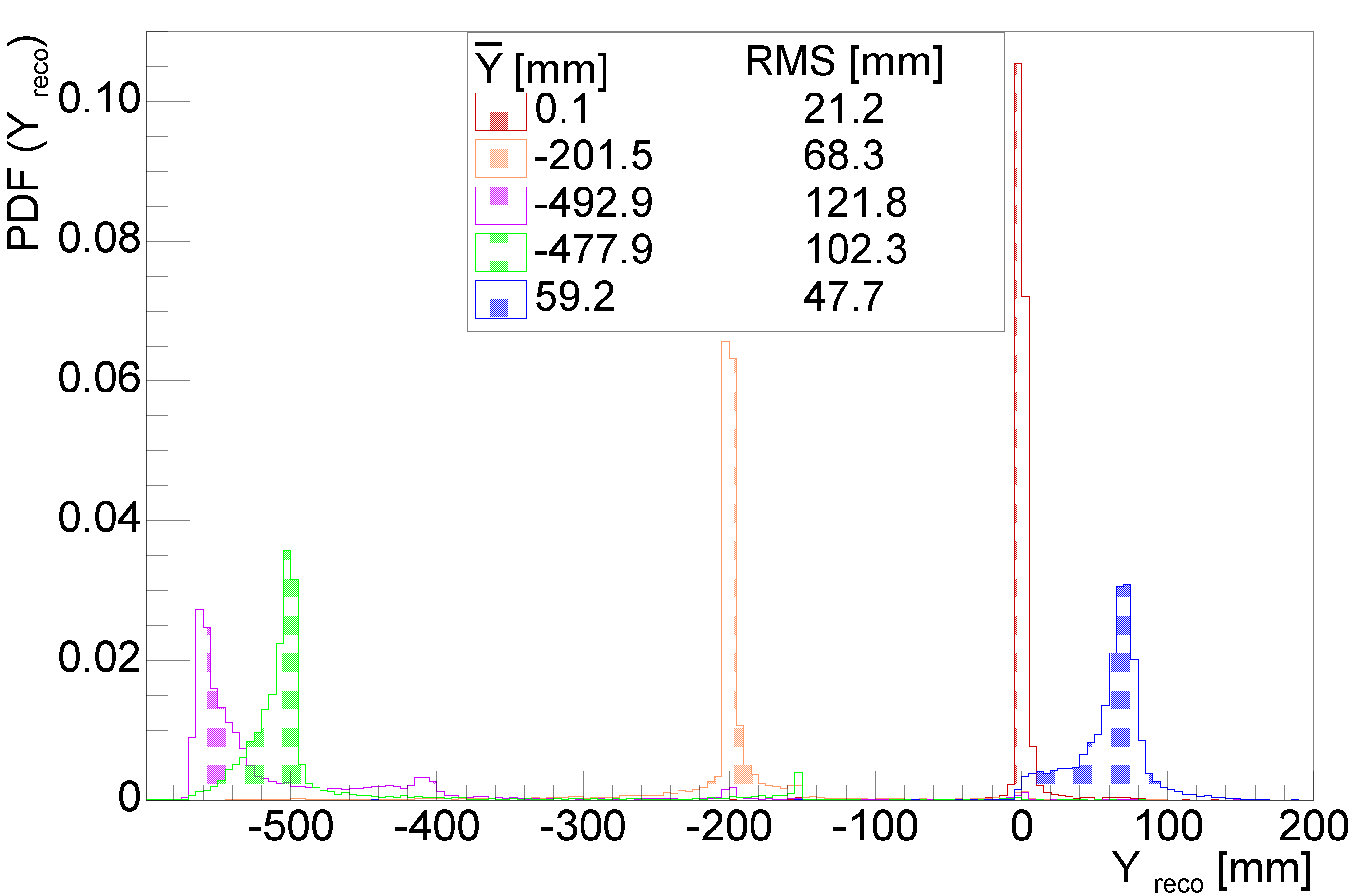}
    \caption{Neural network-based reconstruction of five selected particle crossing points with a beam angle of \SI{0}{\degree} and energy of 1.4\,GeV. The coloured areas in the \textit{top} plots show the probability density of the reconstructed $x$ and $y$ coordinate values, with the true positions indicated by the \textit{blue} arrows. \textit{Top left:} Test beam data. \textit{Top right:} MC simulation. \textit{Bottom left:} Reconstructed $x$ position in test beam data. \textit{Bottom right:} Reconstructed $y$ position in test beam data. The mean and the standard deviation of the estimated coordinates are given in the legend.}
    \label{fig:SpatialReconstruction}
\end{figure}

The \textit{top} plots in Fig.~\ref{fig:SpatialReconstruction} show the two-dimensional positions obtained from the NN analysis for five different particle crossing points with a beam angle of \SI{0}{\degree} and energy of 1.4\,GeV for test beam data (\textit{left}) and MC simulation (\textit{right}). The \textit{bottom} plots in Fig.~\ref{fig:SpatialReconstruction} show the projections of the reconstructed $x$ and $y$ positions in the test beam data.
We observe that the probability density function of the reconstructed positions peak very closely to the true particle crossing point and that the standard deviation for the reconstructed $x$ and $y$ position is at most 14\,cm and hence significantly smaller than the above quoted values for the standard deviations for a uniform distribution. At higher positron energies, these uncertainties are further reduced thanks to the higher integrated yields in the WOMs. While the reconstructed points in the data show for some of the particle crossing point a larger spread than in the simulation, the central part of the reconstructed $x$ distribution is more peaking in data than in the simulation. This is likely coming from the uneven distribution of rust stains on the inner cell walls, which allows the NN to better distinguish between different particle crossing points.

\section{Summary and outlook}
\label{Sec:Summary}

As a prototype detector for the SHiP Surrounding Background Tagger, we constructed a cell (dimensions $120\, \mathrm{cm} \times 80\, \mathrm{cm} \times 25\, \mathrm{cm}$) made from corten steel and filled with a LS made of LAB and PPO that is equipped with two WOMs for light collection of the primary scintillation photons. Each WOM consists of a PMMA tube dip-coated with WLS on its surface, employing an optimised dye and dip-coating procedure. The secondary photons emitted from the WLS are guided via internal total reflection to a ring-shaped array of 40 SiPMs for light detection. The readout of the 40 SiPMs of each WOM is combined into eight channels of five SiPMs each.

The detector cell was tested at the DESY II test beam facility with positrons of energies varying between 1.4\,GeV and 5.4\,GeV. Compared to previous test beam exposures with earlier detector prototypes, the performance of the detector could be significantly improved in several aspects: Using only one WOM, the detection efficiency over the whole detector cell was found to be $\geq$\SI{99.3}{\percent} at \SI{68}{\percent} confidence level, and $\geq$\SI{99.5}{\percent} at \SI{68}{\percent} confidence level when requiring at least one of the two WOMs to register a signal above the dark-count threshold. 
The granularity of the SiPM array coupled to the WOM tube furthermore allows for an innovative approach to gain spatial information on the particle crossing point: After applying a likelihood-based correction estimating the most likely particle crossing point, the energy response over the whole detector cell varies between $+\SI{25}{\percent}$ and $-\SI{10}{\percent}$. Using the signal arrival times in the SiPM arrays coupled to both WOMs, a time resolution of $\mathcal{O}$(1\,ns) could be achieved, also profiting from a likelihood-based correction. A neural network-based analysis taking into account the time response and collected integrated yields in the SiPM readout channels of the two WOMs allowed to reconstruct the particle crossing point with an uncertainty of better than 14\,cm. 

The cell wall reflectivity can still be improved, currently achieving only about \SI{65}{\percent} of the optimal value due to unforeseen rust stains. This problem will be avoided in the future by the application of a dedicated anti-rust primer, which is expected to significantly increase the collected photon yield.

The next step in our R\&D program for the SHiP LS-SBT will be the construction and test of a multi-cell detector comprising $2 \times 2$ cells, studying muons in a dedicated test beam campaign. With such a detector, further information can be inferred w.r.t. the particle crossing point and even its incidence angle when it traverses more than one cell.

\section*{Acknowledgements}
The measurements leading to these results have been performed at the DESY II test beam facility at DESY Hamburg (Germany), a member of the Helmholtz Association (HGF). 
We acknowledge the fruitful discussions with all the individuals from JGU Mainz and DESY involved in establishing the WOM as a sensor for IceCube.
We thank E.~List-Kratochvil, A.~Opitz and, T.~Florian from the HU Berlin Institute of Physics for discussions on methods for measuring and increasing the WOM coating layer thickness and for providing access to the transmission spectrometer and the profilometer for characterisation of coated glass and PMMA slides.
We thank Andrea Miano (University of Naples Federico II) for the figures showing the schematics of the decay vessel structural design.
We are grateful for the skill and effort of the technicians of our collaborating institutions.
We thank the Deutsche Forschungsgemeinschaft (DFG) for funding support within grant 289921825 and the Bundesministerium für Bildung und Forschung (BMBF) for funding support within the High-D consortium.


\begin{thebibliography}{99}

\bibitem{SHiP:2015vad}
M.~Anelli \textit{et al.} [SHiP], ``A facility to Search for Hidden Particles (SHiP) at the CERN SPS,''
arXiv:1504.04956 [physics.ins-det].

\bibitem{SHIP:2021tpn}
C.~Ahdida \textit{et al.} [SHiP], ``The SHiP experiment at the proposed CERN SPS Beam Dump Facility,'' Eur. Phys. J. C \textbf{82} (2022), doi:10.1140/epjc/s10052-022-10346-5.

\bibitem{SHiP-LoI}
C.~Ahdida \textit{et al.} [SHiP], ``BDF/SHiP at the ECN3 high-intensity beam facility (Letter of Intent)," CERN-SPSC-2022-032 SPSC-I-256.

\bibitem{Bastian-Querner:2021uqv}
B.~Bastian-Querner \textit{et al.}, ``The Wavelength-Shifting Optical Module,'' Sensors \textbf{22} (2022) no. 4, doi:10.3390/s22041385  .

\bibitem{Hebecker:2016mrq}
D.~Hebecker \textit{et al.}, ``A Wavelength-shifting Optical Module (WOM) for in-ice neutrino detectors,''
EPJ Web Conf. \textbf{116} (2016), doi:10.1051/epjconf/201611601006 .

\bibitem{IceCube:2021mdb}
J.~Rack-Helleis \textit{et al.} [IceCube], ``The Wavelength-shifting Optical Module (WOM) for the IceCube Upgrade,'' PoS \textbf{ICRC2021} (2021), doi:10.22323/1.395.1038  .

\bibitem{Ehlert:2018pke}
M.~Ehlert \textit{et al.}, ``Proof-of-principle measurements with a liquid-scintillator detector using wavelength-shifting optical modules,'' JINST \textbf{14} (2019) no. 3, doi:10.1088/1748-0221/14/03/P03021 .

\bibitem{Diener:2018qap}
R.~Diener \textit{et al.}, ``The DESY II Test Beam Facility,'' Nucl. Instrum. Meth. A \textbf{922} (2019), doi:10.1016/j .nima.2018.11.133.

\bibitem{KamLAND}
K.~Eguchi \textit{et al.} [KamLAND], ``First results from KamLAND: Evidence for reactor anti-neutrino disappearance,'' Phys. Rev. Lett. \textbf{90} (2003), doi:10.1103/PhysRevLett.90.021802 .

\bibitem{Borexino}
G.~Alimonti \textit{et al.} [Borexino], ``The Borexino detector at the Laboratori Nazionali del Gran Sasso,'' Nucl. Instrum. Meth. A \textbf{600} (2009), doi:10.1016/j.nima.2008.11.076 .

\bibitem{JUNO}
Z.~Djurcic \textit{et al.} [JUNO], ``JUNO Conceptual Design Report,'' arXiv:1508.07166  [physics.ins-det].

\bibitem{SnoPlus}
V.~Albanese \textit{et al.} [SNO+], ``The SNO+ experiment,'' JINST \textbf{16} (2021) no. 8, doi:10.1088/1748-0221/16/08/P08059 .

\bibitem{Agostinelli}
S.~Agostinelli \textit{et al.}, ``Geant4 - a simulation toolkit,'' Nucl. Instrum. Meth. A \textbf{506} (2003), doi:10.1016/S0168-9002(03)01368-8 .

\bibitem{JunoColumn}
Z.~Zhu \textit{et al.}, ``Optical purification pilot plant for JUNO liquid scintillator,'' Nucl. Instrum. Meth. A \textbf{1048} (2023), doi:10.1016/j.nima.2022.167890 .

\bibitem{Schmidt}
J. P.~Schmidt, Studies of Dip-Coating Methods to improve the Light Absorption in Wavelength-Shifting Optical Modules, Master thesis, HU Berlin (2022).

\bibitem{Brinker}
C. J.~Brinker, "Dip coating," in Chemical Solution Deposition of Functional Oxide Thin Films, Springer Vienna, 2013, doi:10.1007/978-3-211-99311-8 \_10.

\bibitem{Hamamatsu-S13360}
https://www.hamamatsu.com/content/dam/hamamatsu-photonics/sites/documents/99\_SALES\_LIBRARY/ssd/s13360\_series\_kapd1052e.pdf.

\bibitem{Hamamatsu-S14160}
https://www.hamamatsu.com/content/dam/hamamatsu-photonics/sites/documents/99\_SALES\_LIBRARY/ssd/s14160\_s14161\_series\_kapd1064e.pdf.
 
\bibitem{MUSIC-ASIC}
S.~G\'{o}mez {\it et al.}, ``MUSIC: An 8 channel readout ASIC for SiPM arrays'', in Proceedings, Optical Sensing and Detection IV, SPIE Photonics Europe, 2016, Brussels, Belgium, vol. 9899, doi:/10.1117/12.2231095 .
 
\bibitem{WaveCatcher}
D.~Breton and J.\,Maalmi, ``WaveCatcher Family User’s Manual,'' \linebreak
URL: https://www.hep.ucl.ac.uk/pbt/wikiData/manuals/WaveCatcher/WaveCatcherFamily$\_$V1.2.pdf

\bibitem{Ernst}
A.~Ernst, Study of the position-dependent detector response of a liquid- scintillator detector instrumented with WOMs and SiPMs using cosmic muons, Bachelor thesis, HU Berlin (2021).

\bibitem{Vagts}
A.~Vagts, Improvement of light yield and spatial resolution of a liquid-scintillator detector equipped with a wavelength-shifting optical module coupled to a SiPM array, Master thesis, HU Berlin (2022).

\bibitem{Eckardt}
C.~Eckardt, Effect of optical coupling between silicon-photomultiplers and a wavelength-shifting optical module in a scintillator detector on angular resolution and light collection, Bachelor thesis, HU Berlin (2023).

\bibitem{Nichia_LED}
Nichia Corporation, ``Specifications for UV LED NSPU510CS''\linebreak
URL: https://led-ld.nichia.co.jp/api/data/spec/led/NSPU510CS-E(2745H)Ubx\%20Improvement.pdf

\bibitem{Alt_2023}
J.~Alt, Readout of Wavelength-shifting Optical Modules, Master thesis, ALU Freiburg (2023).

\bibitem{Tseung}
H.~Wan Chan Tseung and N.~Tolich, ``Ellipsometric measurements of the refractive indices of linear alkylbenzene and EJ-301 scintillators from 210 to 1000 nm,'' Phys. Scr. \textbf{84} (2011), doi:10.1088/0031-8949/84/03/035701 .

\bibitem{Anderson}
M. R.~Anderson \textit{et al.} [SNO+], ``Development, characterisation, and deployment of the SNO+ liquid scintillator,'' JINST \textbf{16} (2021), doi:10.1088/1748-0221/16/05/P05009 .

\bibitem{Marrodan}
T.~Marrod\'{a}n Undagoitia \textit{et al.}, ``Spectroscopy of electron-induced fluorescence in organic liquid scintillators,'' Eur. Phys. J. D \textbf{105} (2010), doi: 10.1140/epjd/e2010-00004-1.

\bibitem{Onken}
D. R.~Onken \textit{et al.}, ``Time response of water-based liquid scintillator from X-ray excitation,'' Mater. Adv. \textbf{1} (2020), doi:10.1039/D0MA00055H .

\bibitem{Sultanova}
N.~Sultanovaa, S.~Kasarovaa and I.~Nikolovb, ``Dispersion Properties of Optical Polymers,'' Acta Physica Polonica A \textbf{116} (2009), doi:10.12693/APhysPolA.116.585.

\bibitem{Bodmer}
M.~Bodmer \textit{et al.}, ``Measurement of Optical Attenuation in Acrylic Light Guides for a Dark Matter Detector,'' JINST \textbf{9} (2014), doi:10.1088/1748-0221/9/02/P02002 .

\bibitem{Clopper-Pearson}
 C.~Clopper and E. S.~Pearson, ``The use of confidence or fiducial limits illustrated in the case of the binomial,'' Biometrika \textbf{6} (1934) no. 4, doi:10.1093/biomet/26.4.404.
 
\bibitem{Jadidi_2023}
M.~Jadidi, Reconstruction of Spatial and Timing Information with the one-cell SBT Prototype using Neural Networks, Master thesis, ALU Freiburg (2023).

\bibitem{Mehta_2019}
P.~Mehta, \textit{et al.}, ``A high-bias, low-variance introduction to Machine Learning for physicists,'' Physics Reports \textbf{810} (2019), doi:10.1016/j.physrep.2019.03.001.

\bibitem{Ganaie_2022}
M. A.~Ganaie, \textit{et al.}, ``Ensemble deep learning: A review,'' Eng. App. AI \textbf{115} (2022), doi:10.1016/j.engappai.2022.105151.

\bibitem{scikit-learn}
F.~Pedregosa \textit{et al.}, ``Scikit-learn: Machine Learning in {P}ython,'' J. Mach. Learn. Res. \textbf{12} (2011), doi:10.5555/1953048.2078195.

\bibitem{abadi2016tensorflow}
M.~Abadi, \textit{et al.}, ``TensorFlow: Large-Scale Machine Learning on Heterogeneous Distributed Systems,'' (2016), doi:10.48550/arXiv.1605.08695, \\ Software available from tensorflow.org.

\end{thebibliography}
\end{document}